%% file: halo_excess_r2.tex
\documentclass[aps,prd,twocolumn,groupedaddress,nofootinbib,floatfix,longbibliography]{revtex4-2}

\usepackage[T1]{fontenc}
\usepackage[utf8]{inputenc}
\usepackage{amsmath, amssymb, amsfonts}
\usepackage{graphicx}
\usepackage{xcolor}
\usepackage[colorlinks=true, citecolor=blue, linkcolor=blue, urlcolor=blue]{hyperref}
\usepackage{bm}
\numberwithin{equation}{section}

\newcommand{\dd}{\mathrm{d}}
\newcommand{\sigv}{\langle\sigma v\rangle}
\newcommand{\mchi}{m_\chi}
\newcommand{\bbar}{b\bar{b}}
\newcommand{\WW}{W^+W^-}
\newcommand{\tautau}{\tau^+\tau^-}
\newcommand{\rhosq}{\rho^2}
\newcommand{\Jpole}{\mathcal{J}_{\rm pole}}
\newcommand{\degree}{^\circ}
\graphicspath{{figures/}}
\newcommand{\Rlo}{2.5}          %
\newcommand{\Rhi}{8.5}          %
\newcommand{\Rbest}{5.2}        %
\newcommand{\RWW}{4.2}          %
\newcommand{\RWWlo}{2.0}        %
\newcommand{\RWWhi}{6.8}        %
\newcommand{\boostbase}{45}     %
\newcommand{\boostlo}{9}        %
\newcommand{\boosthi}{\rtwo{47}}  %
\newif\ifshowstatus \showstatusfalse
\newcommand{\statusbox}[2]{}
\newcommand{\secfinal}{}
\newcommand{\abswaits}[1]{}
\newcommand{\secwaits}[1]{}
\newif\ifmarkchanges
\markchangesfalse
\ifmarkchanges
  \newcommand{\rtwo}[1]{{\color{red}#1}}
  \newenvironment{rtwoblock}{\par\color{red}}{\par}
\else
  \newcommand{\rtwo}[1]{#1}
  \newenvironment{rtwoblock}{\par}{\par}
\fi
\newif\ifshowlatest
\showlatestfalse
\newcommand{\rlatest}[1]{\ifshowlatest{\color{red}#1}\else#1\fi}

\begin{document}

\title{The 20~GeV Galactic Excess at Pixel Level: What It Is, What It Is Not, and What Dark Matter Would Need}
\author{T.~R.~Stenhouse}
\email{trinity.stenhouse.24@ucl.ac.uk}
\author{C.~Ghag}
\author{F.~F.~Deppisch}
\affiliation{Department of Physics and Astronomy, University College London,
Gower Street, London WC1E 6BT, United Kingdom}
\date{\today}

\begin{abstract}
\rtwo{A 20~GeV gamma-ray excess at high Galactic latitudes, reported by Totani as a halo and read as dark matter
annihilation, has been linked to the high-energy nuclear recoil in LUX-ZEPLIN. We reanalyse 17 years of
\textit{Fermi}-LAT data in $0.125\degree$ pixels, across two-dimensional GALPROP foregrounds and two
models of the \textit{Fermi} bubbles. A centrally concentrated excess near 20~GeV appears in every configuration.
Mismatch between the GALPROP foregrounds used to generate mocks does not produce it, and the injected cosmic-ray
electron populations we test do not replace it. It is not, however, a spherical halo alone, since the emission
brightens and softens away from the plane and lies largely inside the bubble outline. On the baseline foreground it is
described by a spherical $\rho^2$ component peaking at 20.8~GeV and a flat-spectrum component filling the bubbles, and
on its own a bubble-interior component fitted without a halo induces about half of the halo signal over 10--40~GeV.
Replacing our bubble templates with those of Cholis, Hu and Zhong lowers the halo amplitude by about 40\% on the
baseline, and a sky built from their templates without a halo reproduces most of the signal. Read as prompt
annihilation, the excess requires a particle of a few hundred GeV annihilating at \boostlo{} to \boosthi{} times the
thermal rate, \Rlo{} to \Rhi{} times above dwarf spheroidal limits for non-informative dwarf priors, and no route we consider is consistent with the
dwarf, relic-abundance, CMB and AMS-02 antiproton constraints together. The foreground range is a lower bound,
obtained with a fixed point-source catalogue and two-dimensional propagation models, so the origin of the excess
remains open. With the Cholis, Hu and Zhong templates the excess is not distinguishable from foreground mismatch, and a
dark matter reading would apply at most to the spherical component, at a rate in tension with the dwarf and
antiproton limits.}
\end{abstract}

\maketitle

\section{Introduction}
\label{sec:introduction}
\secfinal

Weakly interacting massive particles (WIMPs) are among the best-motivated dark
matter candidates. A stable particle with electroweak-scale mass ($10$~GeV--few~TeV)
and weak-scale annihilation cross section is naturally produced as a thermal relic
with the observed dark matter abundance, the so-called WIMP
miracle~\cite{BertoneHooperSilk2005, Steigman2012}. Such particles annihilate in
pairs to Standard Model final states, and in dark matter--dense regions the
resulting gamma-ray flux traces the square of the dark matter density. The inner
Galaxy, the smooth Galactic halo, and dwarf spheroidal galaxies are therefore the
primary targets for indirect dark matter searches~\cite{Feng2010}.

The inner Galaxy is the brightest of these targets and the hardest to model, and it
hosts large-scale non-thermal structure whose history is a direct warning about
featureless residuals. Template regression of WMAP data revealed a centrally
concentrated haze with an anomalously hard spectrum~\cite{Finkbeiner2004,
DoblerFinkbeiner2008}, later confirmed by \textit{Planck}~\cite{PlanckHaze2013}, and
the corresponding gamma-ray residual, the \textit{Fermi} haze, was reported as a
smooth, hard, quasi-elliptical component extending to
$|b|\simeq50\degree$~\cite{Dobler2010}. Su et al.~\cite{Su2010} then applied improved diffuse modelling and resolved that
apparently featureless component into the sharp-edged \textit{Fermi} bubbles, a
pair of hard-spectrum lobes extending above and below the Galactic Centre. The
bubbles were subsequently measured to have an edge width of $3.4\degree$ and a
spectrum well described by a power law of index $1.9\pm0.2$ with an exponential
cutoff at $110\pm50$~GeV, showing no \rtwo{detected} spectral variation above
$|b|=10\degree$~\cite{Ackermann2014}.  The
bubbles are now understood as the inner region of a larger outflow traced in soft
X-rays by eROSITA~\cite{Predehl2020, Sarkar2024}. The precedent is not confined to
the haze. Dixon et al.~\cite{Dixon1998} reported a statistically significant \rtwo{extended} gamma-ray halo in EGRET data nearly three decades ago, \rtwo{and a diffuse model with a larger inverse-Compton halo, consistent with that indication, was later found to reproduce the high-latitude emission~\cite{Strong2000}}. The apparent smoothness of a template-subtracted residual is therefore not by itself evidence of a \rtwo{new emission component}. \rtwo{We
address that concern by fitting at the native pixel scale (Sec.~\ref{sec:global}), which removes the smoothing introduced
by aggregating into cells, and by varying the foreground model across the range of Sec.~\ref{sec:envres}, which is where
the diffuse-model question is tested.}

Gamma-ray observations of the inner Galaxy have for over a decade shown an excess
of $\sim$few-GeV emission above diffuse-background models, the Galactic Centre
Excess (GCE)~\cite{GoodenoughHooper2009, HooperGoodenough2011, HooperLinden2011,
AbazajianKaplinghat2012}. That excess was shown early on to extend well beyond the
innermost degrees. Hooper and Slatyer~\cite{HooperSlatyer2013} divided the
\textit{Fermi} bubbles into latitude bands, \rtwo{reporting a component below $|b|\simeq20\degree$ that, they argue,} no realistic electron spectrum can produce by inverse-Compton (IC) scattering. They described it as approximately spherically
symmetric about the Galactic Centre, with a luminosity per volume consistent with
the square of a generalised Navarro--Frenk--White (NFW) density
profile~\cite{NFW1996}. Daylan et al.~\cite{Daylan2016} scanned the axis ratio
explicitly and found a strong preference for near-unit ellipticity out to at least
$10\degree$. Calore, Cholis and Weniger~\cite{Calore2015} placed a lower limit of
$10.0\degree$ ($95\%$ CL) on the extension under a scan of 60 diffuse models
generated with the GALPROP cosmic-ray propagation code~\cite{GALPROP}. GALPROP
solves the transport equation for Galactic cosmic rays and predicts the resulting
interstellar gamma-ray emission, and the interstellar-emission templates used
throughout this work are also based on it. The morphology has since been measured
more precisely by Di Mauro~\cite{DiMauro2021}, who finds a generalised NFW profile
with $\gamma=1.2$--$1.3$, an ellipsoid axis ratio of $0.8$--$1.2$, and no evolution
of the morphology with energy up to $30$~GeV. \rtwo{A spherical, halo-like residual in the inner Galaxy is therefore not a new claim, and the halo of
Ref.~\cite{Totani2025} extends that description to high latitudes and to higher energy.}

Whether the GCE traces dark matter annihilation in a cuspy halo or an unresolved
population of millisecond pulsars (MSPs) remains
debated~\cite{BartelsKrishWeniger2016}. This ambiguity arises because the inner Galaxy is
crowded with conventional gamma-ray sources and the diffuse model is least certain
exactly where the signal is brightest. A residual of comparable amplitude, relative
to its local background, is recovered in control regions along the Galactic plane
where no dark matter signal is expected~\cite{Ackermann2017}, \rtwo{which sets a floor on the amplitude a residual must exceed before it can be attributed to a genuine excess} rather than to foreground mismodelling. Searches in overlapping high-latitude sky have \rtwo{not detected a corresponding signal}~\cite{Chang2018, Huang2016,Ackermann2012}, and above $10$~GeV
the inner-Galaxy excess has been better described by a boxy stellar-bulge morphology
than by an annihilation profile~\cite{Manconi2024}. \rtwo{Other studies have also found} that a hard leptonic population injected in the inner Galaxy and transported
outwards produces an extended, centrally concentrated IC component that reproduces
both the spectrum and the angular profile of the GCE~\cite{Petrovic2014,
Cholis2015}, with the hadronic analogue developed in
Refs.~\cite{CrockerAharonian2011, CarlsonProfumo2014}. An extended IC component from
an injected population is therefore a specific alternative that must be confronted
directly, and we do so by \rtwo{constructing and fitting} injected-population IC templates (Sec.~\ref{sec:hardic}).

Against this background, the distinguishing feature of the signal reported by
Totani~\cite{Totani2025} is spectral rather than morphological. Analysing fifteen
years of \textit{Fermi} Large Area Telescope (LAT)~\cite{Atwood2009} data over the
region $|\ell|\le 60\degree$, $10\degree \le |b| \le 60\degree$, and modelling the
data with point sources, GALPROP cosmic-ray templates, an isotropic component,
Loop~I, and the \textit{Fermi} bubbles~\cite{Su2010}, Totani isolates a residual,
spherically symmetric, halo-like component with a spectrum that peaks near 20~GeV
and is consistent with zero below 2~GeV and above 200~GeV. The measurements of a
spherical inner-Galaxy excess cited above peak instead at $1$--$3$~GeV, and Hooper
and Slatyer report little or no evidence for emission above $10$~GeV from their own
NFW template~\cite{HooperSlatyer2013}. Emission with this morphology in
the $9.5$--$47.5$~GeV band had been characterised previously~\cite{Linden2016}, but
as a subdominant high-energy tail of a spectrum peaking two decades lower rather
than as an independent component.

The spectrum of Totani~\cite{Totani2025} is best fit by WIMP annihilation with
$\mchi \sim 0.4$--$0.8$~TeV and
$\sigv \sim (5$--$9)\times10^{-25}~\mathrm{cm^3\,s^{-1}}$ for the $\bbar$ and $\WW$ channels together, with a radial profile
compatible with a smooth NFW halo, possibly slightly shallower
toward the centre.
The masses sit well above the $30$--$75$~GeV, near-thermal fits characteristic of the
GCE literature~\cite{Daylan2016, CaloreTails2015}, and the inferred cross section
exceeds both the canonical thermal relic value and the limits from dwarf spheroidal
galaxies (dSphs). Totani argues that it nonetheless remains viable once halo-profile
uncertainties are folded in. The peak is reported at $13$--$19\sigma$ statistical
significance, and is \rtwo{reported} robust to its systematic checks ($\geq9.1\sigma$ under
conservative masking of negative-residual pixels), although statistical and systematic
budgets are not combined into a single figure. 

\begin{rtwoblock}
Since that report the excess has been examined from several directions. Cholis, Hu and Zhong~\cite{Cholis2026}
\rtwo{find that a new model of the \textit{Fermi} bubbles accounts for the halo-like emission, which they consider likely to be bubble emission}, with a flat component, inner bubbles and cocoons, and report that the emission brightens by a factor of 2 to 5 between their latitude ranges.
Abazajian, Li and Macias~\cite{Abazajian2026} refit it across 281 diffuse configurations and find its amplitude and
spectrum foreground dependent. On the particle side, several models produce the required rate through a light
mediator~\cite{Jho2025,Yoshimatsu2026,Yamashita2026} or an electroweak doublet~\cite{NomuraTotani2026}, and the excess
has been connected to the high-energy nuclear recoil event reported by LUX-ZEPLIN (LZ)~\cite{LZ2026,Nomura2026,Yamashita2026b}. Each
of these readings rests on a particular answer to the same questions: where the emission is, how its spectrum and
morphology depend on the foreground and on the model of the bubbles, and how much of it any template set calls halo.

This paper gives a pixel-level account of what the excess is, what it is not, and what any dark matter explanation
would have to do. We fit the data at the native $0.125\degree$ resolution with an exact Poisson likelihood,
energy-dependent forward folding of the point-spread function and pixel-scale masking of resolved sources. \rtwo{The feature persists at the pixel scale, so it is not an artefact of cell aggregation or PSF smearing, and the mask and injection tests of Sec.~\ref{sec:globalresults} bound point-source contamination.} Working at the pixel scale does not break the degeneracy with the diffuse foreground. We repeat the fit across the
two-dimensional GALPROP foregrounds \rtwo{that vary the cosmic-ray and gas parameters of the diffuse model} and with two independent template sets for the
\textit{Fermi} bubbles, our residual-derived templates and those of Cholis, Hu and Zhong \rtwo{(hereafter the CHZ set)}, and use three-dimensional
GALPROP v57 models as an additional check. \rtwo{We refer to this set of configurations the envelope, and the range of results across it the foreground range, which is a lower bound on the foreground uncertainty (Sec.~\ref{sec:conclusions}).} We calibrate what foreground mismatch alone can produce with mock skies
without a halo. We then test the morphology directly, ask whether the excess separates into a spherical part and a
bubble part, confront the injected-electron alternative, and fit dark matter spectra as ranges over the configurations
rather than at one point\rtwo{, expressing the comparison with the dSph limits as the ratio $\mathcal{R}$ of the
required cross section to the dSph upper limit at the same mass (Sec.~\ref{sec:dmmethod})}.

The paper is organised as follows. Section~\ref{sec:methods} describes the data, the templates, the cellwise
replication of the published analysis and the pixelwise fit, and then the foreground models, bubble template sets,
three-dimensional models and null calibration used here\rtwo{, and how a dark matter reading is assessed}. 
\rtwo{Section~\ref{sec:results} first reports the recovery of the excess and the agreement between the cellwise and
pixelwise fits (Sec.~\ref{sec:globalresults}), the excess across the foreground configurations (Sec.~\ref{sec:envres}),
its morphology and its description by two components (Sec.~\ref{sec:morphology}), and its relation to the CHZ set
(Sec.~\ref{sec:chzres}). It then gives the null calibration (Sec.~\ref{sec:nullres}), the latitude dependence
(Sec.~\ref{sec:disk}), the foreground systematics, including the rebuilding of the bubble templates (Sec.~\ref{sec:systematics}), the posterior covariance (Sec.~\ref{sec:covariance}), and tests an astrophysical alternative, an
injected population of cosmic-ray electrons (Sec.~\ref{sec:hardic}). Section~\ref{sec:dm_interpretation} interprets
the excess as dark matter. It gives the annihilation fits and the comparison with the dwarf limits across the
configurations, then considers velocity-dependent annihilation, decay, the antiproton constraint, the requirements on
a candidate, and what the results imply for proposed explanations of the LZ event.}
Section~\ref{sec:conclusions} \rtwo{discusses the results and concludes, and Appendix~\ref{app:dmfits} lists the dark matter fits for every configuration}.

Throughout, we separate the empirical statement, that a centrally concentrated 20~GeV excess is present and robust across the
foreground models tested, from the claims of what it is. No detection is claimed. All statements are statistical and at the
template level unless stated, and $2\Delta\ln\mathcal{L}$ is used as a diagnostic, never as a significance.
\end{rtwoblock}
\section{Methodology}
\label{sec:methods}
\secfinal

\subsection{Data and Templates}
\secfinal
We analyse an extended \textit{Fermi}--LAT Pass~8 dataset
covering 4~August~2008 to 11~March~2026~\cite{Atwood2013}\rtwo{, with P8R3 ULTRACLEANVETO events (event class 1024, FRONT and BACK conversions) and a zenith-angle cut at $100\degree$}.
The region of interest (ROI) follows Ref.~\cite{Totani2025}, $|\ell|\le 60\degree$ and
$10\degree \le |b| \le 60\degree$, in 13 logarithmically spaced energy bins
spanning $1.16$--$1058~\mathrm{GeV}$. \rtwo{Throughout, the signal range, 10--40~GeV, denotes the three of these bins centred at 12.3, 20.8 and 35.1~GeV.} Counts
and exposure cubes are binned on a $0.125\degree$ Cartesian grid in
$(\ell,b)$.  \rtwo{The high-latitude cut at $|b|\geq10\degree$ excludes the Galactic disk, where the diffuse
model is least reliable and source confusion is greatest, also following Ref.~\cite{Totani2025}.}

The sky model comprises a set of spatial templates, which follow
Ref.~\cite{Totani2025} unless otherwise noted. The Galactic diffuse components (gas and IC emission) derive from a GALPROP-based interstellar-emission
model~\cite{Ackermann:2012pya}, in the baseline configuration adopted by
Ref.~\cite{Totani2025}.

The gas template describes
$\pi^0$-decay and bremsstrahlung emission from molecular and atomic gas maps,
with the flux ratio between pion decay and bremsstrahlung fixed to the original
model value so that gas enters as a single template. IC scattering is given by
the GALPROP IC map corresponding to the same baseline diffuse model.

The isotropic component represents the \rtwo{spatially} isotropic diffuse background of
instrumental and extragalactic origin, initialised in MCMC fits at a constant flux of
$10^{-4}$~MeV~cm$^{-2}$~s$^{-1}$~sr$^{-1}$. 

The point-source component is a 4FGL-DR4 source
map constructed from \texttt{gll\_psc\_v35.fit}~\cite{Abdollahi2020,Ballet2023},
and the extended-source catalogue of this same dataset is used to mask circular regions around
extended sources with radius twice the listed semi-major axis. In the
pixel-level analysis we additionally mask bright catalogued point sources, while
fainter catalogue emission remains represented by the point-source template.

Loop~I is modelled as a geometric two-shell structure
with uniform emissivity in each shell, to capture the large-scale radio loops.

\begin{rtwoblock}
The \textit{Fermi} bubbles are represented by residual templates, modelled either as a single flat template or as separated positive and negative residual components (the structured-bubble model). The baseline fit and every member of the envelope that uses our own bubble templates use the structured model, as does the dark matter annihilation result of Ref.~\cite{Totani2025}. Replacing the structured templates by a single flat one lowers the fitted halo coefficient, and is quantified as one of the foreground systematic variants (Sec.~\ref{sec:systematics}). That substitution is superseded by the bubble-interior test of Sec.~\ref{sec:morphology}, which fits a uniform interior component alongside the structured templates, not in their place. The halo--bubble posterior correlation is quantified in Sec.~\ref{sec:covariance}.
\end{rtwoblock}

Finally, the NFW halo is
a spherically symmetric, halo-like template based on an NFW density
profile~\cite{NFW1996}, with the same Galactic-pole normalisation convention as Ref.~\cite{Totani2025}.
We consider emissivity scalings $q_\gamma \propto \rho^p$ with
$p \in \{1, 2, 2.5\}$, where $\rho^2$ corresponds to annihilation in a smooth
halo, $\rho^{2.5}$ to a more centrally
concentrated stress test, and $\rho^1$ to a decay-like morphology.

Each template $P_{m,jk}$ gives the predicted photon count in pixel $j$ at
energy bin $k$ for a unit scaling coefficient $\theta_m = 1$.  The total
model prediction is
\begin{equation}
  \mu_{jk} = \sum_{m=1}^{M} \theta_{m} \, P_{m,jk}.
\end{equation}

\subsection{Replication of Totani: Cellwise Poisson MCMC}
\label{sec:cellwise_mcmc}
\secfinal
We perform the template fit independently in each energy bin after aggregating the ROI into $10\degree \times 10\degree$ spatial cells. Within a cell $c$, the observed pixel counts are summed to give a cell count $d_{ck}$, and each spatial template is likewise integrated over the same pixels to give a cell-level expected-count template, $\bar P_{m,ck}$. The model prediction in cell $c$ and energy bin $k$ is then
\begin{equation}
    \mu_{ck} = \sum_m \theta_{mk}\,\bar P_{m,ck},
\end{equation}
where $\theta_{mk}$ is the fitted normalisation of component $m$ in energy bin $k$. The Poisson log-likelihood for that energy bin is
\begin{equation}
    \ln\mathcal{L}_k
= \sum_c \left[d_{ck}\ln\mu_{ck}-\mu_{ck}\right],
\end{equation}
up to the additive data-only factorial term.

A flat prior is imposed on the template-normalisation vector
($\boldsymbol{\theta}_k=\{\theta_{mk}\}$), \rtwo{this is non-negative for the standard templates and has no other bounds}.
Following Totani~\cite{Totani2025}, the NFW halo coefficient and the negative
\textit{Fermi}-bubble residual-template coefficients are permitted to take negative
values, allowing the posterior to be explored continuously through zero.

The walkers are initialised in a small Gaussian ball around a fiducial starting point
$\boldsymbol{\theta}_{k,0}$.  For components whose templates are already constructed in expected-count units with their nominal catalogue or diffuse model normalisation, namely the point-source component, the GALPROP-derived gas and
IC components, and the isotropic component, we set the initial normalisation to
$\theta_{mk,0}=1$.  All additional residual or phenomenological components,
including Loop I, \textit{Fermi}-bubble residual templates, and the NFW halo template,
are initialised at $\theta_{mk,0}=0$.  The subsequent posterior exploration is
not constrained to remain near these values, and they are used only to initialise the
ensemble before sampling within the allowed prior bounds.

For each energy bin $k$, the posterior
$p(\boldsymbol{\theta}_k \mid \{d_{ck}\})$ is sampled using the affine-invariant ensemble sampler \texttt{emcee} \cite{ForemanMackey2013}, with 64 walkers, up to $3\times10^5$ steps, a
burn-in of $5\times10^4$ steps, and a thinning factor of $10$.  Convergence is
assessed using the integrated autocorrelation time $\tau$. Sampling is
terminated once the minimum $N/\tau$ exceeds the target value of $50$.

From the post-burn, thinned posterior samples $\{\boldsymbol{\theta}_k^{(s)}\}$
we report, for each fitted component $m$, the maximum-likelihood value
$\hat{\theta}_{mk,\mathrm{ML}}$ (the value of $\theta_{mk}$ in the sample with the
highest log-probability), together with the 16th, 50th, and 84th percentiles
$\hat{\theta}_{mk,16}$, $\hat{\theta}_{mk,50}$, $\hat{\theta}_{mk,84}$ of the
marginalised one-dimensional posterior, quoted as the central value and
approximate $1\sigma$ credible interval.

\subsection{Global Pixel-Level Fit}
\label{sec:global}
\secfinal
To explore whether the residual of Ref.~\cite{Totani2025} could be an artefact of the $10\degree$ cellwise aggregation, we perform a global pixel-level template fit and evaluate the likelihood directly on the native $0.125\degree$
pixels.  The fit is carried out independently in each energy bin.  The same
linear template model is used,
\begin{equation}
  \mu_{jk} = \sum_m \theta_{mk} P_{m,jk},
\end{equation}
where $d_{jk}$ and $\mu_{jk}$ are the observed and predicted counts in pixel
$j$ and energy bin $k$, and $P_{m,jk}$ is the counts-space prediction of
template $m$ at unit amplitude.

We adopt a two-tier scheme. A fast deterministic maximum-likelihood fit provides
per-bin point estimates, which seed a production MCMC that delivers the final
per-bin halo spectrum and its marginalised uncertainties. \rtwo{Because it is fast, the deterministic fit is also used for some systematic variations which are repeated many times
(Sec.~\ref{sec:detfit})}.

Two effects that are negligible for $10\degree$ cells are treated explicitly at
pixel scale.
First, the halo and selected diffuse templates are forward-folded
with an energy-dependent LAT PSF, modelled as a King profile~\cite{Ackermann2012ResponseFuncsCalibration} matched to the Pass~8 containment radius~\cite{Atwood2013, Bruel2018}.  The convolution is applied in counts space and
conserves the total counts in each energy plane.  We do not forward-fold the LAT energy dispersion. The Pass~8 energy resolution
at the 20~GeV peak ($\sigma_E/E\simeq8\%$)~\cite{Atwood2013, Bruel2018} is much smaller than the logarithmic
width of the energy bins (a factor $\simeq1.69$ per bin, ${\simeq}70\%$), so
dispersion redistributes counts well within a single bin and neither shifts the
20~GeV peak nor biases the inferred mass at a level comparable to the
foreground-modelling systematics of Sec.~\ref{sec:systematics}. We likewise do
not propagate the systematic uncertainty on the LAT effective area, which the
instrument team quantifies as an energy-dependent band at the few-per-cent level
over the range used here~\cite{LATcaveats}. It enters multiplicatively on the
recovered flux and therefore linearly on $\sigv$ and on the dSph tension,
$\mathcal{R}$, where it is an order of magnitude below the $\pm48\%$ \rtwo{halo} $J$-factor term of Sec.~\ref{sec:tension}. It also cancels in part in
$\mathcal{R}$, since the dSph limits against which the halo cross section is
compared are themselves derived from LAT data, though not identically, as the
two analyses use different event selections and energy ranges.
Second, bright catalogued
4FGL point sources are masked rather than absorbed into a single global
point-source normalisation.  In the production pixelwise runs we use an
energy-dependent bright-source mask, constructed from sources above
$3\times10^{-9}\,\mathrm{ph\,cm^{-2}\,s^{-1}}$ and combined with the standard
extended-source mask.

The Poisson log-likelihood is
\begin{equation}
  \ln\mathcal{L}_k =
  \sum_{j\in{\rm fit}}
  \left[d_{jk}\ln\mu_{jk}-\mu_{jk}\right],
\end{equation}
up to the usual data-only factorial term.  The default disk-excluded production
fits use $|\ell|\leq60\degree$ and $10\degree\leq |b|\leq60\degree$. For comparison
we also run otherwise identical fits including the Galactic disk,
$|b|\leq60\degree$.  In both cases the halo spectrum is evaluated over the same
high-latitude region, $10\degree\leq |b|\leq60\degree$, so that the disk-included
and disk-excluded spectra are directly comparable and match the convention used
for the Totani comparison in Sec.~\ref{sec:globalresults}. The disk therefore
enters as a change to the fit region rather than to the region over which the
halo is reported.

We retain the exact per-pixel Poisson likelihood, rather than a Gaussian or
$\chi^2$ approximation. Counts fall by roughly three and a half orders of magnitude across the band, and
only $5.6\%$ of $0.125\degree$ pixels are occupied at the $20.8$~GeV peak
($42\,199$ of $750\,384$), falling to $0.011\%$ in the highest-energy bin.
Those sparse bins are also the constraining ones. Over half the template-level likelihood preference for the halo arises where fewer than $12\%$ of pixels are occupied, while the photon-rich bins below $5~\mathrm{GeV}$ contribute only \rtwo{$11\%$}. A Gaussian per-pixel error model would therefore fail precisely where the measurement is made, whereas the Poisson form is exact for empty pixels and requires no binning-up, which would wash out the morphology the pixel-level treatment exists to retain.

\subsubsection{Deterministic Global Fit: Initialisation and Systematics}
\label{sec:detfit}
As a fast diagnostic stage, we run a deterministic global maximum-likelihood
fit using the L-BFGS-B algorithm~\cite{Byrd1995,Zhu1997}. In this optimiser only,
the template columns are internally rescaled for numerical conditioning,
\begin{equation}
  \varphi_{mk} = \theta_{mk} S_{mk},
  \qquad
  S_{mk} = \frac{\sum_j P_{m,jk}}{\sum_j d_{jk}},
\end{equation}
with the amplitudes transformed back via $\theta_{mk} = \varphi_{mk} / S_{mk}$. This is a
coordinate change internal to the optimiser, not a physical template normalisation.

\rtwo{
The deterministic fit is used in two ways. Its maximum-likelihood solution provides the starting point for the
production MCMC, shortening burn-in without entering as a prior. It is also used where a fit is repeated over many
configurations and a full MCMC for each would be costly, and results obtained this way are marked as deterministic
where they are quoted. On the baseline, over the 10--40~GeV bins that constrain the halo, its halo coefficient lies
2.3--3.0\% below the MCMC median, and its Fisher uncertainty agrees with the half-width of the MCMC 68\% interval to
within 1.5\%. The two methods differ by 7--11\% in the halo coefficient in the three lowest-energy bins and by more in the two
highest-energy bins, at 482 and 814~GeV, where the halo is weakly constrained.}

\subsubsection{Production Pixelwise MCMC}
\label{sec:mcmc}
The production pixelwise MCMC is sampled with an affine-invariant ensemble sampler.
The fitted amplitudes are assigned flat bounds: standard emission templates are
constrained to be non-negative, while the NFW halo coefficient and the negative
\textit{Fermi}-bubble residual template are allowed to take either sign, as in Sec.~\ref{sec:cellwise_mcmc}.  The chains are
initialised near the deterministic global maximum-likelihood solution of
Sec.~\ref{sec:detfit} when available. If the summed template support in a given
energy bin is not positive, the sampler falls back to the default Totani-style
initialisation detailed in Sec.~\ref{sec:cellwise_mcmc}.

For numerical robustness we require the total, template-summed model prediction
to be positive wherever the data constrain it. We adopt a strict convention, in
which the likelihood sums $d\ln\mu-\mu$ over pixels with $\mu>0$ only. A proposal is rejected outright if any pixel with $d>0$ has $\mu\leq0$, whereas
pixels with $d=0$ and $\mu\leq0$ are dropped from the sum and contribute exactly
zero rather than terminating the walker.

The model is positive everywhere below $20\,\mathrm{GeV}$, and in all but fewer
than $\rtwo{0.04\%}$ of fitted pixels up to ${\sim}300\,\mathrm{GeV}$. The
non-positive fraction rises only in the two highest-energy bins, where the model
is faintest, reaching $0.35\%$ and $0.63\%$ of fitted pixels (2640 and 4754 of
the ${\sim}7.5\times10^{5}$ total), with smaller numbers at $20.8$, $59$, $100$,
$169$ and $285$~GeV. Across every fitted pixel in all 13 energy bins, no pixel with non-zero observed counts has a non-positive model prediction.
The requirement therefore never binds where the data carry information, and \rtwo{does not} shape the halo.

For each energy bin we use 64 walkers, with a maximum of 60000 steps, a burn-in
of 10000 steps, and thinning by a factor of 10.  Autocorrelation times are
monitored during the run, and chains may terminate early once the minimum $N/\tau$
exceeds the target value of 50.

\subsubsection{Uncertainties, Residuals, and the Halo Likelihood Ratio}
\label{sec:uncertainties}
For each template amplitude we report the maximum-likelihood sample and the
central posterior quantiles from the MCMC chain.  These intervals marginalise
over the other fitted template amplitudes within the same energy bin and replace
the local Fisher/Laplace errors used in the deterministic diagnostic fit. After
fitting, we construct residual maps
$\delta_{jk}=d_{jk}-\mu_{jk}(\hat{\boldsymbol{\theta}})$, together with model and
fractional-residual maps, as spatial diagnostics of the fit quality and to
identify structures not captured by the template model.

To quantify the preference for a halo component, we compare the best
log-likelihood in the halo fit with the corresponding no-halo baseline,
\begin{equation}
  \Delta\ln\mathcal{L}_k =
    \ln\mathcal{L}_k
    \!\left(\hat{\boldsymbol{\theta}}^{\,\mathrm{halo}}\right)
    -
    \ln\mathcal{L}_k
    \!\left(\hat{\boldsymbol{\theta}}^{\,\mathrm{no\text{-}halo}}\right).
\end{equation}
We use this as a per-bin diagnostic of the improvement obtained by including the halo template. \rtwo{Over 10--40~GeV we quote $2\sum_k\Delta\ln\mathcal{L}_k$ summed over the three bins of the signal range, written $2\Delta\ln\mathcal{L}$ for short. Where two models are compared directly, $\Delta(-2\ln\mathcal{L})$ is the change in $-2\ln\mathcal{L}$ from the first model to the second, so a negative value favours the second.} \rtwo{If the foreground model were exact and Wilks' regularity conditions~\cite{Wilks1938} held, each per-bin $2\Delta\ln\mathcal{L}_k$ would follow a $\chi^2$ distribution with one degree of freedom, and their sum over the three signal bins a $\chi^2$ distribution with three. Because foreground mismatch alone can produce large values, we do not convert it to a significance, and instead calibrate its scale on simulated skies that contain no halo emission, generated from the no-halo best fit of one foreground model and fitted with a different one with the halo template free (Sec.~\ref{sec:nullcal}).}

\begin{rtwoblock}
\subsection{Foreground Families and Bubble Template Sets}
\label{sec:families}
\secfinal

The pixel-level fit of Sec.~\ref{sec:global} is repeated across the foreground models and
bubble template sets described below, rather than on a single preferred model. 
\paragraph{Two-dimensional GALPROP models.} The two-dimensional GALPROP grid of Ackermann et al.~\cite{Ackermann:2012pya}
defines our baseline (cosmic-ray source distribution, height $z_h$ and radial extent $R$ of the cosmic-ray diffusion
halo, H\,I spin temperature $T_S$, and the $E(B-V)$ dust cut), with the gas and inverse-Compton (IC) templates fitted
separately. The baseline is the member with the Lorimer source distribution, $z_h = 6$~kpc, $R = 30$~kpc,
$T_S = 150$~K and the $E(B-V) = 2$ dust cut, which is the baseline configuration of Ref.~\cite{Totani2025}. \rtwo{The other 13 models of the grid vary the cosmic-ray source distribution (Lorimer, supernova-remnant, Yusifov and OB-star), $z_h$, $R$, $T_S$ and the dust cut, and are listed in Sec.~\ref{sec:systematics}.}

\paragraph{Three-dimensional GALPROP models.} We use three-dimensional GALPROP v57 models of Porter
et al.~\cite{Porter2017}, with three-dimensional gas~\cite{Johannesson:2018bit} and either the standard or a
three-dimensional interstellar radiation field \rtwo{(the R12 and F98 models of Ref.~\cite{Porter2017}), and with 0, 50 or 100\% of the cosmic-ray source luminosity in the spiral arms (SA0, SA50 and SA100)}, computed with GALPROP v57 using the published parameter
files (Sec.~\ref{sec:galprop3d}).

\paragraph{The \textit{Fermi}-LAT interstellar emission model.} The \textit{Fermi}-LAT Galactic interstellar emission
model (GIEM, \texttt{gll\_iem\_v07})~\cite{Abdollahi2020} is fitted alone as a single diffuse component, in place of the gas, IC, Loop~I and
\textit{Fermi}-bubble templates of any construction, together with the isotropic and point-source templates and the
halo. The model already includes non-template components, primarily Loop~I and the \textit{Fermi}
bubbles~\cite{Abdollahi2020}, so no bubble template is added to it. These components were fitted to LAT residuals and may already contain part of the reported excess, so the halo coefficient fitted with the GIEM may understate it. Its row is therefore shown in full (Table~\ref{tab:envelope}) but kept outside the quoted ranges, so that the effect of excluding it is visible.

Two bubble template sets enter. The first is ours: Totani's bubble outline with positive and negative
residual templates, re-extracted against each foreground model. The second is the CHZ set~\cite{Cholis2026}, comprising the Flat Bubbles, Inner Bubbles and Cocoons. Except where
marked otherwise, it is used at its published geometry, rendered on our pixel grid, with the normalisation of each
component free in each energy bin, so that only its morphology enters the fit. We also re-fit their component geometry on our data, on three of the three-dimensional foregrounds. \rtwo{We do not rely on those re-fits as our implementation of their method did not reproduce
their best-fit bubble edges on the two-dimensional control (Sec.~\ref{sec:chzres}), and the three-dimensional foregrounds
do not reproduce the amplitude of a published reference comparison (Sec.~\ref{sec:galprop3d}). Results that use a geometry re-fitted in this work are marked $\dagger$ and are shown only as a check. Every other result with the CHZ set uses its published geometry: on the baseline, the
OB-star model and three of the ten three-dimensional models (SA50 R12, SA0 R12 and SA50 F98). The other seven
three-dimensional models and the other 13 models of the two-dimensional grid are fitted with our bubbles only, the
latter at $|b| \geq 10\degree$ in the ten bins up to 169~GeV, taken from the latitude-cut scan
(Secs.~\ref{sec:disk} and \ref{sec:dmres}).}

Three bubble outlines recur below. The outline of Su et al.~\cite{Su2010} is the original published outline of the
\textit{Fermi} bubbles. Our outline is the geometric outline read off Fig.~1 of Totani~\cite{Totani2025}, from which
our bubble templates are built. Outline E is the edge of the CHZ Flat Bubbles~\cite{Cholis2026} (their Appendix~B),
derived on their two-dimensional foreground. It is not their Inner Bubbles, which are ellipsoids inside of it.

\paragraph{Admissibility.} A foreground or template set is admissible if its morphology is fixed by physics or by
data independent of the 10--40~GeV residual\rtwo{, or if it is tuned to the data and fitted alongside the halo rather than in place of it}. Data-tuned components are treated alike. Our residual-derived
bubble templates and the CHZ geometry are both admitted, since each is tuned to the data at some level and each
is fitted alongside the halo, not in place of it. Nothing is excluded for giving a small halo.

\subsection{Fitting Protocol and Convergence}
\label{sec:protocol}
\secfinal

The combinations of foreground and template set are fitted with the production pixelwise MCMC of
Sec.~\ref{sec:mcmc}, initialised from the deterministic fit of Sec.~\ref{sec:detfit}, with an energy-dependent
point-source mask. With an energy-independent mask instead, the halo coefficient over 10--40~GeV changes by less than
5\%. Each combination is fitted with and without the halo, in all 13 energy bins (the ten lowest for the further grid
models of Sec.~\ref{sec:envres}).

Each new \rtwo{template construction} is first tested on simulated data. A noiseless mock is built from
that combination's best fit with the halo set to its baseline value, and the fit must recover it within $1\sigma$ in
each of the three 10--40~GeV bins. A mock without a halo must likewise return none. Chains are accepted only
when the minimum $N/\tau$ exceeds 50, as in the production sampler of Sec.~\ref{sec:mcmc}.

\subsection{Position of the Sun}
\label{sec:rsun}
\secfinal

The halo templates place the Sun at $R_0 = 8.0$~kpc, following Ref.~\cite{Totani2025}. Every GALPROP model we use, two- and
three-dimensional, places it at 8.5~kpc, as does the CHZ geometry. No other template depends on $R_0$. The Loop~I
shells are Sun-centred, the bubble templates are sky outlines or data residuals, and the isotropic and point-source
templates carry no geometry.

Moving the $\rho^2$ halo to $R_0 = 8.5$~kpc changes the template shape by $-2.4\%$ to $+2.8\%$ per pixel and lowers
$J_{\rm pole}$ by 13.2\% (to $7.755\times10^{14}~M_\odot^2~{\rm kpc}^{-5}$). Refitting the 10--40~GeV bins with it
lowers the halo coefficient by 5.2--5.4\% with our bubbles and 5.3--5.7\% with the CHZ set, and changes
$2\Delta\ln\mathcal{L}$ over 10--40~GeV from 927.2 to 923.4 (ours) and 136.5 to 135.1 (CHZ). The change is small,
so the two-dimensional fits keep $R_0 = 8.0$~kpc for comparability with Ref.~\cite{Totani2025}, and it is carried as a line in the
systematic budget. The three-dimensional fits also keep $R_0 = 8.0$~kpc in the halo template, and the effect of moving every template to
8.5~kpc is given in Sec.~\ref{sec:envres}.

\subsection{Three-Dimensional GALPROP Models and Their Validation}
\label{sec:galprop3d}
\secfinal
The three-dimensional models were computed with GALPROP v57~\cite{Porter2022}, built and run for this work. We
validate the build against two published results.

The first is the set of 2012 maps, which were produced with an earlier release of the code. Run on the 2012 baseline
parameter file, with the placeholder values of its magnetic-field parameters replaced by those of its own field model,
our build reproduces the 2012 maps to within 2\% in every component at $|b|\geq10\degree$ over the GALPROP energy
planes between 10 and 40~GeV (\rtwo{ratio of our maps to the 2012 maps,} pion decay 0.98, bremsstrahlung 1.00, IC 0.98). Our three-dimensional production
parameter files carry the same physical values.

The second is Fig.~8 of Porter et al.~\cite{Porter2017}, which shows the fractional difference
$(\mathrm{SA50}-\mathrm{SA0})/\mathrm{SA0}$ in total intensity at 10.6~MeV, 1.2~GeV and 79~GeV between source density
models that place 0 and 50\% of the injected cosmic-ray luminosity in the spiral arms, computed with GALPROP v56 for
the two-dimensional gas and the standard radiation field. We compute the same quantity with v57, using the isotropic
inverse-Compton cross section where Porter et al.\ use the anisotropic one. The two agree closely in morphology (sign
agreement 0.99 and 0.92, $r = 0.996$ and 0.992 at 1.2 and 79~GeV) but not in amplitude. Our contrast is smaller than theirs at 1.2~GeV, by a median of 2.4 of their contour levels (0.0022 each), and about
1.5 times theirs at 79~GeV, where it lies a median of 16.0 levels above. The 10.6~MeV panel lies below the 32~MeV
lower edge of our energy grid and is not compared. The SA50--SA0 difference at 79~GeV is almost entirely IC emission.
\rtwo{Recomputing both models with the anisotropic cross section (GALPROP v57, coarsest anisotropic grid) leaves the
amplitude discrepancy in place. Our contrast lies 2.5 contour levels below theirs at 1.2~GeV and 17.3 above at
79~GeV. The isotropic cross section therefore does not explain it. The most likely remaining cause is the change of
code release, which we cannot test directly.} Abazajian et al.~\cite{Abazajian2026} likewise find that the
inverse-Compton maps drive the poorer fits of their v57 models. The three-dimensional results are therefore reported as an additional
check, outside the quoted ranges, and each carries this caveat.

\subsection{Calibrating the Likelihood Diagnostic Against Foreground Mismatch}
\label{sec:nullcal}
\secfinal

To quantify the halo amplitude and the $2\Delta\ln\mathcal{L}$ that a mismodelled foreground produces on its own, we fit
mocks that contain no halo. Each mock is the no-halo best fit of one foreground, the generator, and is fitted with a
different foreground, the fitter, with the halo free. The generators are SA50 and SA0 with the R12 radiation field
(three-dimensional, with our bubbles), the OB-star model with a diffusion halo height of 6~kpc (two-dimensional, with our bubbles),
GIEM alone, and the baseline with the CHZ set. The fitters are the baseline with our bubbles and the baseline with the
CHZ set. GIEM as a generator carries a caveat, since it was tuned to the residuals and may already contain part of the excess.

The diagnostic has two steps. The first fits the noiseless mock of every generator and fitter pair in the five bins
from 7.3 to 59~GeV. For each fitter, the pair with the largest induced halo goes to the second step, where the induced
halo is the mean over 10--40~GeV of the posterior median, keeping its sign. The second step fits 50 Poisson
realisations of that pair, with and without the halo, to give the spread of the induced amplitude and of
$2\Delta\ln\mathcal{L}$ that this mismatch produces. We compare the data with both. The generators, fitters and
selection rule were fixed before any mock was built. The results are given in Sec.~\ref{sec:nullres}.

\subsection{Injected Electrons as an Alternative, Across Foregrounds}
\label{sec:ie}
\secfinal

The injected-electron population of Sec.~\ref{sec:hardic} is tested on four combinations, the baseline and SA50 R12,
each with our bubbles and with the CHZ set. Each is fitted twice in the pixelwise setup with the electron template free
and non-negative in every bin, once in place of the halo and once alongside it. For each combination we report the halo
ratio, the halo coefficient with the electrons present divided by its value without them, per bin and over
10--40~GeV, and compare it across the four combinations.

The electron templates are the semi-analytic 3 and 6~kpc templates of Sec.~\ref{sec:hardic}. We also computed the population's IC and bremsstrahlung with GALPROP v57, using the
parameter file of the foreground it is fitted against and adding only a burst source at the Galactic Centre.
Repeating that calculation with time steps five times finer changes the templates by less than 1\% up to 100~GeV for
the baseline and up to 285~GeV for SA50 R12, but by 1.5--24\% and 3--8\% respectively above those energies
(Sec.~\ref{sec:hardic}). The propagated templates are therefore not converged at high energy, and we use the
semi-analytic templates throughout.

The single-normalisation fit of Sec.~\ref{sec:hardic}, in which one amplitude $\theta$ scales the injected
population across all energies, \rlatest{is fitted pixelwise on the baseline and SA50 R12 with both bubble template sets, for the 3~kpc and 6~kpc
templates. It is also run deterministically, which measures how far the deterministic method moves these results.}

\subsection{Reproducing the Analysis of Cholis, Hu and Zhong}
\label{sec:c2d}
\secfinal

To separate differences of method from differences of setup, we repeat \rtwo{the analysis of Cholis, Hu and Zhong~\cite{Cholis2026}} with our implementation of their data selection:
P8R3 CLEAN events (event class 256, response functions P8R3\_CLEAN\_V3), FRONT conversions, 0.275--928~GeV, zenith angle below
$100^\circ$, standard good-time cuts and a rocking angle below
$52^\circ$, 4 August 2008 to 29 April 2025, binned on their $0.1^\circ$ grid ($1200\times1200$,
Galactic, centred on the Galactic Centre) in their 16 energy bins. Their geometry is
then re-fitted with their mask, penalties and final model, and on one three-dimensional foreground.
We compare the sign and size of each published likelihood comparison, and their per-bin component spectra (\rtwo{their} Table IX), with ours. The
outcome is in Sec.~\ref{sec:chzres}.

\subsection{Assessing a Dark Matter Reading}
\label{sec:dmmethod}
\secfinal

Rather than testing for dark matter, we determine what a dark matter interpretation of the halo would require,
through annihilation or decay. For annihilation, prompt spectra from PPPC4DMID~\cite{Cirelli2011} in the $b\bar b$,
$W^+W^-$ and $\tau^+\tau^-$ channels are fitted to the pixelwise $\rho^2$ halo spectrum with a single normalisation at
each mass, and the normalisation is converted to a cross section through the halo $J$-factor toward the Galactic
poles (Sec.~\ref{sec:annihilation}). This cross section is compared with the dSph \rtwo{95\% confidence-level} upper limits of
Refs.~\cite{Hoof2020,McDaniel2024} at the same mass, and their ratio defines the dSph tension $\mathcal{R}$
(Sec.~\ref{sec:tension}). The uncertainty on $\mathcal{R}$ combines the range across the foreground members with the
\rtwo{uncertainty on the halo $J$-factor, since the dSph limits already include that of the dwarf $J$-factors} \rtwo{(Sec.~\ref{sec:tension})}. The fits are repeated across the members (Sec.~\ref{sec:dmres}), and
velocity-dependent annihilation with its CMB bound and the antiproton limits are treated in Secs.~\ref{sec:veldep}
and~\ref{sec:antiprotons}. For decay, prompt PPPC4DMID decay spectra in the same three channels are fitted to the
pixelwise $\rho^1$ halo spectrum, since the decay rate scales with the density rather than its square, and the
normalisation gives a lifetime that is compared with bounds from the isotropic gamma-ray
background~\cite{AndoIshiwata2015,Cohen2017,BlancoHooper2019} (Sec.~\ref{sec:decay}).

\end{rtwoblock}

\section{The 20~GeV Excess: Extraction and Robustness}
\label{sec:results}
\secfinal

\subsection{Recovery and Cellwise--Pixelwise Agreement}
\label{sec:globalresults}
\secfinal

\begin{figure}[htbp]
    \includegraphics[width=\columnwidth]{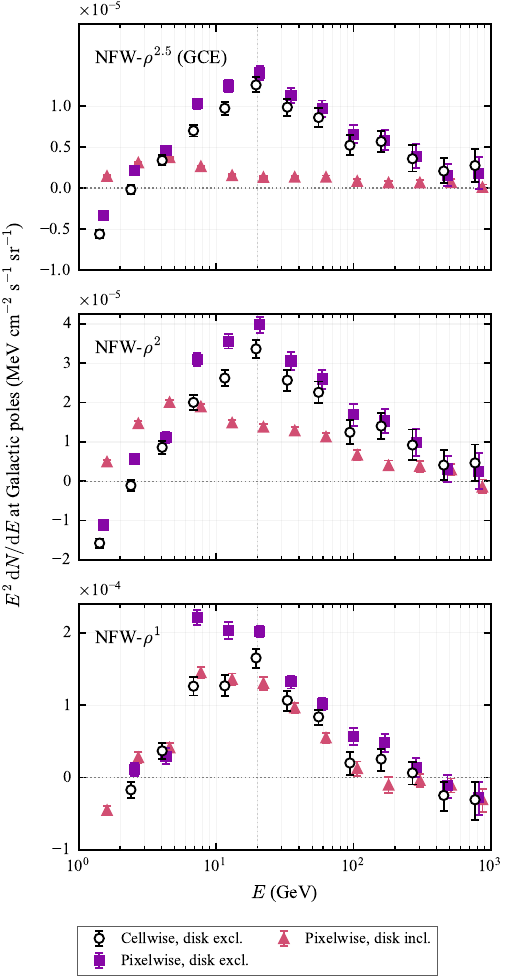}
    \caption{
Comparison of the \rtwo{halo-template} spectra recovered in the Totani-style cellwise fit
(disk-excluded) and in the global pixel-level fit, the latter shown both with
and without the Galactic disk included in the fit. The three panels show
the NFW halo templates with emissivity scalings $\rho^{2.5}$, $\rho^2$, and
$\rho^1$. The disk-excluded global pixel-level fit uses the same high-latitude signal
extraction convention as the cellwise analysis, but evaluates the likelihood
directly on the $0.125\degree$ pixels with PSF-forward-folded templates and
bright-source masking. The persistence of the same broad spectral feature
indicates that the \rtwo{feature fitted by the halo template} is not an artefact of the $10\degree$
cell aggregation. Changes between the disk-excluded and disk-included curves
illustrate the residual degeneracy between the halo and low-latitude diffuse
templates. \rtwo{The top panel is the $\rho^{2.5}$ profile preferred for the Galactic Centre excess (GCE).} Error bars are statistical, \rtwo{the 16--84\% interval of} the MCMC posterior, and the foreground-modelling systematics are treated separately in Sec.~\ref{sec:systematics}.}
    \label{fig:global_overlays}
\end{figure}

\rtwo{In the global pixel-level fit (Fig.~\ref{fig:global_overlays}), the spectrum taken up by the $\rho^2$ halo
template peaks at $E\simeq20.8\,\mathrm{GeV}$}, with a peak amplitude toward the
Galactic poles of
$E^2\,dN/dE\simeq4.0\times10^{-5}\,\mathrm{MeV\,cm^{-2}\,s^{-1}\,sr^{-1}}$. The
$\rho^{2.5}$ scaling peaks at a similar energy, while $\rho^1$ is broader and
more sensitive to the treatment of the diffuse background.

For all three emissivity scalings the pixelwise spectrum reproduces the shape of
the cellwise halo, both our own and that of Ref.~\cite{Totani2025}, including
the low-energy sign change. Its amplitude, however, lies systematically above
the cellwise result.

The two fits read the same counts cube and the same exposure cube, so this is
methodological rather than a difference of data reduction. Varying the PSF
treatment moves the recovered halo by \rtwo{less than $1\%$ over 10--40~GeV} and varying the
source mask by at most \rtwo{$4.4\%$ (see the point-source paragraph below)}, neither
approaching the offset, as expected for a smooth component far from the
source-crowded disk. The remaining candidate is the aggregation scale. The cellwise
likelihood is evaluated on counts summed within each cell, so every template
enters through its cell average, whereas the pixel-level likelihood retains the
variation of each template across the ROI. \rtwo{We therefore expect aggregation to discard the variation within each cell that distinguishes a halo template from the smooth gas and IC components, so that the cellwise fit assigns less flux to the halo template than the pixel-level fit does, and more to those components.}

The size of the effect is consistent with that argument. Taking the per-bin median across
the signal band and rounding to the nearest five per cent, the offset is
$+15\%$ for $\rho^{2.5}$, $+20\%$ for $\rho^{2}$ and $+25\%$ for $\rho^{1}$. We
do not quote it more finely, since the value depends on whether the per-bin
median or the band sum is taken, and for $\rho^{1}$ those differ by ten
percentage points. The ordering is unaffected by that choice. The flattest
scaling, already the most degenerate with the diffuse background, loses the most
to aggregation, and the most centrally concentrated loses the least.
The offset
sits within the combined uncertainty of most individual bins but is coherent
across them, so it propagates directly into the inferred annihilation cross
section (Sec.~\ref{sec:annihilation}), and it is smaller than the
foreground-modelling spread of Sec.~\ref{sec:systematics}.

\begin{rtwoblock}
\paragraph{Point sources.} The point-source model is the 4FGL-DR4 catalogue, held fixed beneath every foreground, so we
test how much of the halo could be catalogued or unresolved source emission. Refitting with six more aggressive source
masks, from a flux threshold of $3\times10^{-9}$~cm$^{-2}$~s$^{-1}$ above 1~GeV down to every 4FGL-DR4 source in the
region masked to $1\degree$ (38.7\% of the region and 48\% of its photons at 20.8~GeV), moves the 10--40~GeV halo by
$-4.4\%$ to $+3.0\%$, and not monotonically with the masked fraction. Sources below the detection threshold, drawn from
the source-count distribution of Ref.~\cite{Zechlin:2016b} and folded with the same PSF as the fit, induce a halo at
20.8~GeV whose 95th percentile is 1.6--2.1\% of the measured amplitude for detection thresholds varied by a factor of
two, 6.6\% for that population scaled by ten, and below 0.1\% for a Galactic population of millisecond pulsars. This
population reproduces the number of catalogued sources at $|b|\geq30\degree$ but falls short of the unresolved
intensity measured at 1--10~GeV~\cite{Zechlin:2016a}, below the bins that carry the excess, and it is applied
isotropically at lower latitudes, where the catalogue holds more bright sources. The catalogue itself was derived with the
standard \textit{Fermi}-LAT diffuse model~\cite{Abdollahi2020,Ballet2023} and is not re-derived for each foreground
model, and the positions and spectra of the sources it retains are not varied. These tests therefore bound the effect
of completeness and of masking, for the populations tested and without PSF mismatch in the injection, but not the exchange of smooth emission against source positions and spectra
(Sec.~\ref{sec:conclusions}). How the treatment of unresolved and faint sources shapes a smooth residual has been
examined for the Galactic Centre excess~\cite{Leane:2019xiy,Leane:2020nmi,Buschmann:2020adf}.
\end{rtwoblock}

\begin{rtwoblock}
\subsection{The Residual Across Foregrounds and Bubble Template Sets}
\label{sec:envres}
\secfinal

Table~\ref{tab:envelope} lists the fitted $\rho^2$ halo coefficient in the three 10--40~GeV bins for every
combination of foreground and bubble template set, together with $2\Delta\ln\mathcal{L}$ over 10--40~GeV against the
same combination without the halo, and Fig.~\ref{fig:envelope} shows the principal members at 20.8~GeV. \rtwo{Every template construction is tested on noiseless mocks built from its own best fit
(Sec.~\ref{sec:protocol}). For the baseline with each bubble set, SA50 R12 with the CHZ set and every
three-dimensional model with our bubbles, the mocks recover the injected halo within $1\sigma$ in each of the three
10--40~GeV bins, and the mocks without a halo return none. The other members reuse one of these tested constructions
on a different foreground and have no mock with the halo of their own. These are the OB-star model, whose mock without
a halo also returns none, the further grid models, which are built and fitted in the same way as the OB-star model
with their own gas, IC and bubble templates, and SA0 R12 and SA50 F98 with the CHZ set.}

The two-dimensional members define the quoted range. \rtwo{They are the baseline and the OB-star model, each with both bubble sets, and the other 13 models of the two-dimensional grid with our bubbles, taken from the latitude-cut scan of Sec.~\ref{sec:disk} at our cut of $|b| \geq 10\degree$.} At 20.8~GeV the halo coefficient ranges from 0.236 (the
baseline with the CHZ set) to \rtwo{0.409 (the supernova-remnant model with $z_h = 10$~kpc} and our bubbles), compared with 0.397 for the baseline with
our bubbles. It is positive in every member, with $2\Delta\ln\mathcal{L}$ over 10--40~GeV between 136.5 and 1236 \rtwo{(a diagnostic, not a significance, Sec.~\ref{sec:nullres})}.
The choice of bubble template set moves the amplitude more than the choice of foreground model, as
Ref.~\cite{Abazajian2026} also finds, where the bubble description changes it by a factor of 1.4--2.1. With our bubbles,
the four two- and three-dimensional foreground families agree to within 10\% (0.381--0.418 at 20.8~GeV, including the
three-dimensional members), whereas with the CHZ set the coefficient falls to 59\% of its
value with our bubbles on the baseline and to about 80\% on the OB-star and three-dimensional models.

The GIEM row (0.206), fitted with the GIEM as the only diffuse component (Sec.~\ref{sec:families}), is shown and kept outside the range \rtwo{because its components were fitted to LAT residuals}.

\begin{figure}
\includegraphics[width=\columnwidth]{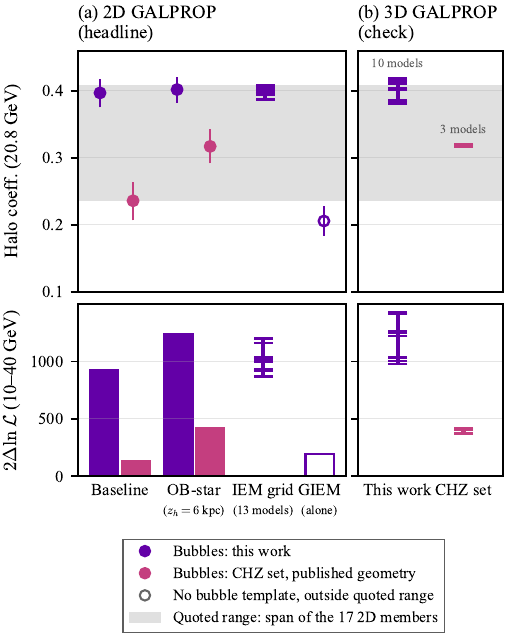}
\caption{\label{fig:envelope} Halo coefficient at 20.8~GeV (top) and $2\Delta\ln\mathcal{L}$ over 10--40~GeV from adding the halo (bottom), for the foreground and bubble-template combinations that enter the analysis. \rtwo{(a)} The two-dimensional GALPROP models, which give the quoted range (grey band). Points are posterior medians with 16--84\% intervals, for the baseline and the OB-star model ($z_h = 6$~kpc), each with the bubbles of this work (purple) and with the CHZ set at its published geometry (pink). \rtwo{The interstellar-emission grid column spans the other 13 models of the two-dimensional grid with our bubbles, one tick per model.} The open circle is the GIEM fitted alone, with no bubble template. It lies outside the quoted range because it was fitted to LAT residuals and a halo fitted on top of it would count the same emission twice. \rtwo{(b)} The three-dimensional GALPROP v57 models, an additional check whose amplitude differs from a published reference (Sec.~\ref{sec:galprop3d}). Each bar spans the minimum to the maximum over the models where that bubble set was run, ten for this work and three for the CHZ set, and each tick is one model. The CHZ set was fitted only at its published geometry, because we could not trust the geometry we re-fitted, and only on three of the ten models (SA50 R12, SA0 R12 and SA50 F98). Members with a re-fitted geometry are not shown.}
\end{figure}

\input{envelope_table}

\paragraph{Three-dimensional foregrounds.} With the GALPROP v57 models and our bubbles, the halo at 20.8~GeV
is 0.381--0.418 across ten models and three
radiation fields, and $2\Delta\ln\mathcal{L}$ is 976--1428. With the CHZ set on three models it is 0.316--0.321, with
$2\Delta\ln\mathcal{L}$ of 371--419. These models reproduce the morphology but not the amplitude of the source-model difference in Fig.~8 of Porter
et al.\ (Sec.~\ref{sec:galprop3d}), which at 79~GeV is almost entirely inverse-Compton emission, so they are reported
as a check and kept outside the quoted range.

\paragraph{CHZ geometry re-fitted to our data$^\dagger$.} Re-fitting the CHZ component geometry on our data, separately
on three three-dimensional foregrounds, gives 0.218 (SA50 R12), 0.227 (SA0 R12) and 0.218
(SA50 F98) at 20.8~GeV, with $2\Delta\ln\mathcal{L}$ of 162, 162 and 149. The three re-fits agree with each other to
4\%. \rtwo{We cannot confirm the re-fitted geometry (Sec.~\ref{sec:families}), so these values lie outside every quoted range and we do not rely on them.}

\paragraph{Energy sidebands.} Without the halo, our bubbles describe the 1.2--5.6~GeV sideband
better than the CHZ set on every foreground, \rtwo{lowering $-2\ln\mathcal{L}$ by} 9430--11483 (the GIEM, which carries no bubble set, is not compared). The CHZ
set describes the 77--1058~GeV sideband slightly better, by 24--76. These comparisons \rtwo{do not by themselves choose between the two sets}. The low sideband carries part of the excess's low-energy shoulder.

\paragraph{Position of the Sun.}

In the three-dimensional fits we also move the Sun to 8.5~kpc in every template, as in the GALPROP models, with our
bubbles re-extracted at that position, and compare with the same fits with the halo template at 8.0~kpc. With our
bubbles, $2\Delta\ln\mathcal{L}$ over 10--40~GeV falls by 6--13\% across the ten models (from 1221 to 1100 for
SA50 R12 and from 1415 to 1233 for SA100 R12), and the halo coefficient at 20.8~GeV, each referred to the halo
template at its own Sun position, falls from 0.381--0.418 to 0.354--0.380, a ratio of 0.91--0.93. With the CHZ set on SA50 R12, $2\Delta\ln\mathcal{L}$ changes by less than 2\% (419.1 to 418.2) and the
coefficient falls from 0.32 to 0.30, a ratio of 0.94. For the three CHZ geometries re-fitted by us$^\dagger$,
$2\Delta\ln\mathcal{L}$ changes by 1.0--1.5\% (to 147.5--159.9) and the coefficient by 6\%.
\end{rtwoblock}

\subsection{Morphology of the Excess}
\label{sec:morphology}
\secfinal
\begin{rtwoblock}
\begin{figure*}
\includegraphics[width=\textwidth]{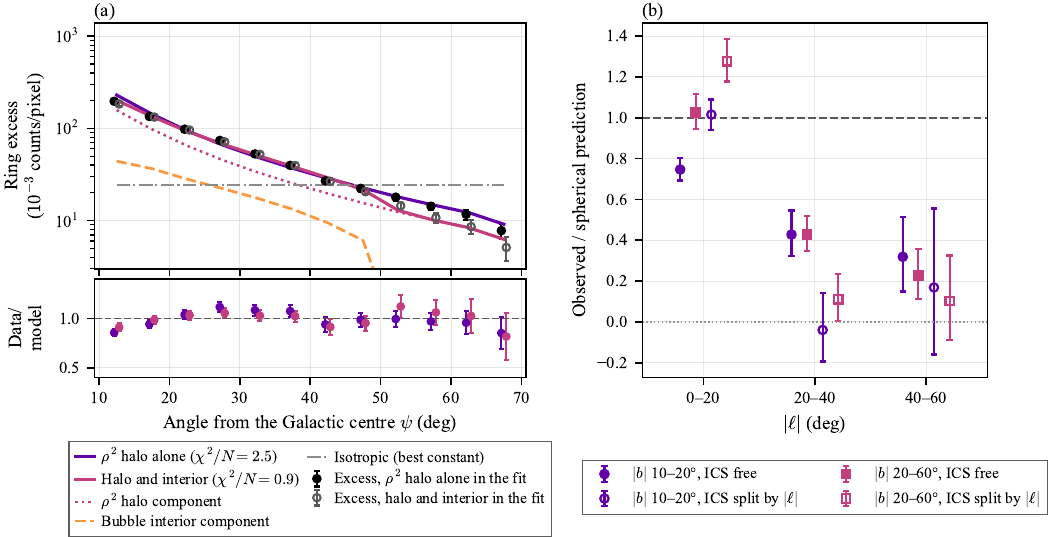}
\caption{\label{fig:morph} \rtwo{(a) Ring profile of the 10--40~GeV excess in rings of $5\degree$ in the angle $\psi$ from the Galactic Centre (top), with the data over the model underneath. The excess is the data minus every fitted component except the signal components. It is shown for the fit with the $\rho^2$ halo alone (black) and for the fit with the halo and the uniform bubble-interior component of Sec.~\ref{sec:morphology} (grey). The lines are the fitted $\rho^2$ halo alone ($\chi^2/N = 2.5$ over 12 rings, purple) and the halo plus interior ($\chi^2/N = 0.9$, pink), with the halo (dotted) and interior (dashed) components of the latter drawn separately and the best constant isotropic emission (dash-dotted, grey). (b) Observed over predicted halo coefficient for the six latitude--longitude pieces, where the prediction is the spherical emissivity fitted to the four $\psi$ shells, with the IC template free (filled) and also split by longitude (open). Dashed: ratio of one. Ratios and their 16--84\% intervals are taken sample by sample from the same posterior draws. Statistical only, template-level.}}
\end{figure*}

\paragraph{Shape of the excess.} We characterise the shape of the excess by its radial profile, by splitting the
$\rho^2$ template into pieces, and by comparing the two hemispheres.

The radial ring profile of the excess is the data minus every fitted component except the halo, taken from the
baseline fit that includes the halo template. It follows the emissivity scaling $q_\gamma\propto\rho^p$ with
$p = 1.64$ when a constant offset is free and $p = 1.98$ without one. The $\rho^2$ halo of the baseline fit describes
the rings to within about 15\% ($\chi^2/N = 2.5$ over 12 rings\rtwo{, where $N$ is the number of rings}), and adding the \rtwo{uniform} bubble-interior component \rtwo{introduced below} to the fit improves this to $\chi^2/N = 0.9$ (Fig.~\ref{fig:morph}\rtwo{a}). A ring profile taken from the residual of a fit
without a halo is steeper, because the foreground templates then take up part of the emission that the halo template takes up when it is included.

We test the morphology directly by dividing the $\rho^2$ template into spatial pieces with independent
normalisations, fitted jointly with all other components. The first split uses four shells in the angle $\psi$ from the Galactic Centre, where $\cos\psi = \cos l\cos b$, with
edges at $\psi = 20\degree$, $30\degree$ and $45\degree$ chosen so that each shell holds a similar share (21--28\%) of
the predicted halo counts at 20.8~GeV. In these shells the pieces follow $\rho^2$ averaged over direction, with the
outer three at 1.22, 1.22 and 1.12 times the innermost over 10--40~GeV. Split by latitude, however, the emission
fitted by the halo template brightens away from the plane, \rtwo{with the foregrounds shared between the pieces (Sec.~\ref{sec:disk} fits each band alone),} by $1.35\,[1.26, 1.43]$ \rtwo{(posterior median and 16--84\% interval)} at $|b| = 20$--$40\degree$ and
$1.72\,[1.57, 1.86]$ at $40$--$60\degree$ relative to $10$--$20\degree$, and its spectrum softens, peaking at 20.8~GeV at $10$--$20\degree$ but at 7.3--12.3~GeV at $40$--$60\degree$, where relative to
$10$--$20\degree$ the halo coefficient is 2.42 times larger at 7.3~GeV but 1.50 times at 20.8~GeV. Split further by longitude, the pieces at
$|l| > 20\degree$ fall to 0.23--0.43 of the spherical prediction, and the six latitude--longitude pieces (the three latitude bands, each split at $|l| = 20\degree$) together \rtwo{depart from} a single $\rho^2$ normalisation \rtwo{by} $\chi^2 = 85.1$ for 6 degrees of freedom (Fig.~\ref{fig:morph}\rtwo{b}),
as do the quadrant and rotated partitions of Ref.~\cite{Abazajian2026}.
Freeing the IC template in longitude as well leaves this structure in place ($\chi^2 = 99.9/6$, open symbols in
Fig.~\ref{fig:morph}), although within each longitude column the halo and IC coefficients are strongly
anticorrelated ($r = -0.6$ to $-0.93$), so the split between them there is poorly determined. With Loop~I free in
sign, the latitude and longitude pattern is unchanged.

\rtwo{Over 10--40~GeV the halo coefficient differs between north and south by less than $2\sigma$.} At $|l| < 20\degree$ it is $0.51$ in the north and $0.46$ in the south (each $\pm0.02$), and it is about $0.26$
in both at $|l| > 20\degree$.

The excess is therefore centrally concentrated but not described by a single spherical template. Its high-latitude, $|l| < 20\degree$ piece lies
largely within the \textit{Fermi} bubbles: about 82\% of that piece's predicted counts fall inside the outline of
Ref.~\cite{Su2010} (81\% for our outline and 86\% for outline E~\cite{Cholis2026}). \rtwo{A fit restricted to the sky outside the bubbles is not a clean test with our templates. The residual-derived
bubble templates extend over the whole region of interest, so a fit that drops them assigns their emission outside
the outline to the halo template. On a noiseless mock built from the baseline fit, dropping them raises the fitted
halo coefficient by 1.1 to 1.7 times its statistical uncertainty at 12.3--35.1~GeV. Abazajian et
al.~\cite{Abazajian2026} find the component outside the footprint of the flat \textit{Fermi}
bubbles~\cite{Ackermann2014} at nominal significances of about 4--16$\sigma$ over 5.6--130~GeV, falling to
1--7$\sigma$ when the enlarged boundaries of the CHZ set are also excluded.} 

\paragraph{Shape under the CHZ set.} We repeat this latitude, angle and longitude split with the CHZ set, on the two-dimensional baseline and on SA50 R12. The halo coefficient is
positive in every latitude band. Over 10--40~GeV at most one posterior sample in $4.7\times10^{4}$ lies at or below
zero in any band, except at $|b| = 40$--$60\degree$ on the baseline, where 26 do (a fraction $6\times10^{-4}$).

Relative to the amplitude with our bubbles on the baseline, the amplitude summed over 10--40~GeV at
$|b| = 10$--$20\degree$, $20$--$40\degree$ and $40$--$60\degree$ is 0.54, 0.47 and 0.56 with the CHZ set on the
baseline, and 0.87, 0.68 and 0.76 with the CHZ set on SA50 R12. It rises with latitude, by factors of 1.20 and 1.80
relative to $|b| = 10$--$20\degree$ on the baseline, consistent with those found with our bubbles, and by smaller
factors of 1.07 and 1.51 on SA50 R12. \rtwo{The split departs from} a spherical emissivity on \rtwo{both foregrounds}
($\chi^2 = 19.7$ and 24.6 for 6 degrees of freedom, $p = 3\times10^{-3}$ and $4\times10^{-4}$), although the departure
is smaller than with our bubbles and on the baseline the $|b| = 10$--$20\degree$ row alone is consistent with one
($p = 0.14$). In the ring profile with the halo in the fit, the emissivity power is 1.78 [1.64, 1.90] with a constant
offset and 2.00 without one on the baseline, and 1.86 [1.78, 1.92] and 2.04 on SA50 R12, and on both foregrounds
$\rho^2$ describes the profile far better than isotropic emission ($\chi^2/\nu = 4.2$ against 51 on the baseline and
2.6 against 131 on SA50 R12\rtwo{, with $\nu = 10$ degrees of freedom}).

The spectrum differs between the two foregrounds. On the baseline, the halo coefficient peaks at 20.8~GeV at
$|b| = 10$--$20\degree$, is flat from 7.3 to 20.8~GeV at $20$--$40\degree$ and falls from 7.3 to 35~GeV at
$40$--$60\degree$. On SA50 R12 it peaks at 20.8~GeV in no band. It is highest at 7.3~GeV in every band, flat to
20.8~GeV at $|b| = 10$--$20\degree$ and falls from 7.3 to 35~GeV at higher latitudes. Two caveats apply. SA50 R12 does
not reproduce the amplitude of the published reference comparison (Sec.~\ref{sec:galprop3d}), and our implementation
of the CHZ method does not recover their published bubble edges (Sec.~\ref{sec:chzres}).

\paragraph{What a spherical halo leaves behind.} A spherical halo leaves similar residuals whether it is described by
the single $\rho^2$ template or by the four $\psi$ shells with free coefficients (Table~\ref{tab:sphres},
Fig.~\ref{fig:sphres}). Including either reduces the 20.8~GeV residual inside outline E from $+9.3\%$ of the total
emission to $+3.0\%$ ($\rho^2$) and $+2.9\%$ ($\psi$ shells), by about two thirds, and removes almost all of it at
$|b| = 10$--$20\degree$. The outline of Su et al.\ gives the same residuals as outline E to within 0.3 percentage
points. The remaining residual follows the bubble outline, with an excess of 3--5\% at $|b| = 20$--$60\degree$,
$|l| < 20\degree$ and a deficit of similar size beside it at $|l| = 20$--$40\degree$, strongest at the southern cap of
the bubbles. Freeing the radial profile changes these residuals by 0.1 to 1.6 percentage points. The excess at
$|b| = 20$--$60\degree$, $|l| < 20\degree$ falls from 4.3\% to 3.2\% but remains, and the deficit beside it deepens,
so the departure from sphericity is mainly angular rather than radial. 

\begin{table}
\caption{\label{tab:sphres} Residual (data minus model) at 20.8~GeV in per cent of the total emission, for the baseline
fit without a halo, with the $\rho^2$ halo, and with a spherical halo of free radial profile ($\psi$ shells). In
brackets, $|d-m|/\sqrt{d}$ for the counts summed over each region, a statistical-only measure of the residual rather
than a detection significance.}
\begin{ruledtabular}
\begin{tabular}{lccc}
Region \rtwo{(deg)} & No halo & $\rho^2$ & $\psi$ shells \\
\hline
$|b|$ 20--60, $|l| < 20$ & $+6.4$ (6.9) & $+4.3$ (4.7) & $+3.2$ (3.5) \\
$|b|$ 20--60, $|l|$ 20--40 & $-2.0$ (1.9) & $-2.9$ (2.8) & $-3.6$ (3.5) \\
$|b|$ 10--20, $|l| < 20$ & $+10.1$ (8.0) & $-2.3$ (2.1) & $-0.8$ (0.7) \\
Inside outline E & $+9.3$ (11.5) & $+3.0$ (3.9) & $+2.9$ (3.8) \\
Outside outline E & $-5.3$ (9.4) & $-1.8$ (3.1) & $-1.8$ (3.0) \\
\end{tabular}
\end{ruledtabular}
\end{table}

\begin{figure*}
\includegraphics[width=\textwidth]{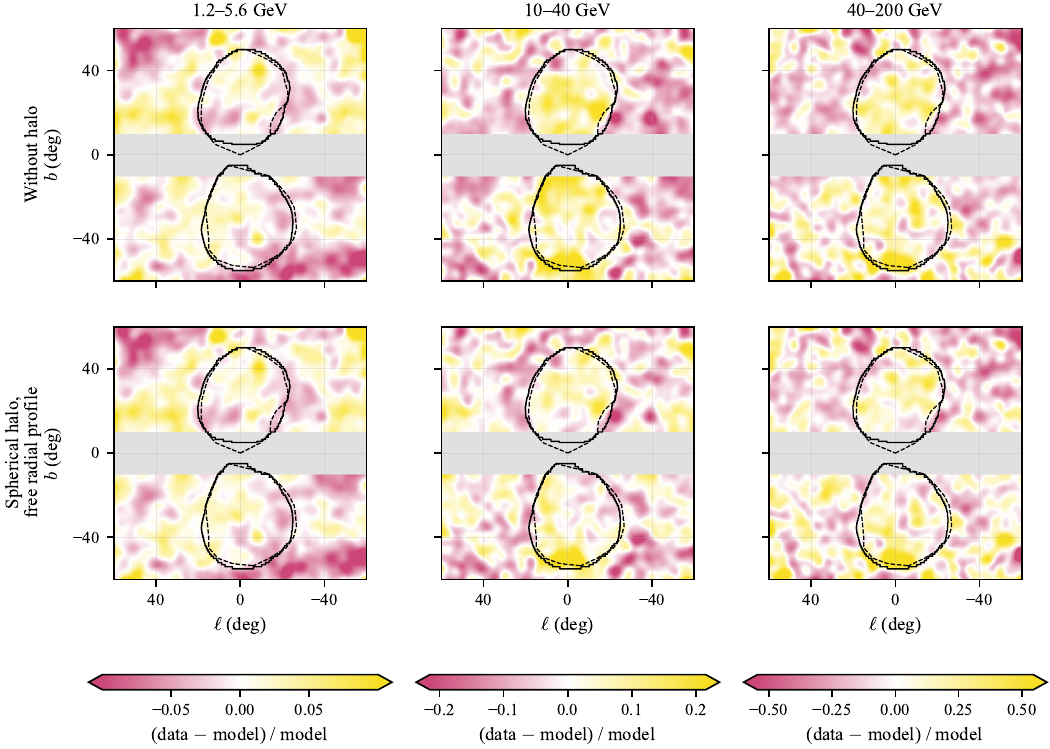}
\caption{\label{fig:sphres} Fractional residual $(\mathrm{data}-\mathrm{model})/\mathrm{model}$, smoothed by
$2\degree$, without a halo (top) and with a spherical halo of free radial profile (bottom), at 1.2--5.6, 10--40 and
40--200~GeV. Solid: outline E, the edge of the CHZ Flat Bubbles~\cite{Cholis2026}. Dashed: Su et al.~\cite{Su2010}. One colour scale per band\rtwo{, running to the 98th percentile of the absolute no-halo residual at $|b| \geq 10\degree$, with arrowheads on the colour bars marking values beyond it. The band $|b| < 10\degree$ is masked and shown in grey}. Statistical only.}%
\end{figure*}

\paragraph{A spherical halo plus a bubble interior.} %
To see whether that remainder is a separate component, we add to the spherical halo one template that is uniform inside
a bubble outline. We use two, our outline and outline E (Sec.~\ref{sec:families}). The coefficient is free in sign in every energy bin.
On a noiseless mock built from the baseline fit, which contains the halo but no interior component, the fit recovers
the halo coefficient used to build the mock within $1\sigma$ and returns the new component within $1\sigma$ of zero
over 10--40~GeV, so \rtwo{on this mock} the interior template does not take flux from the halo template.
\rtwo{The converse test starts from a sky with the interior and no halo. A noiseless mock built from the baseline fit in
which the interior of our outline replaces the halo, fitted as the data are, with the halo template and no interior
template, gives a halo coefficient that is flat over 12.3--35.1~GeV within its statistical uncertainty of about 0.02
(0.168, 0.173 and 0.166 in the pixelwise MCMC, largest at 20.8~GeV by 0.005). Over 10--40~GeV it is 48\% of the coefficient fitted to the data. With
outline E the induced coefficient reaches 55\%, and the CHZ-set mock of Sec.~\ref{sec:nullres}, fitted in every energy
bin, reaches 93\% (deterministic fits). In both the largest value is at 12.3~GeV, by less than their uncertainty. An interior with no halo therefore induces about half of the halo
coefficient in the data with no peak beyond its uncertainty (its largest value is at 20.8~GeV with our outline and at
12.3~GeV with outline E), while in the data the 20.8~GeV coefficient exceeds those at 12.3 and 35.1~GeV by 1.5 and
2.9$\sigma$. These comparisons are
statistical only and at template level.}

The fit prefers both components (Figs.~\ref{fig:bint} and \ref{fig:bintspec}). With the interior added, the halo coefficient
falls to 68\% (our outline) and 61\% (outline E) of its 10--40~GeV value and the halo spectrum still peaks at 20.8~GeV. The
interior component has a flat spectrum, $E^2\,dN/dE \simeq (\rtwo{1.1}$--$2.5)\times10^{-4}$~MeV~cm$^{-2}$~s$^{-1}$~sr$^{-1}$
inside the outline from 7.3 to about 170~GeV, and turns negative at 4.3~GeV, the energy at which our bubble templates
are built. Between 7.3 and 59~GeV, adding the interior improves $2\Delta\ln\mathcal{L}$ by 33--188 per bin, and adding
the halo to a fit that already contains the interior improves it by 37--151. Over 10--40~GeV the halo's
$2\Delta\ln\mathcal{L}$ falls from 927 to 334 (our outline) and 270 (outline E) once the interior is present.

With both components, no region has a residual above 1.7 times its Poisson error at 20.8~GeV. For our outline and outline E
respectively, the residuals are $+1.2\%$ and $+0.2\%$ inside the outline, $+1.2\%$ and $+0.9\%$ at
$|b| = 20$--$60\degree$, $|\ell| < 20\degree$, $-0.5\%$ and $-0.8\%$ beside it, and $-1.2\%$ and $-1.3\%$ at
$|b| = 10$--$20\degree$, $|\ell| < 20\degree$. By contrast, the interior without a halo leaves $+6.5\%$ (5.3 times its Poisson error)
for our outline and $+5.5\%$ for outline E at $|b| = 10$--$20\degree$, $|\ell| < 20\degree$, where a spherical halo is
brightest. A halo of free radial profile behaves in the same way, with every shell losing a similar fraction.

\rtwo{The same mock check, repeated with our outline on the OB-star model and on every three-dimensional model,
gives the same outcome. On the OB-star model and all ten three-dimensional models (an additional check, Sec.~\ref{sec:galprop3d}) the halo coefficient
falls to 78--81\% of its value over 10--40~GeV once the interior is added with our outline, against 68\% on the baseline with the same outline. Adding the
interior improves $2\Delta\ln\mathcal{L}$ by 52--118 per bin at 10--40~GeV, and adding the halo to a fit that already contains the interior
improves $2\Delta\ln\mathcal{L}$ by 75--386 per bin.}

The excess is therefore best represented, within our template set, as two components. The first is spherical, centrally
concentrated and peaked at 20~GeV. The second has the flat spectrum of the bubbles, which our bubble templates, built
from the residual at 4.3~GeV, do not follow at higher energy. The second is consistent with the effect that Abazajian
et al.\ find when they rebuild such templates (Sec.~\ref{sec:systematics}), and its spectrum resembles that of the bubbles and has not been fitted with a dark matter channel here. If the second component is bubble emission, a dark matter interpretation would need to account for the first component only, which
Sec.~\ref{sec:dm_interpretation} fits alongside the full excess.

\begin{figure}
\includegraphics[width=\columnwidth]{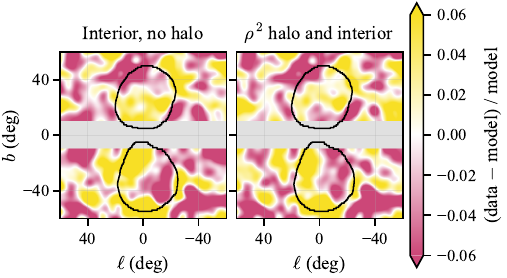}
\caption{\label{fig:bint} Fractional residual at 10--40~GeV, smoothed by $3\degree$, for the baseline fit \rtwo{with a
uniform interior inside our outline and no halo (left) and with the spherical $\rho^2$ halo and the interior (right).
Fig.~\ref{fig:sphres} (middle column) shows the same band without the interior, at $2\degree$ smoothing and with the
spherical halo of free radial profile.} Solid:
outline E, the edge of the CHZ Flat Bubbles. \rtwo{The band $|b| < 10\degree$ is masked and shown in grey. Arrowheads on the colour bar mark values beyond its limits.} Statistical only, template-level.}
\end{figure}

\begin{figure}
\includegraphics[width=\columnwidth]{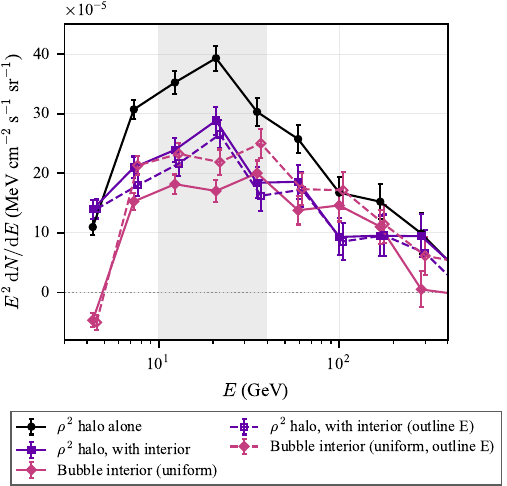}
\caption{\label{fig:bintspec} Mean $E^2\,dN/dE$ inside outline E of the spherical $\rho^2$ halo fitted alone (black),
and of the halo (purple) and the uniform bubble interior (pink) fitted together, for our outline, the geometric bubble outline of Totani~\cite{Totani2025} (solid), and outline E, the edge of the CHZ Flat Bubbles (open, dashed). The shaded band is 10--40~GeV. \rtwo{Error bars are the 16--84\% posterior interval. Negative medians are drawn where they fall, below the dotted zero line.} Statistical only.}
\end{figure}
\end{rtwoblock}

\subsection{Comparison With Cholis, Hu and Zhong}
\label{sec:chzres}
\secfinal
\begin{rtwoblock}

\paragraph{Reproducing their analysis.}
\label{sec:c2dres}
None of the published likelihood comparisons \rtwo{of Ref.~\cite{Cholis2026}} is reproduced in both sign and size (Sec.~\ref{sec:c2d}). The edge comparison
reverses sign, since their outline (outline E) is preferred over that of Su et al.~\cite{Su2010} by $\Delta(-2\ln\mathcal{L}) =
-444$ and $-572$ in their seven- and eight-parameter models but disfavoured in ours by $+156$ and $+212$, and a fit in
our own setup gives the same reversal ($+292$ against $-572$). The three template comparisons keep their sign but reach
less than half the published size, since adding the Cocoons, replacing their halo-like template (a generalised NFW
profile squared, the counterpart of our halo template) by the Inner Bubbles, and adding the halo-like template back to
their final model improve our fit by 41\%, 38\% and 49\% of their values ($-1634$ against $-3998$, $-1927$ against
$-5018$, $-58$ against $-120$). Their per-bin component spectra (\rtwo{their} Table IX) are reproduced
within $2\sigma$ in 41 of 48 entries, with the Cocoons at low energy accounting for most of the rest.

Two further checks \rtwo{show that} our renderings of their templates \rtwo{do not account for the difference}. The first repeats the fit with a Gaussian
prior holding Loop~I near its published normalisation (widths 0.10, 0.25 and 0.50), and the second replaces our
renderings of their Flat, Inner and Cocoon bubbles, and of Loop~I, by the public CHZ maps~\cite{Cholis2026}. In both,
the template improvements stay at 38--49\% of theirs, the edge comparison keeps the opposite sign, and 41 (first check)
and 41 to 44 (second check) of the 48 entries of their Table~IX are reproduced within $2\sigma$. Nor does a scan of the
geometry recover theirs, since on the baseline foreground the scans with each of three bubble outlines (outline E, that of Su et al.\ and a hand-drawn one) all run to the limit of 100 that we imposed on the
Cocoon south-to-north brightness ratio.

The difference from their results may
lie in details of the implementation that the published description does not fix, and results that use a geometry
re-fitted by us are marked $\dagger$ for that reason (Sec.~\ref{sec:families}).

\subsubsection{Separability of the Halo Template From the CHZ Bubbles}
\label{sec:sepres}

Without a halo template, the CHZ set describes the 10--40~GeV data better than our structured bubbles on every
foreground, \rtwo{lowering $-2\ln\mathcal{L}$ by} 635--675 on the two-dimensional foregrounds and 263--490 on the
three-dimensional ones, with one more free normalisation per bin (three bubble templates against two). With a halo
template in both fits the order reverses, and our bubbles fit better by 142--373 on every foreground (the GIEM, which
carries no bubble set, is not compared), because without a halo template the CHZ components take up emission that a $\rho^2$ template otherwise describes.

With the CHZ set, the halo is still well constrained (fractional width 0.08--0.12 at 20.8~GeV, against 0.05 with
our bubbles), but it shares flux with their components. The halo is anticorrelated with their Flat Bubbles and with IC (Sec.~\ref{sec:covariance}), and the summed squared correlation \rtwo{(the sum of the squared posterior correlation coefficients between the halo and each bubble template)} is 0.63--0.77, against 0.02--0.39 with ours. Adding the halo lowers the coefficient of their Flat Bubbles from about 5.2 to
2.9 and that of IC by about a third, and raises that of their Inner Bubbles from about 0.01 to 0.24. With the three geometries
re-fitted by us$^\dagger$ the summed squared correlation is 0.32--0.75.

\subsection{\rtwo{Halo Coefficient} Induced by Foreground Mismatch}
\label{sec:nullres}
\secfinal

The noiseless mocks of Sec.~\ref{sec:nullcal}, built without a halo, measure how much of the halo coefficient
foreground mismatch alone can produce. Each mock is generated from one foreground and fitted with the baseline with our bubbles or with the CHZ set
(Sec.~\ref{sec:nullcal}), and we quote the mean halo coefficient over
10--40~GeV \rtwo{(Fig.~\ref{fig:nullmocks} in Appendix~\ref{app:nullmocks})}. When a GALPROP foreground (SA50 R12, SA0 R12 or the OB-star model) generates the mock, the coefficient is
$-0.072$ to $-0.002$ in both fitters, so mismatch between GALPROP foregrounds produces none of it. \rtwo{Within these mocks,} a different description of the bubbles produces most of it. A mock generated from the baseline with the CHZ set gives 0.329 when
fitted with our bubbles, 93\% of the 0.353 measured in the data, with $2\Delta\ln\mathcal{L} = 784$ from adding the
halo against 927 in the data. A mock generated from the GIEM, whose bubble and Loop~I components were fitted to LAT
residuals, gives 0.283 with our bubbles (80\% of the data) and 0.184 with the CHZ set, as much as the 0.181 measured
in the data with that set. Within these mocks, foreground mismatch therefore produces up to 93\% of the halo
coefficient fitted with our bubbles and all of that fitted with the CHZ set.

With Poisson noise included, the data lie at the edge of what the CHZ sky produces in our fitter. The data amplitude
exceeds 48 of the 50 realisations and the data's $2\Delta\ln\mathcal{L}$ exceeds their 95th percentile (885), but the
difference of 0.024 in amplitude is comparable to its statistical uncertainty ($1.2$--$2.0\sigma$ depending on the
uncertainty adopted). In the CHZ fitter the data amplitude lies above only 21 of the 50 realisations and its
$2\Delta\ln\mathcal{L}$ (136.5) below their 95th percentile (171). The halo coefficient fitted with our bubbles is
therefore only marginally distinguishable from emission described by the CHZ Inner Bubbles and Cocoons. Because the mocks sample
only the generators listed, these fractions are a lower bound on what foreground mismatch can produce.

\end{rtwoblock}

\subsection{Latitude Dependence and the Galactic Centre Excess}
\label{sec:disk}
\secfinal
\begin{rtwoblock}
\paragraph{Latitude dependence.} Fitting each latitude band on its own, with every foreground refitted, the
halo above $20\degree$ peaks at 7.3~GeV and its 10--40~GeV amplitude rises to 1.60 and 2.55 times that at
$10$--$20\degree$. With the foregrounds shared between bands about half of this rise remains (1.35 and 1.72,
Sec.~\ref{sec:morphology}). The rise reported in Ref.~\cite{Cholis2026}, a factor of 2 to 5 between their latitude
ranges, is reproduced only when each range is fitted alone. With shared foregrounds 8--30\% of it survives \rtwo{over 2--40~GeV, and 15--30\% over 10--40~GeV}.
\end{rtwoblock}
Including the Galactic disk in the fit changes the recovered spectrum
qualitatively. With the disk excluded, the global pixel-level fit returns \rtwo{the 20~GeV excess in the halo template}, and
with the disk included, a few-GeV component dominates and the spectrum rolls over
to a peak \rtwo{at a few GeV} (Fig.~\ref{fig:global_overlays}). The likelihood preference of the halo template in the signal band is essentially
unaffected, the pixelwise MCMC giving
$2\sum_k\Delta\ln\mathcal{L}_k\simeq920$ over $10$--$40\,\mathrm{GeV}$ with the
disk included against $\simeq930$ without it, and the deterministic global fit
giving $\simeq850$ and $\simeq870$ for the same quantities\rtwo{, but its coefficient at 20.8~GeV falls from 0.397 to 0.137, about a third of its value with the disk excluded}. \rtwo{The additional low-energy preference appears where the disk reintroduces the inner-Galaxy excess, and the two ROI conventions are consistent with two components, a few-GeV, low-latitude excess with the spectral character of the GCE and a separate $\sim20$~GeV high-latitude component.}

\begin{rtwoblock}
\paragraph{Latitude cut.} Our fits exclude $|b| < 10\degree$. To see how the halo depends on that choice we
repeated the pixelwise fit for the fourteen foreground models of the grid and the baseline, fifteen in all, with the cut lowered from \rtwo{$10\degree$} to
$1\degree$, for the $\rho^2$ and $\rho^{2.5}$ templates (Fig.~\ref{fig:diskcut} shows $\rho^2$). The halo amplitude falls as the
plane is admitted. At $|b| \geq 1\degree$ the 20.8~GeV coefficient is \rtwo{0.39}--\rtwo{0.58} of its value at \rtwo{$|b| \geq 10\degree$}
for $\rho^2$ (median \rtwo{0.42}) and \rtwo{0.18}--\rtwo{0.30} for $\rho^{2.5}$, which is not shown, and the spectral peak leaves the 20.8~GeV bin in every
model at some cut. As the halo coefficient falls, the coefficient of the inverse Compton template rises to \rtwo{1.80} times its value at
\rtwo{$|b| \geq 10\degree$}. The two do not become degenerate. Their correlation stays between \rtwo{$-0.44$} and \rtwo{$-0.50$}, the summed squared
correlation of the halo with all other templates falls from \rtwo{0.42} to 0.30, and the halo's fractional uncertainty barely
changes (\rtwo{0.053} to 0.057). At \rtwo{$|b| \geq 10\degree$, the cut of our fits,} the fifteen models agree on the halo amplitude to \rtwo{6}\%, and
at $|b| \geq 1\degree$ they span 48\%. The spread is ordered by the cosmic-ray source distribution, from
OB-star (\rtwo{$+0.44$} relative to the baseline at the same cut, between the two cuts) through SNR (\rtwo{$+0.24$} to \rtwo{$+0.29$}) and
Yusifov ($+0.03$ to \rtwo{$+0.09$}) to Lorimer (\rtwo{$-0.03$} to \rtwo{$+0.04$}), and the same ordering sets the cut at which the peak
leaves the 20.8~GeV bin. How the halo coefficient changes at a lower cut is thus largely set by the assumed inner-Galaxy cosmic-ray distribution. The scan shows that the spread grows, not which source distribution is correct.
The two halo profiles also trade places by energy rather than by cut. The steeper $\rho^{2.5}$ template \rtwo{gains appreciably only in the three lowest-energy bins (centred at
$1.5$--$4.3$~GeV)}, where the GeV excess of the inner Galaxy enters as the plane is admitted: on the baseline
$2[\ln\mathcal{L}(\rho^{2.5}) - \ln\mathcal{L}(\rho^{2})]$ summed over \rtwo{those bins} \rtwo{goes from $-17$} at \rtwo{$|b| \geq 10\degree$}
to $+978$ at $|b| \geq 1\degree$. At $10$--$40$~GeV, $\rho^2$ is preferred at every cut and for every foreground model with all three bins fitted,
by \rtwo{116} to \rtwo{741}, and \rtwo{on the baseline} the crossover stays between $4.3$ and $7.3$~GeV at every cut.

\begin{figure}
\includegraphics[width=\columnwidth]{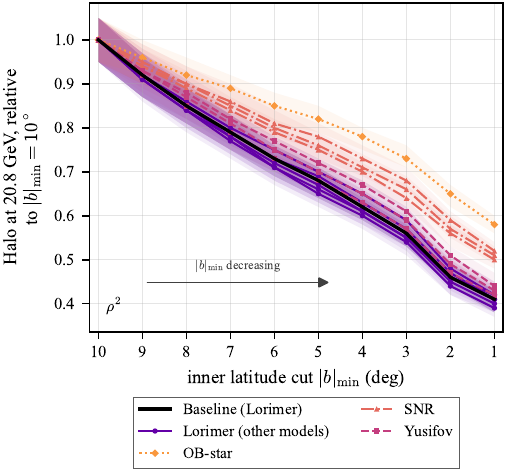}
\caption{\label{fig:diskcut} Halo coefficient at 20.8~GeV for the $\rho^2$ template against the inner latitude cut
$|b|_{\rm min}$, relative to its value at \rtwo{$|b|_{\rm min} = 10\degree$}, for the fourteen foreground models of the grid and
the baseline, coloured by the cosmic-ray source distribution of each model (black: the baseline). Pixelwise
posteriors, bands 16--84\%. Statistical only.}
\end{figure}
\end{rtwoblock}

Three features distinguish the 20~GeV \rtwo{excess} from the GCE and count against a common, inner-Galaxy point-source (e.g.\ MSP) origin. Spatially, it is
recovered over the full high-latitude ROI, $10\degree\le|b|\le60\degree$, with a
\rtwo{smooth $\rhosq$ template, although the emission is not spherical over that range (Sec.~\ref{sec:morphology}),} whereas the GCE is
concentrated within the inner $\sim\!10^\circ$ and prefers a steeper
$\rho^{2.5}$ morphology~\cite{Totani2025}. Spectrally, its 20~GeV peak lies an order of
magnitude above the few-GeV peak of the
GCE~\cite{Calore2015,BartelsKrishWeniger2016} and well above the exponential
cutoff at $E_c\sim1$--$4$~GeV that characterises $\gamma$-ray pulsar and MSP
spectra~\cite{Abdo2013_2PC}, which suggests that the two are not the same population.
Methodologically, the high-latitude region is less exposed to the source
confusion that leaves the inner-Galaxy GCE ambiguous between dark matter and
unresolved MSPs, and a smooth diffuse component is the natural expectation at these latitudes
rather than a clumpy unresolved-source population, or a transported leptonic
population, which Sec.~\ref{sec:hardic} tests directly. \rtwo{For the diffuse models with the most centrally concentrated cosmic-ray source
distributions the halo template also takes up flux at a few GeV near the inner edge of the ROI, and the separation is
correspondingly less clean (Sec.~\ref{sec:systematics}).}

\subsection{Foreground Systematics}
\label{sec:systematics}
\secfinal
\begin{rtwoblock}
\paragraph{Foreground and bubble models.} Within the 10--40~GeV band the posterior fixes the halo coefficient to
about 5\% per bin, while the choice of foreground and bubble templates moves it by a factor of 1.7, from 0.236 to \rtwo{0.409}
at 20.8~GeV across the two-dimensional members of Sec.~\ref{sec:envres}. The extraction is therefore limited by the
foreground model rather than by the photon statistics. Figure~\ref{fig:envdm} shows what that range does to the halo
spectrum and to its dark matter reading. The two-dimensional members of Sec.~\ref{sec:envres} give $\mathcal{R}_{b\bar b}$ from
\rtwo{2.77} to \rtwo{6.96} about the baseline value of \rtwo{5.17}, a factor of $0.53$--$1.35$, and $\mathcal{R}_{WW}$ from \rtwo{2.25} to \rtwo{5.57}
about \rtwo{4.23}. \rtwo{These spreads are the foreground term of the tension bands built in Sec.~\ref{sec:tension}. In both channels
the lowest value comes from the baseline with the CHZ templates and the highest from the OB-star model with ours.}

\begin{figure*}[tp]
\includegraphics[width=\textwidth]{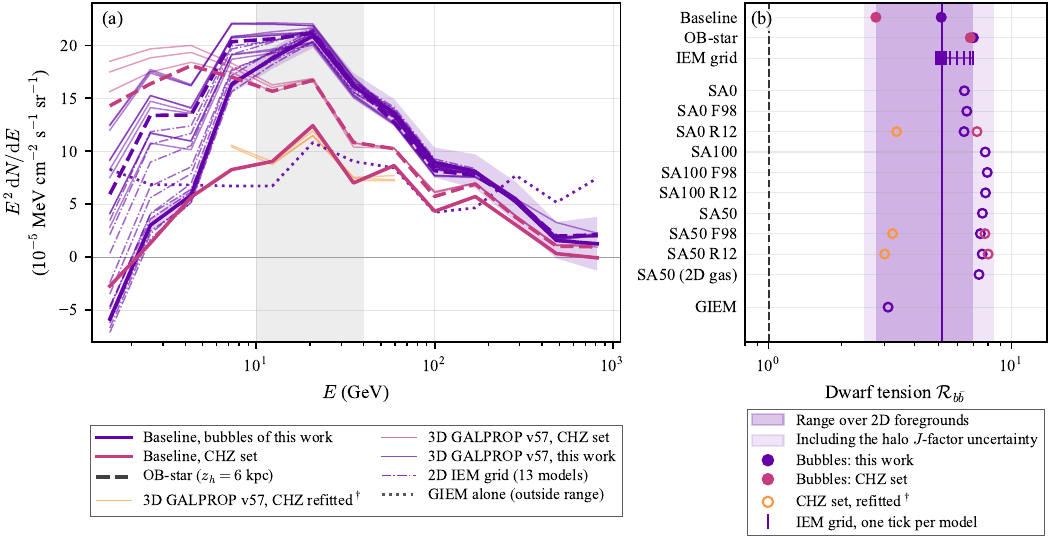}
\caption{\label{fig:envdm} Foreground systematics of the halo and of its dark matter reading. \rtwo{(a)} The $\rho^2$ halo
spectrum, $E^2\,\mathrm{d}N/\mathrm{d}E$ averaged over the region of interest, for the baseline foreground with the
bubble templates of this work (solid, with its 16--84\% band) and with the CHZ templates at their published
geometry, for the OB-star model with $z_h = 6$~kpc with each set (dashed), for the three-dimensional GALPROP v57 models
\rtwo{with our bubbles, and with the CHZ set on three of them} (thin lines, a cross-check whose amplitude does not match the published reference comparison, Sec.~\ref{sec:galprop3d}), and for the
GIEM \rtwo{fitted without a bubble template} (dotted, shown but kept outside the quoted range, Sec.~\ref{sec:families}), \rtwo{and for the 13 further grid models with our bubbles (thin dash-dot lines, to 169~GeV)}. The grey band marks
10--40~GeV. \rtwo{(b)} The dwarf-spheroidal tension $\mathcal{R}_{b\bar b}$ of a prompt $b\bar b$ fit to each member's
spectrum, \rtwo{one row per foreground model with one marker per bubble set, and one row for the further grid models with a tick per model}. Filled symbols are the two-dimensional members that define the range, open symbols the three-dimensional
cross-checks and the GIEM\rtwo{, with the model labels of Sec.~\ref{sec:families}}. The darker band is the range over the two-dimensional members, the lighter band adds the halo $J$-factor \rtwo{uncertainty} in quadrature, and the vertical line marks the baseline\rtwo{, with the dashed line at $\mathcal{R}_{b\bar b}=1$}. Colour gives the bubble set \rtwo{in both panels}: purple for this work, pink for the CHZ set at its published geometry, and orange for the CHZ geometry re-fitted by us ($\dagger$, \rtwo{not relied on and outside every range, Sec.~\ref{sec:families}}). Statistical uncertainties only, at template level.}
\end{figure*}

\paragraph{The interstellar-emission grid.} \rtwo{The two-dimensional members are} drawn from a grid of 14 interstellar-emission models
with physically varied inputs, following Ackermann et al.~\cite{Ackermann:2012pya}. The grid spans four cosmic-ray source
distributions (Lorimer, supernova-remnant, Yusifov and OB-star), diffusion halo half-heights $z_h = 4$, 6, 8 and
10~kpc, radial boundaries $R = 20$ and 30~kpc, spin temperatures $T_S = 150$~K and $10^5$~K, and $E(B-V)$ cuts of 2
and 5, with the gas and IC templates regenerated and the bubble templates re-extracted for every model. Within this two-dimensional basis the cosmic-ray model moves the halo amplitude by a few per cent. What the grid moves is the spectral shape, and with it the fitted mass. \rtwo{In the pixelwise fits of Table~\ref{tab:dmfits} the fitted mass is lower for the supernova-remnant and OB-star
distributions than for the Lorimer and Yusifov ones.}

The models with the most centrally concentrated cosmic-ray source distributions return a halo spectrum with additional
flux below 5~GeV that is essentially absent from the baseline and the Lorimer models. Integrated over 1.5--4.3~GeV
relative to the 10--40~GeV band it is 0.05 for the baseline and 0.57 for the OB-star model. Split by latitude, the excess
over baseline is $+0.38$ at $|b| = 10$--$20\degree$, $+0.26$ at 20--40$\degree$ and $+0.06$ at 40--60$\degree$, so it
sits at the inner edge of the region of interest, where a residual of inner-Galaxy origin would be expected. It is
consistent with the halo template absorbing emission these diffuse models under-produce at mid-latitude rather than with
a property of the halo. Refitting the annihilation spectrum without the low-energy bins reduces the spread in best-fit
$\bbar$ mass across the grid from a factor of 2.15 to 1.23, and the seven Lorimer models then span 614--644~GeV across
both radial boundaries, both spin temperatures, both dust cuts and halo heights from 4 to 10~kpc.

\paragraph{Template composition.} Single changes to the template set, made one at a time on the baseline, are checks on the envelope. Dropping the Loop~I templates leaves $\mathcal{R}_{b\bar b} \simeq 4.9$. Replacing the structured
bubble templates by a single flat one returns 0.718 of the baseline halo at the peak bin and gives
$\mathcal{R}_{b\bar b} \simeq \rtwo{3.4}$ and $\mathcal{R}_{WW} \simeq 2.6$. \rtwo{It is superseded by the bubble-interior test of Sec.~\ref{sec:morphology}.} Reweighting the IC latitude gradient to bracket the vertical scale height of the radiation
field moves $\mathcal{R}_{b\bar b}$ by $\times0.91$--$1.05$ and $\mathcal{R}_{WW}$ by $\times0.93$--$1.01$. Dropping the bubble templates without
replacement raises the halo \rtwo{flux integrated over 5--100~GeV to $\times1.86$} of baseline, which we treat as a stress test rather than a model.

\paragraph{The GIEM.} With the GIEM as the only diffuse component (Sec.~\ref{sec:families}), the halo coefficient is 0.206 at 20.8~GeV, about
half of the 0.397 of the baseline with our bubbles. Its $\bbar$ fit gives $\mathcal{R}_{b\bar b} = 3.1$, inside the range, so including it would not move the lower edge of the band. Under the GIEM the excess still peaks in the 20.8~GeV bin, but its spectrum differs from a featureless constant by only \rtwo{$\chi^2 = 25.8$ for 12 degrees of freedom ($p = 0.011$)}, unlike that of the GALPROP models fitted with statistical errors. This goodness-of-fit test against a
constant is distinct from the template-preference statistic of Sec.~\ref{sec:morphology}. With the GIEM, the spectral
peak on which the annihilation reading rests is only marginally preferred.
\end{rtwoblock}

\paragraph{Symmetry and consistency.} Splitting the ROI into northern ($b>0$) and southern ($b<0$) hemispheres returns consistent spectra \rtwo{for the halo template over the full ROI}. The peak-bin amplitudes agree to $1.4\sigma$, with at most
a mild southern softening, as expected for emission symmetric about the plane and
contrary to \rtwo{Loop~I, which lies mainly in the north, and to the cocoon in the southern \textit{Fermi} bubble}. Splitting the ROI by Galactic quadrant tests the same symmetry
azimuthally. Dropping any one quadrant and refitting the remaining three moves
the dSph tension by a factor $0.90$--$1.23$ relative to baseline, a narrower
spread than the foreground envelope of Sec.~\ref{sec:tension}, with best-fit
masses spanning $396$--$830$~GeV. Fits to a single quadrant are not constraining,
since a quarter of the photons cannot fix the spectral shape, and we do not use
them. An injection-null test, in which the baseline templates are fit to a
mock whose truth contains a deliberately distorted IC morphology
but no halo, recovers \rtwo{about} a $2\sigma$ halo amplitude, roughly $10\%$
of the observed \rtwo{ROI-averaged $2.0\times10^{-4}\,\mathrm{MeV\,cm^{-2}\,s^{-1}\,sr^{-1}}$} peak. \rtwo{A different description of the bubbles can induce a larger halo coefficient (Sec.~\ref{sec:nullres}).} Two closure tests on deterministic mock counts bracket these
checks. Fitting a mock built from the no-halo global model with the baseline
template set returns a halo amplitude of $1.4\%$ of the measured one at the peak
bin and $0.5\%$ summed across the signal band, so \rtwo{on this mock} a self-consistent foreground does not manufacture a halo \rtwo{coefficient}. Injecting the baseline $\rhosq$ halo into the same
mock and refitting recovers it at $1.009$ of the injected amplitude at the peak
bin and $1.003$ across the signal band, so, on a mock built from the same templates, the extraction is unbiased at that level, below the foreground range quoted above.

\begin{rtwoblock}
\paragraph{Rebuilding the bubble templates.}
\label{sec:rebuild}
Rebuilding Totani's residual-derived bubble templates inside simulated skies has been shown to produce a spurious NFW
component on skies without a halo and, for an injected halo, to recover 93\% at 20.8~GeV but 42\% at the 4.3~GeV
construction energy~\cite{Abazajian2026}. Our templates are residual-derived in the same way, so we repeated this test
on them. Rebuilt from the data, they reproduce the originals to within 2\% in total
counts, with a pixel correlation above 0.9999. On skies built from the baseline with our bubbles and no halo, rebuilt
templates induce a halo coefficient at 20.8~GeV of 0.035 in the noiseless mock and $0.058\pm0.016$ over five Poisson
realisations, compared with 0.000 and 0.010 for fixed templates. With the measured halo injected, they recover
0.98--1.02 of it over 10--40~GeV but 0.55 at 4.3~GeV over five Poisson realisations (0.91 in the noiseless mock), in line with Ref.~\cite{Abazajian2026}. Rebuilding therefore adds 0.035--0.048 to the halo coefficient induced at 20.8~GeV with our
bubbles, 9--12\% of the 0.397 fitted to the data. On a sky
built with the CHZ set and no halo, it lowers the induced halo coefficient over 10--40~GeV from 0.329 (Sec.~\ref{sec:nullres}) to 0.22--0.26.
\end{rtwoblock}

\subsection{Posterior Covariance Between the Fitted Templates}
\label{sec:covariance}
\secfinal

\begin{rtwoblock}
\begin{figure*}[htbp]
\includegraphics[width=\textwidth]{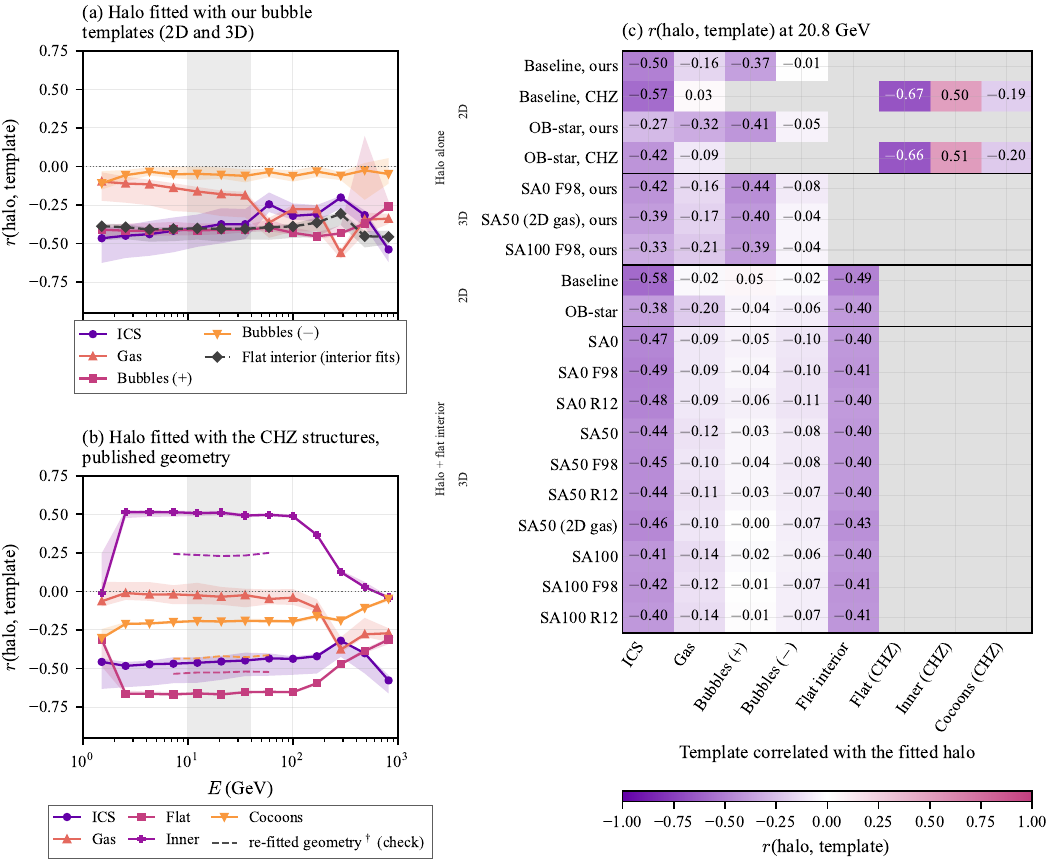}
\caption{Posterior correlation between the halo normalisation and the templates it trades flux with, from the pixelwise
MCMC (Sec.~\ref{sec:mcmc}), for every foreground model of the envelope (Sec.~\ref{sec:envres}). \rlatest{The correlation
coefficient $r$ with the halo is shown for the IC and gas templates, our positive and negative bubble templates, the flat
bubble-interior component of Sec.~\ref{sec:morphology}, and the Flat, Inner and Cocoon structures of the bubble model of
Ref.~\cite{Cholis2026}. \emph{(a)} The energy dependence of $r$ for the fits with our bubbles (the flat interior from the
fits that include it, dashed) and \emph{(b)} for the fits with the CHZ structures at their published geometry. Dashed lines
in (b) are the medians of the three re-fitted geometries ($\dagger$), which have only the five bins from 7.3 to 59~GeV.
Lines are the median over the two- and three-dimensional foreground models together and bands the full range. Panels (a)
and (b) use every fit, including those only in Table~\ref{tab:covariance_rows}. The shaded region is the $10$--$40$~GeV
signal band. \emph{(c)} $r$ at the $20.8$~GeV peak bin. Rows are the halo fits of the two-dimensional foreground models \rtwo{(labels as in Sec.~\ref{sec:families})} with our bubbles (``ours'') or with the CHZ structures at their published geometry
(``CHZ''), \rtwo{three of the ten three-dimensional models with our bubbles (SA0 F98, SA50 with two-dimensional gas and SA100 F98, the three whose correlations differ most), and the fits with the flat interior component. The other three-dimensional fits, including all those with the CHZ structures (published or re-fitted by us, $\dagger$, a check only), are in Table~\ref{tab:covariance_rows} of Appendix~\ref{app:covrows}. Lines separate, from the top, the two- and three-dimensional (2D, 3D, grey labels) fits with the halo alone from the two- and three-dimensional fits with the flat interior component added.} Grey cells: template not in that fit. Colour runs
from $r=-1$ (purple) through zero (white) to $+1$ (pink).} The isotropic,
point-source and Loop~I templates are omitted: their correlation with the halo is at most $0.2$ in magnitude in
every fit and bin.}
\label{fig:covariance}
\end{figure*}

The pixelwise MCMC varies all template normalisations simultaneously, so the halo uncertainty quoted above already
marginalises over the other components. The fits also give the posterior covariance between them, which we show for
every foreground model of the envelope rather than for the baseline alone (Fig.~\ref{fig:covariance}).

The halo coefficient is correlated mainly with two kinds of component (Fig.~\ref{fig:covariance} \rtwo{and Table~\ref{tab:covariance_rows}}). The first is
inverse Compton, with a correlation at 20.8~GeV of $-0.27$ to $-0.57$ across the foreground models that is roughly
constant with energy. The second is whichever template describes the bubble interior: the positive bubble template
with our bubbles ($-0.37$ to $-0.44$), the Flat component with the CHZ set ($-0.65$ to $-0.67$), and, when it is added
to fits with our bubbles, the flat interior component ($-0.40$ to $-0.49$), which then takes over this role from the
positive template. The halo is therefore not independent of the inverse-Compton and bubble-interior components, and
the division of flux between them is consistent with the halo amplitude moving between foreground models
(Sec.~\ref{sec:envres}).

The isotropic, point-source and Loop~I templates are correlated with the halo at no more than 0.2 in magnitude in every fit and every bin
(Fig.~\ref{fig:covariance}). This concerns covariance between templates within one source model
and one diffuse model, and does not bound the sensitivity of the halo to the source model, which
Sec.~\ref{sec:globalresults} addresses. On the baseline the isotropic and Loop~I~A templates are strongly
anticorrelated with each other ($r = -0.93$), a degeneracy that does not involve the halo and is consistent with the
halo amplitude being recovered unchanged when the Loop~I templates are removed. This degeneracy does not affect the
dwarf tension of Sec.~\ref{sec:tension}, which depends on the halo amplitude and the $J$-factors. The halo's correlations enter the tension only through the spread of the halo amplitude across foreground models
noted above.
\end{rtwoblock}

\subsection{An Injected Inner-Galaxy Electron Population}
\label{sec:hardic}
\secfinal

The alternative that the \rtwo{foreground-mismatch} checks \rtwo{above} do not address is astrophysical rather
than instrumental. A hard leptonic population injected in the inner Galaxy and
transported outwards produces IC emission that is smooth, hard and
centrally concentrated, and which peaks at tens of GeV, a description that
matches the \rtwo{morphology of the excess} qualitatively and which no template in the baseline model
can absorb by construction. Scenarios of this kind reproduce both the spectrum
and the angular profile of the GCE~\cite{Petrovic2014, Cholis2015}, and the
hadronic analogue has been developed
similarly~\cite{CrockerAharonian2011, CarlsonProfumo2014}. We therefore
construct such a component explicitly and fit it alongside the halo.

\subsubsection{Templates}

Electrons are injected in the inner Galaxy and propagated with a Syrovatskii
Green's function~\cite{Syrovatskii1959}, using transport parameters matched to the GALPROP model
underlying our own baseline, namely $D_0 = 7.98\times10^{28}\,
\mathrm{cm^2\,s^{-1}}$ at $4$~GV, $\delta = 0.33$ and a diffusion halo
half-height $z_h = 6$~kpc with image-charge boundaries. The resulting electron
distribution is scattered on the full interstellar radiation field with an exact
Klein--Nishina kernel~\cite{BlumenthalGould1970}. Klein--Nishina treatment is necessary rather than
optional here, since at 20~GeV the starlight channel has
$4\gamma\epsilon/m_ec^2 \approx 0.9$ for a target photon energy $\epsilon\approx1$~eV and the Thomson approximation would
misplace the spectral peak~\cite{Moderski2005}.

Two $\lambda$-calibrated templates are built, impulsive injections of
$10^{52}$~erg at a single point at the Galactic Centre with an injection index
$\Gamma = 2.2$, calibrated to diffusion--loss lengths
$\lambda(0.7\,\mathrm{TeV}) = 3$ and $6$~kpc by scaling $D_0$ to $2.4$ and
$9.6$ times the GALPROP value, since at the baseline coefficient the
loss-limited length in this radiation field is ${\sim}2$~kpc. They span
compact to widely extended and are deliberately generous to the scenario in
transport. A further variant replaces the full radiation
field with the CMB alone. It is retained as a labelled positive control rather
than as a physical model. With a uniform target field it gives a more halo-like morphology than the other injected templates, at an injected energy far beyond any Galactic source.

The radiation field on which these electrons scatter carries an inverse-square
bulge and inner-disc term, $U = L/[4\pi c\,(r^2 + r_c^2)]$, with an optical
luminosity of $10^{10}\,L_\odot$, a dust luminosity of $4\times10^{9}\,L_\odot$
and a core radius $r_c = 1$~kpc, together with synchrotron losses at $3\,\mu$G.
These normalisations are order-of-magnitude anchors, compared against the
ISRF maps of Porter et al.~\cite{Porter2017} rather than fitted.
Each template is forward folded with the energy-dependent PSF and fitted using
the same pixel grid, mask and energy binning as every other component, so that
per-bin likelihoods are directly comparable.

A thin double-exponential radiation field with no bulge component was also
tried. In that field the interstellar radiation energy density above
${\sim}1$~kpc falls essentially to the CMB value, a 20~GeV photon requires a
${\sim}2.6$~TeV parent electron whose cooling length confines the emission
below our high-latitude ROI, and the injected component is rejected \rtwo{for that reason rather than for its morphology}. All results reported here use the
bulge-inclusive field.

\subsubsection{Results}
\begin{rtwoblock}
Fitted in place of the halo, the 6 and 3~kpc templates describe the excess worse than $\rho^2$, while the CMB-only control, which needs an injected energy far beyond any Galactic source, fits the ring profile slightly better than $\rho^2$ ($\chi^2/\nu = 3.42$ against 3.85). On the excess left by
the fit without a halo, the pixelwise replacement fits give $\chi^2/\nu = 8.74$ for the 6~kpc template, 16.4 for 3~kpc
and 3.85 for $\rho^2$ (deterministic fits 7.77, 15.6 and 3.24), with the CMB-only control at 3.42. Fitted alongside the halo in the four envelope members, the injected templates leave the halo coefficient over
10--40~GeV at \rtwo{0.99--1.00 of its value at the best fit with our bubbles (0.986 and 0.995) and 0.96--0.97 with the
CHZ set (0.957 and 0.971). At the posterior median these are 0.958 and 0.946, and 0.830 and 0.851, lower because the
electron coefficient is bounded at zero, so that its posterior median lies well above its best fit.} The injected population has not been tested together with the bubble-interior component of Sec.~\ref{sec:morphology}.

\rtwo{Two further fits use the semi-analytic templates\rlatest{, the 6~kpc and 3~kpc templates and}
twelve templates of age 0.3--10~Myr and injection index 1.8--2.6. In the replacement fit, run on the baseline with our
bubbles, each template is fitted in place of the halo and compared with the fit without a halo. Over 10--40~GeV the $\rho^2$ halo improves $2\Delta\ln\mathcal{L}$ by 874 (deterministic fits, which understate the uncertainties), the 6~kpc template by 687 and the 3~kpc template by 341, so the templates fall short of the halo by 187 and 533.}

\rtwo{\rlatest{In the single-normalisation fit one amplitude $\theta$, in units of $10^{52}$~erg, is imposed across the
energy bins, so that each template carries its own emitted spectrum. We fit it pixelwise for the cases of
Sec.~\ref{sec:ie}, leaving out the 814~GeV bin, where both templates predict no counts, and compare each fit with the
fit without the injected template. With our bubbles on the baseline the fit prefers $\theta = 0$ for the 6~kpc
template (68\% interval 0--0.065), which changes $2\Delta\ln\mathcal{L}$ by $-1.5$, $-5.7$ and $-32.1$ at
$\theta = 0.1$, 0.3 and 1, and $\theta = 0.06$ (0.015--0.115) for the 3~kpc template, with
$2\Delta\ln\mathcal{L} = 1.4$. On SA50 R12 with our bubbles it prefers $\theta = 1.71$ (1.49--1.94) for the 6~kpc
template and 0.17 (0.155--0.19) for the 3~kpc template, with $2\Delta\ln\mathcal{L} = 54.4$ and 26.0. With the CHZ
set it prefers, for the 6~kpc template, $\theta = 0$ (0--0.195) on the baseline and $\theta = 0.96$ (0.70--1.22) on
SA50 R12 ($2\Delta\ln\mathcal{L} = 13.6$), and for the 3~kpc template $\theta = 0.38$ (0.33--0.44) and 0.11
(0.05--0.16), with $2\Delta\ln\mathcal{L} = 46.1$ and 3.6. Where the preferred $\theta$ is not zero, the halo
coefficient in each of the three 10--40~GeV bins is 0.75--0.99 of its value without the injected template, with the
lowest values for the 3~kpc template on the baseline with the CHZ set (0.75--0.85) and the 6~kpc template on SA50 R12
with our bubbles (0.77--0.82). In each of these six cases the bins from 1.5 to 4.3~GeV together favour the
population, the bins at 7.3, 12.3 and 20.8~GeV each disfavour it, and summed over the 10--40~GeV bins
$2\Delta\ln\mathcal{L}$ lies between $-25.0$ and $-1.7$.

The same eight fits were also run deterministically, on a grid of $\theta$ from 0.1 to 8. Compared over the bins
the pixelwise fits use, the deterministic best $\theta$ lies within one grid step of the pixelwise value in every case,
and at $\theta = 1$ the halo coefficient, as a fraction of its value without the injected template, differs by at
most 0.024 (0.019 with our bubbles). With the CHZ set the deterministic $2\Delta\ln\mathcal{L}$ is lower by 0.2--2.2
at $\theta = 0.1$ and 0.3 and by 1.7--5.7 at $\theta = 1$. With our bubbles it is higher by 6.2--33.2 at $\theta = 1$
and by up to 11.3 at $\theta = 0.1$ and 0.3 (lower by 1.6 in one case), so these deterministic fits penalise a large
population less than the pixelwise fits do. With our bubbles the deterministic fitter also holds the predicted counts
at a small positive floor wherever they would fall to zero or below, at any $\theta$ of the grid and in the fit
without the injected template alike, in at most 43 of 750\,384 pixels (0.006\%) at
20.8~GeV and in up to 0.8\% of the pixels in the bins from 59 to 482~GeV, and never with the CHZ set.
The twelve age and index templates are fitted
deterministically only, on the baseline, and the two bubble sets were scanned on different grids of $\theta$. With our bubbles, on a grid
from 0.03 to 128, they have a positive best $2\Delta\ln\mathcal{L}$ of 0.4--16.7, at $\theta$ between 0.03 and 64,
with the halo coefficient averaged over the 10--40~GeV bins at 0.96--1.00 of its value without the template. With the
CHZ set, on the grid from 0.1 to 8, nine of them have a positive best $2\Delta\ln\mathcal{L}$ of 23.6--50.4, at
$\theta$ between 0.1 and 4, with that average at 0.80--0.95, and the three 10~Myr templates prefer $\theta = 0$. For
those nine the bins from 1.5 to 7.3~GeV together favour the population and the 10--40~GeV bins together disfavour it. These deterministic fits understate the uncertainties.}}

The version in which the population's inverse Compton and bremsstrahlung emission is computed with GALPROP v57
(Sec.~\ref{sec:ie}) is not converged in the time step at high energy, so \rtwo{the results use the semi-analytic templates}.
\end{rtwoblock}

\section{Dark Matter Interpretation}
\label{sec:dm_interpretation}
\secfinal

\subsection{Annihilation Fit and Cross Section}
\label{sec:annihilation}
\secfinal

\begin{figure}[htbp]
    \includegraphics[width=\columnwidth]{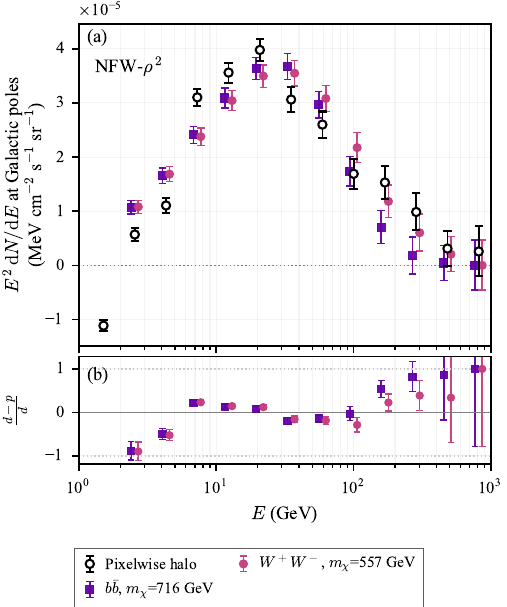}
    \caption{
Spectral fits of prompt PPPC annihilation models to the pixelwise (disk-excluded) $\rho^2$ halo
spectrum.  \rtwo{Panel (a)} shows the extracted halo spectrum together with the
best-fitting $W^+W^-$ and $b\bar b$ annihilation spectra, each fitted with a
single normalisation.  \rtwo{Panel (b)} shows the residuals relative to the best-fitting spectra\rtwo{, $(d-p)/d$, where $d$ is the extracted halo spectrum and $p$ the fitted prediction. The 1.5~GeV bin, where the halo coefficient is negative, is not fitted (Sec.~\ref{sec:dmres})}.  These fits test spectral compatibility only as the PPPC
templates are not refitted spatially at each mass. Error bars are statistical, \rtwo{the 16--84\% interval of} the MCMC posterior, and do not include the foreground-modelling
systematics (Sec.~\ref{sec:systematics}), which are expected to dominate the goodness of fit
($\chi^2/\nu\simeq7.5$ for $\bbar$ and $7.4$ for $\WW$).
}
    \label{fig:annihilation}
\end{figure}
We fit prompt gamma-ray spectra from dark matter annihilation to the global NFW-$\rhosq$
halo. For a self-conjugate WIMP of mass $\mchi$ annihilating with cross section
$\sigv$, the differential flux per unit solid angle toward the poles is
\begin{equation}
\frac{\dd\Phi}{\dd E\,\dd\Omega}
= \frac{\sigv}{8\pi\,\mchi^2}\,\Jpole\,\frac{\dd N}{\dd E}(E;\mchi),
\label{eq:flux}
\end{equation}
where $\dd N/ \dd E$ is the per-annihilation photon yield, taken from
PPPC4DMID~\cite{Cirelli2011}, and $\Jpole = \int_{\rm los}\rho^2\,d\ell \approx
4.0\times10^{21}~\mathrm{GeV^2\,cm^{-5}}$ is the line-of-sight $\rhosq$ integral
toward the Galactic poles for the adopted halo. For each channel we scan $\mchi$,
fit a single normalisation (hence $\sigv$) to the halo spectrum by weighted least
squares, and rank masses by $\chi^2$.

Both the $\bbar$ and $\WW$ annihilation channels reproduce the 20~GeV spectral peak \rtwo{(Fig.~\ref{fig:annihilation})}. For the $\WW$ channel, the fit yields $\mchi \simeq 0.56$~TeV with $\sigv \simeq \rtwo{1.04}\times10^{-24}~\mathrm{cm^3\,s^{-1}}$ (pixelwise posterior, while the deterministic global fit gives $\rtwo{1.01}\times10^{-24}~\mathrm{cm^3\,s^{-1}}$ at the same mass, and the dwarf ratios below use the \rtwo{pixelwise} value). For the $\bbar$ channel, the best-fit parameters are $\mchi \simeq 0.72$~TeV and $\sigv \simeq \rtwo{9.9}\times10^{-25}~\mathrm{cm^3\,s^{-1}}$. While these masses fall within the $0.4$--$0.8$~TeV range reported in Ref.~\cite{Totani2025}, our extracted cross sections sit modestly above their $(5$--$9)\times10^{-25}~\mathrm{cm^3\,s^{-1}}$ bracket. Both remain well above the canonical thermal relic value of $\sigv_{\rm th}\simeq 2.2\times10^{-26}~\mathrm{cm^3\,s^{-1}}$~\cite{Steigman2012}. \rtwo{Part of this offset is normalisation and part is mass. We use an identical NFW-$\rhosq$ profile and pole line-of-sight integral ($\Jpole = 8.93\times10^{14}~M_\odot^2\,\mathrm{kpc^{-5}}$), and the pixelwise halo amplitude lies about 20\% above the cellwise one, ours and that of Ref.~\cite{Totani2025} (Sec.~\ref{sec:globalresults}). Because $\sigv$ scales as $\mchi^2$ at a fixed pole flux (Eq.~\ref{eq:flux}), the rest of the offset is driven by our} $\chi^2$ minimum resting near the upper boundary of the mass band. Refitting the normalisation at $\mchi\simeq0.5$~TeV, where both the $\mchi^2$ factor and the yield change, returns $\sigv\simeq6\times10^{-25}~\mathrm{cm^3\,s^{-1}}$ in the $\bbar$ channel, reproducing the previously reported range. 

We caution that the absolute goodness of fit is poor. At baseline, $\chi^2/\nu \simeq 7.5$ for $b\bar{b}$ and $\simeq 7.4$ for $W^+W^-$, and across the two-dimensional members of the envelope \rtwo{fitted over all bins} the corresponding ranges are 1.8--12.0 and 1.6--8.6. \rtwo{The $\tau^+\tau^-$ channel fits worse than $b\bar b$ and $W^+W^-$ for every foreground model and bubble
template set, with $\chi^2/\nu = 3.8$--40.5 across the other two-dimensional members fitted over all bins.} Because the fit is performed directly on the pixelwise MCMC halo spectrum, with statistical errors taken from the full posterior, this high $\chi^2$ is not an underestimate of the statistical errors. Rather, it is consistent with unmodelled, spatially coherent systematic residuals from the diffuse foreground model, which are
quantified by the foreground-systematics suite in Sec.~\ref{sec:systematics}. The high $\chi^2$ is driven by the statistical weight of the \rtwo{photon-rich bins at low energy}, where small fractional residuals in the diffuse model translate into large deviations. The poor global goodness-of-fit is therefore consistent with the rigidity of a single-component spectral shape against a complex low-energy background, and is not by itself a test of whether the primary anomaly is captured. Comparisons between channels remain meaningful within that limitation, since
each is fitted to the same halo spectrum with the same errors. On that basis our
fit does not reproduce the $\tautau$ solution of Ref.~\cite{Totani2025}. That
channel prefers $\mchi\simeq70$~GeV, an order of magnitude below our $\bbar$ and
$\WW$ solutions, and reaches only $\chi^2/\nu\simeq14.6$, roughly twice the
baseline value and worse than \rtwo{every two-dimensional member} in either channel.
We therefore quote $\bbar$ and $\WW$ throughout.

\rtwo{At the published latitude cut, the} spectral peak sits in the $20.8$~GeV bin for every \rtwo{two-dimensional} foreground model fitted with our bubble templates, and the
fitted mass is well localised at fixed foreground model. \rtwo{It is not well determined across the envelope. The
two-dimensional members give $149$--$746$~GeV for $\bbar$ and $139$--$577$~GeV for $\WW$ (Table~\ref{tab:dmfits}), the
lower end set by the OB-star model with the CHZ templates, whose halo spectrum is nearly flat below
20.8~GeV rather than peaked.} \rtwo{Refitting every member over the bins they share narrows the two-dimensional range (Sec.~\ref{sec:dmres}).} The mass quoted above is therefore conditional on the baseline foreground model. \rtwo{On the baseline the two bubble
template sets give similar masses, 716 and 697~GeV for $\bbar$, but the tension derived from them differs, and it}
carries the band of Sec.~\ref{sec:tension}.

\subsection{Dwarf-Spheroidal Tension and its $J$-Factor Budget}
\label{sec:tension}
\secfinal
We quantify the tension against dSph annihilation limits as the ratio
\begin{equation}
\mathcal{R}(\mchi) \equiv
\frac{\sigv_{\rm halo}(\mchi)}{\sigv^{\rm UL}_{\rm dSph}(\mchi)} ,
\end{equation}
using two independent limit sets, the 11-year global analysis of Hoof,
Geringer-Sameth and Trotta~\cite{Hoof2020} and the 14-year stacked analysis of
McDaniel \textit{et al.}~\cite{McDaniel2024}, both of which build on the
\textit{Fermi}--LAT dwarf program~\cite{Ackermann2015}. An updated legacy
analysis of the dwarf sample has recently appeared~\cite{Circiello2026}. Its
limits are stronger than those of Ref.~\cite{Hoof2020} at our best-fit masses,
so adopting them would raise the tension rather than relieve it, by roughly
$30\%$ in $\mathcal{R}_{b\bar b}$ on our own reconstruction of their stacked
profiles. We do not adopt them here because that analysis reports the $\bbar$
and $\tautau$ channels only, so the $\WW$ leg could not be treated on the same
footing, and we note it as the natural successor set for any re-evaluation of
the tension in both channels.
At the best-fit masses, the Hoof
$\bbar$ limit is $\sigv^{\rm UL}\simeq 1.9\times10^{-25}~\mathrm{cm^3\,s^{-1}}$,
giving $\mathcal{R}_{b\bar b}\simeq\Rbest$, and the $\WW$ channel gives $\mathcal{R}_{WW}\simeq\RWW$.
The McDaniel limits \rtwo{for $\bbar$} yield a
comparable tension. These values are consistent with the $\sim3$--$5\times$
tension quoted in Ref.~\cite{Totani2025}.

\rtwo{One astrophysical $J$-factor uncertainty enters} $\mathcal{R}$ in addition to the
foreground-modelling range established in Sec.~\ref{sec:systematics}
($\mathcal{R}_{b\bar b}\times0.53$--$1.35$). That foreground contribution is the extremal
range over the two-dimensional members, not a quadrature sum over them,
so no assumption of independence is made among the correlated large-scale
components. The members cross each foreground with each bubble set, but the remaining
variations of Sec.~\ref{sec:systematics} are not crossed with them, so the range may
under-cover where two act in the same direction. \rtwo{That} term is the halo $J$-factor:
$\mathcal{R}\propto1/J_{\rm halo}$, with $J_{\rm halo}$ uncertain at the
$\simeq\!48\%$ level, set almost entirely by the local density
$\rho_\odot=0.42\pm0.10~\mathrm{GeV\,cm^{-3}}$ (the scale radius and inner slope
contribute negligibly at the pole normalisation). This is the dominant uncertainty on the halo $J$-factor, as also stressed in Ref.~\cite{Totani2025}. \rtwo{The dwarf $J$-factor uncertainties are already included in the limits of Refs.~\cite{Hoof2020,McDaniel2024}, so no separate dwarf term is added. Quadrature is applied only between the foreground envelope and the halo $J$-factor, which are independent of each other. Because both act multiplicatively on $\mathcal{R}$, they are combined in quadrature in the logarithm, the band being $\mathcal{R} \times \exp \left\{ \pm \left[ \ln^2 f_{fg} + \ln^2(1.48) \right]^{1/2} \right\}$}\rtwo{, where $f_{fg}$ is the lower or upper foreground factor. The result is a band, not a confidence interval}. This gives $\mathcal{R}_{b\bar b}\simeq\Rlo$--$\Rhi$ about a central value of $\Rbest$. Repeating the propagation for the $\WW$ channel gives
$\mathcal{R}_{WW}\simeq\RWWlo$--$\RWWhi$ about a central value of $\RWW$. Its
foreground leg spans \rtwo{$\times0.53$--$1.32$ against $\times0.53$--$1.35$} for
$\bbar$, with \rtwo{the baseline fitted with the CHZ templates} setting the lower edge in both.
The band is therefore not a feature of a single annihilation channel, and the $\WW$ channel
sits closer to the dSph limit throughout. Once the foreground range and the \rtwo{halo} $J$-factor \rtwo{uncertainty are} included, a prompt $s$-wave interpretation remains in tension with the
dSph limits by a factor of \RWWlo{} to \Rlo{} at the most favourable edge, so for the non-informative dwarf priors
of Refs.~\cite{Hoof2020,McDaniel2024} the dwarf limits are not evaded within the uncertainties we propagate.
\rtwo{The dwarf $J$-factors themselves depend on the prior adopted for the ultrafaint satellites. Priors informed by
subhalo formation and tidal stripping weaken the limits of Ref.~\cite{Hoof2020} by a factor of 2 to
7~\cite{Ando:2020yyk}, which would bring the lower edge of the band below unity. The dSph tension is therefore robust
to the uncertainties we propagate but not to the choice of dwarf prior.}

The tension does not close when the annihilation mass is profiled over
its statistically allowed range. \rtwo{The fit is poor (Sec.~\ref{sec:annihilation}), so we quote mass intervals as inflated statistical intervals: $\Delta\chi^2 \leq 3.84$ after the errors are scaled by $S_{\rm err} = \sqrt{\chi^2/\nu}$, a PDG-style scale factor of the kind applied to discrepant measurements~\cite{ParticleDataGroup:2024cfk}.
This widens the interval but cannot move it, it treats the excess scatter as uncorrelated between bins, and it is not
a 95\% confidence level. The intervals are $408$--$726$~GeV for $\WW$ and $537$--$975$~GeV for $b\bar b$. Within them the
minimum of $\mathcal{R}$ is $4.9$ for $b\bar b$ and $3.8$ for $\WW$ (near $490$~GeV).} The required cross section grows with mass
roughly as fast as the dSph limit weakens, so no accessible mass brings
$\mathcal{R}$ below unity. Mass freedom alone therefore does not reconcile a
prompt $s$-wave annihilation interpretation of the 20~GeV halo with the dSph
constraints, for either limit set. If the excess is dark matter, the tension, for the non-informative dwarf priors adopted, must be addressed in the particle model.

\begin{rtwoblock}
\subsection{Dark Matter Fits Across the Envelope}
\label{sec:dmres}
\secfinal

\rtwo{Table~\ref{tab:dmfits} lists the $b\bar b$, $W^+W^-$ and $\tau^+\tau^-$ fits for every member.} Across the
two-dimensional members, the $b\bar b$ fits give $m_\chi = 149$--\rtwo{746}~GeV and $\mathcal{R}_{b\bar b} = 2.77$--6.96.
The bands with the halo $J$-factor \rtwo{uncertainty} are those of Sec.~\ref{sec:tension}, about 5.17 for $b\bar b$ and about 4.23 for $W^+W^-$, the pixelwise value that anchors the envelope. For $W^+W^-$, $m_\chi = 139$--577~GeV. For the $W^+W^-$:$ZZ$ mix of a scalar doublet,
$2c_W^4\simeq1.2$, $\mathcal{R}$ spans $\rtwo{[2.1, 6.7]}$ about \rtwo{the pixelwise value} 4.34, and for $\tau^+\tau^-$ it spans $\rtwo{[1.6, 5.2]}$ at
$m_\chi = 20$--\rtwo{80}~GeV. For the non-informative dwarf priors adopted, the tension with the dwarf limits persists in every member and channel. Placing the Sun at
8.5~kpc raises $\sigma v$ and $\mathcal{R}$ by 9\% at fixed mass.

The three-dimensional members, reported as an additional check, give lower masses and a higher tension
(Table~\ref{tab:dmfits}). The lowest masses, 110--130~GeV, come from the three-dimensional members with the CHZ set.
They do not lie at an edge of the mass grid (20--2000~GeV in steps of 9.95~GeV) but are poor fits, with
$\chi^2 \approx 50$--74 in the 10--40~GeV bins alone, because on these foregrounds the halo spectrum
fitted with the CHZ set is flat from 7.3 to 20.8~GeV rather than peaked (Sec.~\ref{sec:morphology}). We therefore
interpret them as a failure of the prompt-annihilation reading for those members rather than as a preferred mass.

Each member is fitted over the bins in which its halo coefficient is positive, since no annihilation spectrum can
follow a negative one. The baseline with our bubbles loses the 1.5~GeV bin, where its coefficient lies about $10\sigma$ below zero, and the
baseline with the CHZ set also loses the 814~GeV bin, where its coefficient is consistent with zero
(Table~\ref{tab:dmfits} gives the number of bins fitted for every member). \rtwo{The 13 further grid models have the ten bins up to 169~GeV from the latitude-cut scan, and ten of them lose the 1.5~GeV bin. Refitting the complete members over those ten bins moves the best $\bbar$ and $\WW$ mass by at most one step of the mass grid and $\mathcal{R}$ by at most 2\%, so the grid models enter the mass, rate and tension ranges (the $\tau^+\tau^-$ and $W^+W^-$:$ZZ$ ranges were not tested in this way), but their $\chi^2/\nu$ is not compared with that of the members fitted over all bins.} The rows with the CHZ geometry re-fitted by us cover only the five bins from 7.3 to 59~GeV. Refitting every member over the bins they share, 7.3--59~GeV,
brings the masses together, with $m_\chi = 348$--557~GeV and $\mathcal{R}_{b\bar b} = 3.2$--6.8 for the
two-dimensional members, $m_\chi = 318$--398~GeV for the three-dimensional ones, and $\chi^2/\nu = \rtwo{1.2}$--5.1. The
110--130~GeV values do not persist, and the baseline with our bubbles moves from 716 to 498~GeV, which shows that the
full-range mass is pulled up by the bins above 59~GeV.

\paragraph{With a bubble-interior component.} Adding the uniform bubble-interior component of
Sec.~\ref{sec:morphology} changes the halo spectrum and therefore its dark matter fit. For $b\bar b$, the halo of
that fit gives $m_\chi = 647$ (627)~GeV, $\sigma v = 6.2$ $(5.4)\times10^{-25}$~cm$^3$~s$^{-1}$ and
$\mathcal{R}_{b\bar b} = 3.5$ (3.3) with our outline (outline E), with $\chi^2/\nu = 4.9$ (3.6). The same setup
without the interior gives $m_\chi = 716$~GeV, $\sigma v = 9.8\times10^{-25}$~cm$^3$~s$^{-1}$,
$\mathcal{R}_{b\bar b} = 5.1$ and $\chi^2/\nu = 7.6$, consistent with the headline fit of Sec.~\ref{sec:annihilation}
($9.9\times10^{-25}$~cm$^3$~s$^{-1}$ and $\mathcal{R}_{b\bar b} = \Rbest$). For $W^+W^-$, $\mathcal{R}_{WW} = 2.8$
(2.5), and for $\tau^+\tau^-$, $m_\chi = 60$~GeV with $\mathcal{R}_{\tau\tau} = 2.4$ (2.2). Separating the bubble
component therefore lowers the dwarf tension by about a third without removing it. Because the $b\bar b$ cross
section falls to 0.63 (0.55) of its value without the interior, the antiproton rate falls in proportion\rtwo{. At $(5.4$--$6.2)\times10^{-25}~\mathrm{cm^3\,s^{-1}}$ it still lies 27--31 times above the $b\bar b$ antiproton bound of Ref.~\cite{WangDuan2025} at these masses. No CMB or positron limit has been evaluated for these fits, and the other constraints of Secs.~\ref{sec:antiprotons} and~\ref{sec:veldep} are quoted for the fits without the interior only.}

\end{rtwoblock}

\subsection{Velocity-Dependent Annihilation}
\label{sec:veldep}
\secfinal

\begin{figure*}[htbp]
\includegraphics[width=\textwidth]{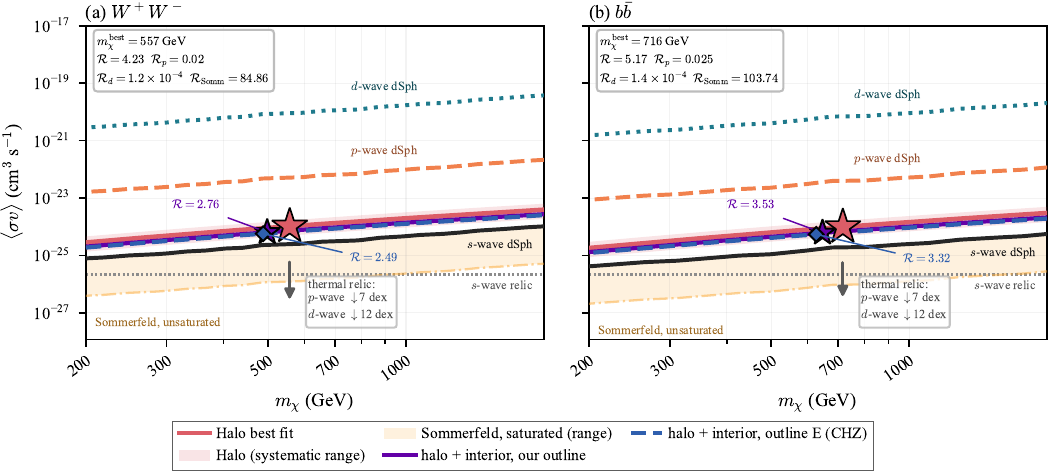}
\caption{Velocity-dependent dSph tension comparison for the global $\rho^2$
halo best fit, in the $W^+W^-$ \rtwo{(a)} and $b\bar b$ \rtwo{(b)} channels. \rtwo{The dwarf limits and the thermal relic line are labelled on the curves.} The halo
best-fit cross section $\langle\sigma v\rangle(m_\chi)$ (red solid) is shown with
its systematic normalisation band, propagated separately for each channel and
corresponding to $\mathcal{R}_{WW}\simeq\RWWlo$--$\RWWhi$ in \rtwo{(a)} and
$\mathcal{R}_{b\bar b}\simeq\Rlo$--$\Rhi$ in \rtwo{(b)} (Sec.~\ref{sec:tension},
Fig.~\ref{fig:envdm}). In both channels the lower edge stays above the
operative $s$-wave limit. The best-fit mass is marked by a star
($m_\chi\simeq0.56$~TeV for $W^+W^-$, $0.72$~TeV for $b\bar b$). The purple and blue lines and markers show the same quantities for the halo fitted together with the uniform bubble-interior component of Sec.~\ref{sec:morphology}, for our outline (purple solid line, star) and for outline E, the edge of the CHZ Flat Bubbles (blue dashed line, diamond). Their best-fit masses are 0.50 and 0.49~TeV for $W^+W^-$ and 0.65 and 0.63~TeV for $b\bar b$, and their $s$-wave tensions $\mathcal{R}$, annotated beside the markers, are 2.76 and 2.49 for $W^+W^-$ and 3.53 and 3.32 for $b\bar b$. The systematic band is shown for the halo alone only.
Dwarf-spheroidal upper limits are shown for four velocity laws: the $s$-wave limit
of Hoof et al.~\cite{Hoof2020} (black solid) and the $p$-wave (dashed), $d$-wave
(dotted) and Sommerfeld effective-$J$-factor limits of Boddy et
al.~\cite{Boddy2019,Boddy2017}. The unsaturated Coulomb-limit Sommerfeld curve
(faint dash-dot) is the most constrained, limiting case. The $1/v$
enhancement saturates below a critical velocity, so the physical saturated Sommerfeld
limit lies in the shaded band between that curve and the $s$-wave curve (reached at
full saturation). The black $s$-wave curve is therefore the operative dSph limit
for the surviving resonant-Sommerfeld interpretation. The tension
$\mathcal{R}=(\sigma v)_0^{\rm halo}/(\sigma v)_{\rm dSph}^{\rm UL}$ at the best-fit
mass is annotated for each law, with $\mathcal{R}_{\rm Somm}$ the unsaturated Coulomb-limit
value rather than the operative (saturated) one. The horizontal grey dotted line
marks the canonical thermal $s$-wave relic cross section,
$\langle\sigma v\rangle_{\rm th}\simeq2.2\times10^{-26}\,\mathrm{cm^3\,s^{-1}}$. The
downward arrow gives the present-day halo rate that a thermal relic of each
velocity-suppressed law would produce with its coupling fixed at freeze-out
($p$-wave and $d$-wave lie ${\sim}7$ and ${\sim}12$ dex below, respectively).}
\label{fig:veldep}
\end{figure*}

The dSph tension of Sec.~\ref{sec:tension} assumes the same $\langle\sigma v
\rangle$ in the Milky-Way halo and in the dwarfs. If the annihilation rate is
velocity dependent this need not hold, as the two systems probe different
characteristic velocities, and the standard $J$-factor must be replaced by an
effective $J$-factor that weights $\rho^2$ by the velocity dependence of the
cross section~\cite{Boddy2019,Boddy2017}. We write $\langle\sigma v\rangle=
(\sigma v)_0\,(v/c)^{2n}$ for $s$-, $p$- and $d$-wave annihilation
($n=0,1,2$), and $\langle\sigma v\rangle=(\sigma v)_0\,S(v)$ for a Sommerfeld-enhanced cross section ($S\propto1/v$ in the Coulomb regime). The Milky-Way
relative-velocity moments are obtained from a Jeans model of the adopted NFW
profile (one-dimensional dispersion, $\sigma_v\simeq 131~\mathrm{km\,s^{-1}}$ toward the poles),
and the dwarf moments are taken from the stacked posteriors of
Ref.~\cite{Boddy2019}.

Figure~\ref{fig:veldep} compares the halo best-fit with the resulting dSph
limits for each velocity law. We quote the tension $\mathcal{R}=(\sigma v)_0^{
\rm halo}/(\sigma v)_{\rm dSph}^{\rm UL}$ at the best-fit mass. For $p$- and
$d$-wave, the cold dwarfs ($v\sim$ tens of $\mathrm{km\,s^{-1}}$) annihilate far
less efficiently than the halo ($v\sim10^{-3}c$), so the dSph limits weaken: $\mathcal{R}_p\simeq0.02$ and $\mathcal{R}_d\simeq10^{-4}$.
Velocity-suppressed annihilation therefore relieves the dSph tension, by one-to-two ($p$-wave) and roughly
four ($d$-wave) orders of magnitude.

\rtwo{These ratios carry the foreground envelope and the halo $J$-factor term of Sec.~\ref{sec:tension}. Our $p$- and $d$-wave limits are the $s$-wave limit of Ref.~\cite{Hoof2020} rescaled by the velocity-moment ratio of Ref.~\cite{Boddy2019}, so the dwarf $J$-factor uncertainty is already included in them, as for the $s$-wave. Evaluating the two terms at their most conservative limits raises $\mathcal{R}$ by at most a factor of $1.6$ in quadrature, the same multiplier that sets the upper edge of the $s$-wave band, or $2.0$ if they are instead combined multiplicatively.}
Consequently, $\mathcal{R}_p$ reaches a maximum of $\simeq0.04$, or $\simeq\rtwo{0.05}$ under the multiplicative combination, remaining more than an order of magnitude below unity in either case.
We note that the velocity-dependent effective $J$-factors rely on higher moments of the dwarf phase-space distributions than the standard $s$-wave $J$-factors, and are therefore inherently less constrained. The stated uncertainty band therefore represents a lower bound on the total $p$- and $d$-wave systematic budget.

\rtwo{The binding constraint on these scenarios is therefore the relic abundance, not the dwarf limits.} For a $p$-wave coupling
$\langle\sigma v\rangle = a\,\langle(v/c)^2\rangle$, the relic density fixes $a$
at freeze-out, where $\langle(v/c)^2\rangle_{\rm fo}\simeq 6/x_f\simeq0.24$ for
$x_f\equiv m_\chi/T_{\rm fo}\simeq25$. Matching the observed abundance then
requires $a\,\langle(v/c)^2\rangle_{\rm fo}\simeq\langle\sigma v\rangle_{\rm th}
\simeq2.2\times10^{-26}\,\mathrm{cm^3\,s^{-1}}$~\cite{Steigman2012}. In the halo
the Jeans-modelled moment is
$\langle(v/c)^2\rangle_{\rm halo}\simeq1.0\times10^{-6}$, so the predicted
present-day rate is $\langle\sigma v\rangle_{\rm halo}=
\langle\sigma v\rangle_{\rm th}\,\langle(v/c)^2\rangle_{\rm halo}/
\langle(v/c)^2\rangle_{\rm fo}\simeq9\times10^{-32}\,\mathrm{cm^3\,s^{-1}}$.
This lies about seven orders of magnitude below the
$\sim10^{-24}\,\mathrm{cm^3\,s^{-1}}$ required to reproduce the 20~GeV halo. The $d$-wave case follows the same arithmetic with the velocity factor squared,
giving $\langle\sigma v\rangle_{\rm halo}\simeq4\times10^{-37}\,
\mathrm{cm^3\,s^{-1}}$ and a shortfall of some twelve orders of magnitude. A
thermal single-species $p$- or $d$-wave origin is thus excluded by relic
abundance even though it satisfies the dSph limits. \rtwo{A non-thermal abundance removes the relic problem but not the size of the coupling. A halo rate of
$10^{-24}\,\mathrm{cm^3\,s^{-1}}$ with $\langle(v/c)^2\rangle\simeq10^{-6}$ needs a $p$-wave coefficient near
$10^{-18}\,\mathrm{cm^3\,s^{-1}}$, which for $\sigv\sim\pi\alpha_{\rm eff}^2/m_\chi^2$ at these masses means a coupling $\alpha_{\rm eff}$ of order 100.}

A low-velocity enhancement, from Sommerfeld exchange of a light
mediator~\cite{FengKaplinghatYu2010}, can raise the present-day halo $\sigv$ to
the required ${\sim}10^{-24}\,\mathrm{cm^3\,s^{-1}}$ while preserving the
thermal relic density set at freeze-out. No resonance is required for this. At
the baseline foreground model the boost needed is
$\sigv_{\rm halo}/\sigv_{\rm th}\simeq
(\rtwo{9.9}\times10^{-25})/(2.2\times10^{-26})\approx\boostbase$. That figure is
conditional on the baseline normalisation. Carrying the foreground-modelling
envelope of Sec.~\ref{sec:systematics} through to $\sigv$ gives
$\approx\boostlo$--$\boosthi$ across the envelope, so the requirement is a range
rather than a single number, and at its lower end the enhancement needed is $\approx\boostlo$. The figure is essentially channel-independent, the two channels giving
${\approx}\boostbase$ and ${\approx}47$ at their baseline normalisations.

The relic-consistent Sommerfeld benchmarks of Ref.~\cite{FengKaplinghatYu2010}
give maximal enhancements of ${\sim}30$ and ${\sim}90$ at $\mchi=0.3$ and
$1$~TeV, so a boost of ${\approx}\boostbase$ corresponds to
$\mchi\approx0.5$~TeV, coincident with our $\WW$ best fit, and the whole
$\boostlo$--$\boosthi$ range sits at or below the $1$~TeV benchmark. \rtwo{These are
maximal values under optimistic freeze-out assumptions, including resonances, for a 250~MeV mediator annihilating to
mediator pairs. For a mediator at or above the 0.6~GeV floor derived below the attainable enhancement is
lower~\cite{FengKaplinghatYu2010}, so the reconciliation of the halo rate with the relic abundance is less comfortable
than these benchmarks suggest.} \rtwo{A light mediator that also carries the annihilation products gives cascade rather than prompt spectra, and those fit the halo worse (below), so this route requires the annihilation to proceed to prompt Standard Model final states, with the light state supplying only the enhancement.} A Breit--Wigner
resonance~\cite{IbeMurayamaYanagida2009,GuoWu2009} reaches comparable
$\mathcal{O}(100)$ halo boosts for TeV-scale dark matter by a different route.

This reconciliation operates through the relic abundance, not by weakening the
dSph limit. Since the tension scales as
$\mathcal{R}\propto\langle S\rangle_{\rm dwarf}/\langle S\rangle_{\rm halo}$,
standard Sommerfeld or below-threshold Breit--Wigner enhancement grows
monotonically toward low velocity and cannot reduce $\mathcal{R}$ below its
$s$-wave value, because the colder dwarfs are boosted at least as much as
the halo for any monotonic enhancement. Only a resonance placed above threshold, whose enhancement peaks at a
finite velocity and then falls for colder systems, could invert that ordering
($\langle S\rangle_{\rm halo}>\langle S\rangle_{\rm dwarf}$) and in principle
bring the dSph tension to unity, at the cost of tuning the pole to the halo
velocity scale. Of the velocity laws considered, that is the only one that could
reconcile the halo normalisation, the dSph limits and the relic abundance
together. Two realisations of that resonance exist and they differ in what
they add to the theory. Sommerfeld enhancement arises from a light $t$-channel
mediator, and is resonant when the mediator mass sits at a bound-state pole. A
Breit--Wigner enhancement arises from an $s$-channel state with
$m_{\rm med}\approx2m_\chi$, carries no light mediator at all, and takes its
velocity dependence from the width and offset of the pole. The saturation
argument that follows applies to the first, and the second is returned to
below.

Reproducing the halo normalisation with a thermal relic abundance therefore requires the annihilator to inhabit
a structured dark sector rather than to be a vanilla thermal WIMP. In the
Sommerfeld realisation, a light mediator supplies the enhancement. The enhancement saturates once the de Broglie
wavelength of the relative motion, $(m_\chi v)^{-1}$, no longer exceeds the
force range $m_\phi^{-1}$, which occurs at
$v_{\rm sat}/c \simeq m_\phi/m_\chi$ up to an $\mathcal{O}(1)$ coefficient and
independently of the coupling~\cite{ArkaniHamed2009,FengKaplinghatYu2010}. The
saturation velocity must lie at or above the Galactic scale
($v_{\rm halo}\simeq10^{-3}c$, the three-dimensional relative-velocity rms $\sqrt{6}\,\sigma_v\simeq320~\mathrm{km\,s^{-1}}$ for the $\sigma_v\simeq131~\mathrm{km\,s^{-1}}$ of Sec.~\ref{sec:veldep}), since for any lower value the colder dwarfs
saturate while the halo is still in the $1/v$ regime. This leaves the dwarfs
boosted more than the halo, worsening the tension rather than leaving it at its
$s$-wave value. That condition places a floor on the mediator mass at $m_\phi \gtrsim m_\chi v_{\rm halo}/c$, of
order $0.6$~GeV for the masses recovered here. The floor is reached at the minimum coupling. Setting $v_{\rm sat} = v_{\rm halo}$, the boost condition
$\pi\alpha_\chi c/v_{\rm halo} \simeq \boostbase$ gives
$\alpha_\chi\simeq1.4\times10^{-2}$, and a heavier mediator satisfies the same condition at proportionally larger $\alpha_\chi$.

The CMB imposes the same condition by an independent
route. A monotonic $1/v$ enhancement is some five orders of magnitude larger at
recombination ($v\sim10^{-8}c$) than in the halo, so an unsaturated boost of
$\approx\boostbase$ is excluded by the Planck limit on annihilation energy
injection, $p_{\rm ann}=f_{\rm eff}\sigv/\mchi<3.2\times10^{-28}\,
\mathrm{cm^3\,s^{-1}\,GeV^{-1}}$~\cite{Planck2018}. Saturation at or above the
halo velocity removes that hierarchy, and the bound then applies to the boosted
rate directly. At our masses, with $f_{\rm eff}=\rtwo{0.207}$ ($\bbar$, 716~GeV) and $\rtwo{0.177}$
($\WW$, 557~GeV) from the tabulated deposition efficiencies of
Ref.~\cite{Slatyer2016}, it gives
$\sigv<1.1\times10^{-24}\,\mathrm{cm^3\,s^{-1}}$ for $\bbar$ and
$1.0\times10^{-24}\,\mathrm{cm^3\,s^{-1}}$ for $\WW$. The $\bbar$ baseline of
$\rtwo{9.9}\times10^{-25}\,\mathrm{cm^3\,s^{-1}}$ sits about $10\%$ below its bound,
while the $\WW$ baseline of $\rtwo{1.04}\times10^{-24}\,\mathrm{cm^3\,s^{-1}}$ sits $\rtwo{0.3}$ to $\rtwo{3.1\%}$ above its own, depending on which fit's normalisation is used, within the uncertainty on $f_{\rm eff}$ and on the halo normalisation. An above-threshold resonance relaxes the bound
in both channels, since the enhancement falls for colder systems and the
recombination-epoch rate is suppressed. For the Sommerfeld realisation the CMB
thus points to the mediator scale the dwarf comparison already selects, and it provides a near-term test of the saturated alternative, which is
the Sommerfeld case, and of the boosted rate in either realisation.

The saturation floor is the specific requirement the Sommerfeld realisation
carries, over and above the modest boost itself, and placing the mediator on a
bound-state pole so that the enhancement peaks at the halo velocity demands a
near-threshold coincidence of scales on top of it. \rtwo{The Breit--Wigner realisation has no floor but requires a mediator within a small fraction of $2m_\chi$ with its pole tuned so that the enhancement peaks at the halo velocity, which is a coincidence of the same order.} Light-mediator dark sectors are well
studied~\cite{ArkaniHamed2009}, and both realisations are an addition to the minimal picture, so the dark matter interpretation of the 20~GeV halo does
not point to simple thermal annihilation. That model cost must be weighed
alongside the astrophysical and systematic uncertainties above.

\begin{rtwoblock}
\paragraph{Light mediators and cascade spectra.} If the annihilation products are pairs of a light mediator $\phi$,
the mediator changes the spectrum as well as the rate, and annihilation proceeds by the cascade
$\chi\chi\to\phi\phi\to\mathrm{SM}$. For $m_\phi\simeq0.6$~GeV only leptonic and light hadronic final states are
open, since $\phi\to b\bar b$ requires $m_\phi>2m_b\simeq8.4$~GeV and $\phi\to W^+W^-$ requires
$m_\phi>2m_W\simeq160$~GeV. We refit the halo spectrum with one-step cascade spectra through a 0.6~GeV mediator. With
the mediator removed, the fit reproduces the prompt result ($m_\chi = 716.5$~GeV, $\chi^2 = 82.6$ for 11 degrees of
freedom), and every cascade fits worse, by $\Delta\chi^2 = 31.4$ ($\pi^+\pi^-$), 34.4 ($\mu^+\mu^-$), 49.2 (dark
photon), 67.9 ($e^+e^-$) and about 229 ($\pi^0\pi^0$). Because the boosted mediators spread each decay product over a
broad range of energies, the cascade spectra are flatter than the prompt ones. In the $\pi^+\pi^-$, $\mu^+\mu^-$ and
dark-photon channels the preferred mass falls to $m_\chi = 115$--212~GeV and the required boost rises to 186--759,
compared with \boostbase\ for prompt $b\bar b$, while the dwarf tension, evaluated with a common dwarf likelihood, is
unchanged at 0.94--0.97 of its prompt value. The $e^+e^-$ and $\pi^0\pi^0$ cascades, which fit worst, are excluded
from these ranges. A relic-fixed, non-resonant Sommerfeld enhancement falls about two orders of magnitude short of the
required boost, so a light mediator that carries the annihilation products does not reconcile the halo rate with the
relic abundance. This analysis omits the inverse-Compton emission of the cascade electrons and a full Sommerfeld
calculation, and does not exclude a resonant enhancement. The extracted halo spectrum and its morphology are
unaffected, since they come from the spatial fit, which does not depend on the annihilation spectrum assumed.
\end{rtwoblock}

This structure is not peculiar to our normalisation. Independent model-building
studies of the excess reach it from different starting points. Three invoke a
mediator light compared with the annihilator, which is what supplies the
low-velocity enhancement, through a Dirac candidate with a light
scalar~\cite{Jho2025}, secluded annihilation into $\WW$~\cite{Yoshimatsu2026},
and a parity-violating Higgs-portal spin-1 candidate~\cite{Yamashita2026}. An
electroweak-doublet candidate with a small inelastic splitting reaches a
comparable rate by the same means, a new light state rather than the weak gauge
bosons, which are too short-ranged at these masses~\cite{NomuraTotani2026}. Our own saturation floor, of order $0.6$~GeV, sits at the light end of that range, and
our one-step cascade fits, made for a 0.6~GeV mediator, disfavour that end when the mediator also carries the annihilation products. A
resonant realisation reaches the same rate without a light state, the
enhancement coming instead from a Breit--Wigner pole near
$2m_\chi$~\cite{Murayama2025}, so the published realisations span a mediator
scale from sub-GeV to $\mathcal{O}(10)$~GeV and one case with no light
mediator at all.

\subsection{Decay Templates}
\label{sec:decay}
\secfinal

We also compare the $\rho^1$ halo spectrum with decaying dark matter spectra. The \rtwo{best fits lie at} masses of order
$0.7$--$1.0\,\mathrm{TeV}$ for the $\WW$ and $\bbar$ channels, with lifetimes of
order $5$--$7\times10^{26}\,\mathrm{s}$, whilst the $\tautau$ channel gives a poorer fit.
As in the annihilation case, a single prompt PPPC channel
\rtwo{does not capture all of the structure in the extracted halo, so the decay reading is a morphology-and-lifetime comparison and an alternative emissivity scaling, not a complete description. It evades the dSph limits but not the diffuse-background degeneracies.} A lifetime this short is moreover disfavoured by
the isotropic gamma-ray background, which requires
$\tau\gtrsim10^{28}\,\mathrm{s}$ for TeV-scale dark matter decaying to hadronic or
leptonic channels~\cite{AndoIshiwata2015,Cohen2017,BlancoHooper2019}, \rtwo{more than an order of magnitude above the lifetime required to reproduce the halo.}

\subsection{Antiproton Constraints}
\label{sec:antiprotons}
\secfinal

The cosmic-ray antiproton flux measured by AMS-02 constrains hadronic dark matter
annihilation independently of the gamma-ray data. Wang and Duan~\cite{WangDuan2025}
evaluate that constraint against the 20~GeV halo excess under prompt annihilation
into Standard Model final states, and find that the implied antiproton flux
exceeds the AMS-02 measurement \rtwo{at 100~GeV, in units of its uncertainty,} by $15\sigma$ in the $\bbar$ channel and
$18\sigma$ in the $\WW$ channel at the annihilation parameters of
Ref.~\cite{Totani2025}. Taken at face value, this disfavours a minimal prompt
$s$-wave interpretation of the excess.

Two considerations bear on how the bound applies to the present result. First, it
is evaluated within a single cosmic-ray propagation setup and at a single halo
normalisation. The implied antiproton flux scales with the annihilation rate, and
\rtwo{the required annihilation cross section} varies by a factor of about five across the envelope
of Sec.~\ref{sec:systematics}, so the severity of the constraint varies with it.
The bound remains strong at the baseline normalisation.

\begin{rtwoblock}
Second, the bound arises from annihilation within a few kiloparsecs of the Sun, the region that also dominates the
pole $J$-factor and has the same velocity distribution, so no velocity dependence of the rate relieves it. A
substructure boost raises the antiproton and gamma-ray rates together~\cite{Stref:2016uzb}, and a $p$-wave or
resonant enhancement that lowers the dwarf tension leaves the predicted antiproton flux tied to the gamma-ray
normalisation. Only the final state can change it. Leptonic channels produce no antiprotons, but the
$\tau^+\tau^-$ fit is poor ($\chi^2/\nu\simeq14.6$ at $m_\chi\simeq70$~GeV, Sec.~\ref{sec:annihilation}). A mediator
lighter than $2m_p\simeq1.88$~GeV closes antiproton production in a cascade, but cascade spectra fit the halo worse
than prompt annihilation in every channel tested (Sec.~\ref{sec:veldep}). Mediators of order 10~GeV yield a softer
antiproton spectrum, at energies where solar-modulation uncertainties are larger, and have been argued to evade the
constraint~\cite{Jho2025}, but they too give cascade rather than prompt photon spectra.

The antiproton bound is therefore the most restrictive constraint we consider. The analysis of Ref.~\cite{WangDuan2025} gives $\sigv<2\times10^{-26}~\mathrm{cm^3\,s^{-1}}$ for $b\bar b$, an order of magnitude below the lowest
normalisation in the envelope ($2.1\times10^{-25}~\mathrm{cm^3\,s^{-1}}$), and analyses that marginalise over
propagation and AMS-02 error correlations reach comparable limits near the TeV scale~\cite{Calore:2022stf,Cuoco:2016eej}.
Propagation and halo-profile choices move the bound by factors of a few, not by the factor of ten or more that the
prompt channels would require. The reported antiproton excess near 10--20~GeV, which falls below $1\sigma$ once
correlated errors are included~\cite{Heisig:2020nse}, would point to a 20--100~GeV particle annihilating at about the
thermal rate and does not accommodate the halo.
\end{rtwoblock}

\begin{rtwoblock}
\subsection{What a Dark Matter Candidate Would Need}
\label{sec:requirements}
\secfinal

If the excess is interpreted as dark matter, it places requirements on any candidate, which we test against five constraints\rtwo{, on the photon spectrum, the dwarf limits, the relic abundance, the CMB and the antiproton data}. The photon spectrum must peak at 20~GeV, as prompt annihilation to $b\bar b$ or $W^+W^-$ at a few hundred
GeV does. The present-day rate in the Milky Way must be $\sigv\simeq(2$--$10)\times10^{-25}~\mathrm{cm^3\,s^{-1}}$
across the envelope, \boostlo{} to \boosthi{} times the thermal value. The rate in the halo must exceed that in the dwarfs by at least a
factor of \Rlo{} for $b\bar b$ and \RWWlo{} for $W^+W^-$ at the most favourable edge of the budget \rtwo{for non-informative dwarf priors}, and by about five at the baseline. If only the spherical component of Sec.~\ref{sec:morphology} is dark matter, it carries 61--68\% of the
amplitude on the baseline (78--81\% on the other foreground models tested), its best-fit $b\bar b$ mass is
627--647~GeV, and the baseline tension falls to $\mathcal{R}_{b\bar b}\simeq3.3$--3.5 but remains. \rtwo{Of the other constraints, only the antiproton bound is compared for this component (Sec.~\ref{sec:dmres}).} Table~\ref{tab:routes} compares the routes considered in this section
with the five constraints.

A further constraint applies if the halo is cuspy. Searches in the inner few degrees with H.E.S.S.~\cite{HESS:2022ygk}
and with the \textit{Fermi} continuum~\cite{Dessert:2022evk,Safdi:2025sfs} limit $W^+W^-$ and $b\bar b$ annihilation at
0.5--1~TeV to well below $10^{-25}~\mathrm{cm^3\,s^{-1}}$ for NFW or Einasto profiles\rtwo{, with the H.E.S.S.\ limit for
$W^+W^-$ and an Einasto profile reaching $3.7\times10^{-26}~\mathrm{cm^3\,s^{-1}}$ at 1.5~TeV}, and Ref.~\cite{GarciaCely2016}
finds that H.E.S.S.\ limits the boost of an inert doublet to about 10. Every prompt route in the table therefore
requires an inner core of order a kiloparsec. This leaves the pole $J$-factor, and with it the halo normalisation,
unchanged, since the scale radius and inner slope contribute negligibly there (Sec.~\ref{sec:tension}), and it is
consistent with the absence of a single NFW normalisation across the sky~\cite{Abazajian2026}.
\rtwo{Radio observations of the Large Magellanic Cloud exclude the thermal rate for annihilation to quarks below
480~GeV~\cite{Regis:2021glv}, under assumptions on its magnetic field and cosmic-ray transport. The halo requires
\boostlo{} to \boosthi{} times that rate, so below 480~GeV the $b\bar b$ reading is excluded by at least a factor of \boostlo{} under the same
assumptions, and above it the radio limit has not been evaluated here.}

\begin{table*}
\caption{\label{tab:routes} \rtwo{Routes to a dark matter reading of the halo spectrum and the quantities that test them,
at the baseline unless a range is given. Spectrum: goodness of fit to the halo spectrum (Secs.~\ref{sec:annihilation}
and~\ref{sec:veldep}). dSph: the dwarf tension $\mathcal{R}$ (Sec.~\ref{sec:tension}). Relic: the present-day rate
relative to the thermal value, or for the $p$- and $d$-wave rows the thermal rate as a fraction of the required one, and
for the saturated Sommerfeld row the enhancement available relative to the enhancement required.
CMB: the rate relative to the Planck bound on energy injection (Sec.~\ref{sec:veldep}). $\bar p$: the lowest $b\bar b$ rate in the envelope
relative to the AMS-02 antiproton bound of Ref.~\cite{WangDuan2025} (Sec.~\ref{sec:antiprotons}). n.e.: not
evaluated. n.a.: not applicable. Where two values are given they are for $b\bar b$ and $W^+W^-$, or for the $p$- and $d$-wave rates. The light-mediator row has $m_\phi<2m_p$. Notes on each row, including the markers $\ddagger$, $\S$ and $\P$, follow in the text.}}
\begin{ruledtabular}
\begin{tabular}{llllll}
Route & Spectrum $\chi^2/\nu$ & dSph $\mathcal{R}$ & Relic & CMB & $\bar p$ \\
\hline
Prompt $s$-wave, thermal & 7.5, 7.4 & \Rlo--\Rhi & \boostlo--\boosthi & 0.89, 1.03 & $\geq10$ \\
Prompt $s$-wave, non-thermal & 7.5, 7.4 & \Rlo--\Rhi & by construction & 0.89, 1.03 & $\geq10$ \\
$p$- or $d$-wave & 7.5, 7.4 & $\lesssim0.05$ & $10^{-7}$, $4\times10^{-13}$ & velocity-suppressed & $\geq10$ \\
Saturated Sommerfeld & 7.5, 7.4 & \Rlo--\Rhi & 1.23, 0.91 & 0.89, 1.03 & $\geq10$ \\
Resonance, prompt & 7.5, 7.4 & possible with tuning & possible with tuning & n.e. & $\geq10$ \\
Light-mediator cascade & $\Delta\chi^2\geq31$ worse & 0.94--0.97 of prompt & 186--759 & $>1^\P$ & none$^\ddagger$ \\
Prompt $\tau^+\tau^-$ & 14.6 & 1.6--5.2 & 5.7 & n.e. & none$^\ddagger$ \\
Decay ($\rho^1$) & 19.8--24.0$^\S$ & no tension & n.a. & n.e. & n.e. \\
\end{tabular}
\end{ruledtabular}
\end{table*}

\rtwo{Table~\ref{tab:routes} gives the quantity behind each entry. The $\bar p$ column compares the lowest $b\bar b$ rate in
the envelope with the bound of Ref.~\cite{WangDuan2025}. At the rates of Ref.~\cite{Totani2025} the implied antiproton
flux exceeds the AMS-02 measurement at 100~GeV by 15--18$\sigma$ in units of its uncertainty. No $W^+W^-$ bound is
quoted. $^\ddagger$Leptonic final states produce no antiprotons but produce positrons. AMS-02 positron data exclude the
thermal rate for $\tau^+\tau^-$ below 60~GeV~\cite{John:2021ugy}, so the
$\tau^+\tau^-$ rate ($\sigv\simeq1.25\times10^{-25}~\mathrm{cm^3\,s^{-1}}$ at 70~GeV) is probably excluded. \rtwo{The cascade
rates are tested instead by the CMB bound ($\P$) and by the spectrum, which they fit worse than prompt annihilation by
$\Delta\chi^2\geq31$.} $^\S$The decay fit
is to the $\rho^1$ spectrum and is not comparable with the annihilation fits, and decay is in addition disfavoured by
the isotropic gamma-ray background (Sec.~\ref{sec:decay}). Every prompt row assumes an inner core of order a
kiloparsec, since inner-Galaxy limits exclude the required rate for a cuspy profile. $^\P$The cascade rates exceed the Planck bound for $f_{\rm eff}\gtrsim0.004$--0.009. The resonance row must also be
checked against big-bang nucleosynthesis, since a resonance tuned to the halo velocity is crossed near
$T\sim1$~MeV~\cite{FengKaplinghatYu2010,Hisano:2009rc}.}

\rtwo{No route is consistent with all five constraints.} \rtwo{For every route in Table~\ref{tab:routes} that we evaluate, the final states that reproduce the 20~GeV peak
produce antiprotons, which AMS-02 constrains, and the leptonic final states fit the spectrum worse and produce
positrons, which AMS-02 also constrains.} Velocity dependence can satisfy the dwarf, relic and CMB constraints together, provided the annihilation
rate rises with velocity between the dwarfs and the halo and falls again at recombination. Published realisations
include a narrow near-threshold resonance~\cite{An:2012uu}, selective $p$-wave Sommerfeld enhancement in multi-state
dark sectors, with one benchmark at 780~GeV and $4.6\times10^{-24}~\mathrm{cm^3\,s^{-1}}$~\cite{Das:2019vxm}, and
broad $p$-wave resonances that need no fine tuning~\cite{Yoshimatsu:2026zze}. A present-day rate well above the
thermal value can also arise from a change in the annihilation after freeze-out~\cite{Huang:2026rtz,Hoskinson:2026cdp,Choquette:2016xsw}.
None of these changes the antiproton yield per annihilation (Sec.~\ref{sec:antiprotons}), so a candidate \rtwo{consistent with} all five constraints would also need a final state that reproduces the 20~GeV peak with far
fewer antiprotons than $b\bar b$ or $W^+W^-$. Neither the prompt nor the cascade channels we tested supply one. On present evidence, no single argument here excludes a dark matter interpretation, but none of the routes tested
supports it\rtwo{, including the spherical component alone, whose $b\bar b$ cross section of
$(5.4$--$6.2)\times10^{-25}~\mathrm{cm^3\,s^{-1}}$ (Sec.~\ref{sec:dmres}) still lies more than an order of magnitude
above the antiproton bound of Ref.~\cite{WangDuan2025}}.

\subsection{The LZ Event}
\label{sec:lz}
\secfinal

\rtwo{The LZ Collaboration reports a single nuclear-recoil event at $248$~keV in a $2.84$~tonne-year exposure, with a
global significance of $2.6\sigma$ after the look-elsewhere effect and a local significance of
$3.4\sigma$~\cite{LZ2026}. In the electroweak-doublet reading, the inelastic splitting the event needs is the structure
that model already required to evade elastic scattering~\cite{NomuraTotani2026}, and its mass strip of $440$ to
$600$~GeV~\cite{Nomura2026} overlaps the $\WW$ interval preferred here, $557$~GeV with an inflated statistical interval
of $408$ to $726$~GeV (Sec.~\ref{sec:tension}). For a scalar doublet the $W^+W^-$:$ZZ$ mix $2c_W^4\simeq1.2$~\cite{GarciaCely2016}
gives $607$ to $617$~GeV and $\mathcal{R}\simeq4.1$ in the deterministic mass scan, just above that strip, and
$587$~GeV with $\mathcal{R}\simeq4.34$ in the pixelwise fit (Sec.~\ref{sec:dmres}), inside it, so the doublet reading must meet the constraints of
Table~\ref{tab:routes} at that rate, and its rate lies at the CMB bound we derive at that mass. IceCube limits on solar capture require a larger splitting than the
$\simeq350$~keV the event prefers unless the local velocity distribution has a high-speed tail~\cite{Luengas:2026tlf}.
The higgsino readings of the event, at $1$ to $1.1$~TeV~\cite{Freese2026,Wu2026}, lie outside the mass range the halo
spectrum allows (Sec.~\ref{sec:tension}). The doublet reading is testable through its predicted annual modulation, further LZ exposure and
solar capture. We do not combine the two observations into a joint significance, and no detection is claimed.}
\end{rtwoblock}

\section{Discussion and Conclusions}
\label{sec:conclusions}
\secfinal
\begin{rtwoblock}
We have presented a pixel-level reanalysis of the 20~GeV Galactic halo excess. \rtwo{Building on a reproduction of Totani's analysis, we fit the data at the native $0.125\degree$ resolution with an
exact Poisson likelihood and energy-dependent forward folding of the point-spread function, and repeat the fit across
the two-dimensional GALPROP foregrounds, two independent models of the \textit{Fermi} bubbles and, as an additional
check, three-dimensional GALPROP v57 models.} Mock skies without a halo calibrate how much of the signal foreground mismatch can produce,
rebuilding the residual-derived bubble templates inside those skies quantifies the bias of their construction, and
the dark matter interpretation is tested against the dwarf, relic-abundance, CMB and AMS-02 constraints across the
full foreground range. The results divide into what the excess is, what it is not, and what a dark matter explanation
would need.

\paragraph{What it is.} A centrally concentrated excess is present in every configuration and\rtwo{, with our bubble templates, peaks in the 20.8~GeV bin in every two-dimensional member}.
At that bin its coefficient spans 0.236--\rtwo{0.409} across the two-dimensional members and 0.381--0.418 in the
three-dimensional checks with our bubbles. Mask changes move it by less than 5\%, mismatch between the \rtwo{three GALPROP foregrounds used to generate mocks} does not produce it (mock skies without a halo induce $-0.07$ to 0), and \rtwo{the injected cosmic-ray electron populations we test do not} replace it. As a replacement they fit worse than $\rho^2$, and fitted
alongside it they leave \rtwo{99--100\% of the halo coefficient at the best fit with our bubbles and 96--97\% with the
CHZ set (95--96\% and 83--85\% at the posterior median)} when free in
each bin, and 73--98\% with one amplitude across all energies.

\paragraph{What it is not.} \rtwo{The excess is not described by a spherical halo alone. With the foregrounds shared
between latitude bands, its amplitude relative to $|b| = 10$--$20\degree$ rises by factors of 1.35 at
$20$--$40\degree$ and 1.72 at $40$--$60\degree$, its spectrum softens with latitude, and the emission lies largely
inside the bubble outline. Nor is it cleanly separable from the bubbles. With the CHZ set, the halo coefficient is
59--82\% of its value with our bubbles, and anticorrelates with their Flat Bubbles, while a sky built from the baseline
foreground with their templates and no halo induces a halo coefficient of 0.329 in our fitter, against 0.353 in the
data. With the CHZ set, the data lie inside the distribution of $2\Delta\ln\mathcal{L}$ from skies without a halo
(136.5 against a 95th percentile of 171). With our bubbles they lie above it (927 against 885), but the amplitude
exceeds the 0.329 induced by the CHZ-set sky by only 1.2--2.0$\sigma$, depending on how the uncertainties are
counted. With our bubbles, the excess therefore lies at the edge of what foreground mismatch can produce, and with the
CHZ set, it is not distinguishable from it. Rebuilding our bubble templates within the mocks, as for the data, adds
0.035--0.048 to the induced halo coefficient at 20.8~GeV and lowers the value induced by the CHZ-set sky to
0.22--0.26. A uniform bubble interior with no halo, fitted as the data are, induces 48\% (our outline) and 55\%
(outline E) of the halo coefficient over 10--40~GeV, with no peak beyond its uncertainty (Sec.~\ref{sec:morphology}).}
\paragraph{Two components.} On the baseline foreground the excess \rtwo{is described by} a spherical component that follows
$\rho^2$, peaks at 20.8~GeV and carries 61--68\% of the amplitude the $\rho^2$ template has without the bubble-interior component, and a component with a flat
spectrum uniform inside the bubble outline. Together they leave \rtwo{regional residuals of at most 1.7 times their Poisson error (Sec.~\ref{sec:morphology})} at 20.8~GeV, and neither alone does. The bubble component turns negative at 4.3~GeV, where our bubble templates are built, which is what a
residual-derived template that does not follow the bubbles' spectrum would produce. \rtwo{On the OB-star model and all ten three-dimensional models the same decomposition holds, with the spherical component
at 78--81\% of the amplitude it has without the interior.}

\paragraph{What a dark matter explanation would need.} Interpreted as prompt annihilation, the excess \rtwo{is fitted by} a particle with a mass of a few hundred GeV, with $m_\chi = 716$~GeV for $b\bar b$ on the baseline and 149--\rtwo{746}~GeV
across the two-dimensional members, and a dwarf tension $\mathcal{R}_{b\bar b}$ between \Rlo{} and \Rhi{} once the \rtwo{halo} $J$-factor \rtwo{uncertainty is} included. For the spherical component alone, the baseline fit gives
$m_\chi = 647$ (627)~GeV and $\mathcal{R}_{b\bar b} = 3.5$ (3.3) with our outline (outline E). The tension is present
in every configuration and channel for non-informative dwarf priors, while priors informed by tidal stripping would
weaken the dwarf limits by a factor of 2 to 7. Velocity-suppressed annihilation removes the tension but falls short of
the relic abundance by 7 ($p$-wave) and 12 ($d$-wave) orders of magnitude. A Sommerfeld enhancement \rtwo{that saturates at or above the halo velocity} supplies the required boost \rtwo{for $b\bar b$ and about 90\% of it for $W^+W^-$ (Table~\ref{tab:routes})} but does not change the ratio of the halo and dwarf rates, and a light mediator, which produces
cascade rather than prompt spectra, fits the halo spectrum worse in every channel tested. An annihilation rate that is
resonantly enhanced at the halo velocity can \rtwo{in principle satisfy the dwarf and relic constraints with tuning, with its CMB bound not evaluated here (Table~\ref{tab:routes})}, but no velocity
dependence changes the antiproton yield per annihilation, so the AMS-02 bound still applies, and the final states that
produce fewer antiprotons fit the spectrum worse (Sec.~\ref{sec:requirements}). For the scalar doublet proposed in
connection with the LZ event, the $W^+W^-$:$ZZ$ mix $2c_W^4$ gives a best-fit mass of \rtwo{607--617~GeV in the deterministic scan and 587~GeV in the pixelwise fit, at the upper end of the
440--600~GeV range} proposed for that event, and IceCube limits on solar capture disfavour the splitting the event
prefers unless the local velocity distribution has a high-speed tail.

\rtwo{If the flat-spectrum emission inside the bubble outline belongs to the bubbles, a dark matter interpretation would have to explain the spherical component at a rate above the dwarf upper limits ($\mathcal{R}_{b\bar b}\simeq3.3$--3.5 on the baseline).}

These conclusions are conditional on the foreground basis. The quoted range is spanned by two-dimensional,
azimuthally symmetric propagation models normalised to local cosmic-ray data, with ten three-dimensional GALPROP
models as an additional check, and is obtained with the fixed 4FGL-DR4 source model, so it is a lower bound on the
foreground uncertainty. The fitted halo coefficient could move further under a line-of-sight decomposition of the
inner-Galaxy foreground, with gas and inverse-Compton components fitted in galactocentric
annuli~\cite{Ajello2016,Ackermann2017}, under gas maps built with models of gas flow in the barred inner
Galaxy~\cite{Macias:2016nev,Pohl:2022nnd}, under foreground templates with spatially resolved
freedom~\cite{Storm:2017arh,Bartels:2017vsx,Song:2024iup}, and with a source catalogue re-derived for each foreground
model~\cite{Ajello2016}. The last of these requires a full source search over the region for each model, which is not
attempted here, and it remains an open limitation of the present result. Template suites that sample a broader
propagation space~\cite{Cholis:2021rpp} would provide a further test. Whether a 20~GeV excess at these latitudes is
still described by a spherical template after such a decomposition, with three-dimensional gas and a re-derived
source population, is therefore untested, and a change of likelihood alone would not settle it. Within the foreground models tested, a single $\rho^2$ halo template recovers the excess in every configuration, with
a coefficient at 20.8~GeV of 0.236--0.409 across the two-dimensional members and, with our bubble templates, a
spectrum that peaks in that bin, but it leaves a residual that follows the bubble outline. \rtwo{Adding a flat-spectrum component uniform inside the outline removes that residual (see ``Two components'' above), so within our template set the excess is described by these two components rather than by a single spherical halo.}

\rtwo{A direct astrophysical test is a model of the \textit{Fermi} bubbles whose morphology, and not only its
normalisation, is fitted jointly with the halo at every energy. In the fits here and in Ref.~\cite{Abazajian2026},
the bubble templates have a fixed morphology and only their normalisation varies with energy.} On the particle side,
the dwarf limits, the CMB bound on energy injection, the AMS-02 antiproton and positron data and, for a cuspy profile,
the inner-Galaxy limits constrain the annihilation rate directly, while the LZ exposure and its predicted annual
modulation test the doublet interpretation. \rtwo{The excess is present in every foreground configuration tested, but with the CHZ bubble templates it is not distinguishable from foreground mismatch. Its morphology does not follow a single annihilation profile, and a dark matter reading would apply at most to the spherical component, at a rate in tension with the dwarf and antiproton limits.}
\end{rtwoblock}

\section*{Use of AI tools}
\rtwo{AI tools assisted with
writing and checking analysis code, setting up and monitoring computations, and drafting and editing text. All AI-assisted code, results and
text were checked by the authors, who take full responsibility for the content.}

\begin{acknowledgments}
T.R.S. is supported by a UK Science and Technology Facilities Council (STFC) studentship (grant number 187357). C.G. acknowledges support from the STFC via the Consolidated Grant UKRI2851. F.F.D. acknowledges support from STFC via the Consolidated Grant ST/X000613.
We thank T.~Totani for helpful correspondence regarding the replication of his
analysis. \rtwo{We thank D.~Hooper for discussions regarding the latitude cut, and R. Crocker for discussions regarding the injected electron templates.} C.G. is a member of the LZ Collaboration. The LZ result discussed in Sec.~\ref{sec:lz} is cited from the public preprint only. This work made use of publicly available data and software. The
\textit{Fermi}--LAT data were obtained free of charge from the NASA Fermi Science
Support Center and reduced with the \texttt{fermitools} package, and the Galactic
interstellar-emission templates were generated using the freely available online
GALPROP tools. We further acknowledge the use of PPPC4DMID for the dark matter
annihilation and decay spectra and the \texttt{emcee} ensemble sampler for the
posterior analysis.
\end{acknowledgments}

\section*{Data Availability}
\secfinal
This work uses publicly available \textit{Fermi}--LAT Pass~8 data,
obtained free of charge from the NASA/GSFC Fermi Science Support Center
(\url{https://fermi.gsfc.nasa.gov/ssc/}), alongside the 4FGL-DR4 point-source
catalogue~\cite{Abdollahi2020,Ballet2023} and the GALPROP-based Galactic
interstellar-emission model~\cite{Ackermann:2012pya}, both publicly distributed.
The GALDEF configuration files for the 14-model interstellar-emission grid
are those released as supplementary material to Ref.~\cite{Ackermann:2012pya},
and the gas and inverse-Compton templates used here are regenerated from them by
the code below.
The prompt annihilation and decay spectra are taken from
PPPC4DMID~\cite{Cirelli2011}. The dSph $J$-factors and
velocity-dependent effective $J$-factors are taken from the published analyses of
Refs.~\cite{Hoof2020,Boddy2019,Boddy2017}. \rtwo{The analysis code and instructions for downloading and processing data to reproduce this analysis are available at
\url{https://github.com/trinitystenhouse/Totani-Reanalysis.git} and archived at Zenodo, \href{https://doi.org/10.5281/zenodo.24626882}{doi:10.5281/zenodo.24626882}.} 
\appendix
\begin{rtwoblock}
\section{Dark matter fits per foreground model and bubble template set}
\label{app:dmfits}
Table~\ref{tab:dmfits} lists the dark matter fit to the halo spectrum of every member of the envelope, for the three
prompt annihilation channels of Sec.~\ref{sec:dmres}.
\input{dm_table}

\section{Halo coefficient induced in the null mocks}
\label{app:nullmocks}
Figure~\ref{fig:nullmocks} shows the halo coefficient and the likelihood gain induced in the noiseless and Poisson
mocks of Sec.~\ref{sec:nullres}, built without a halo.
\begin{figure*}
\includegraphics[width=\textwidth]{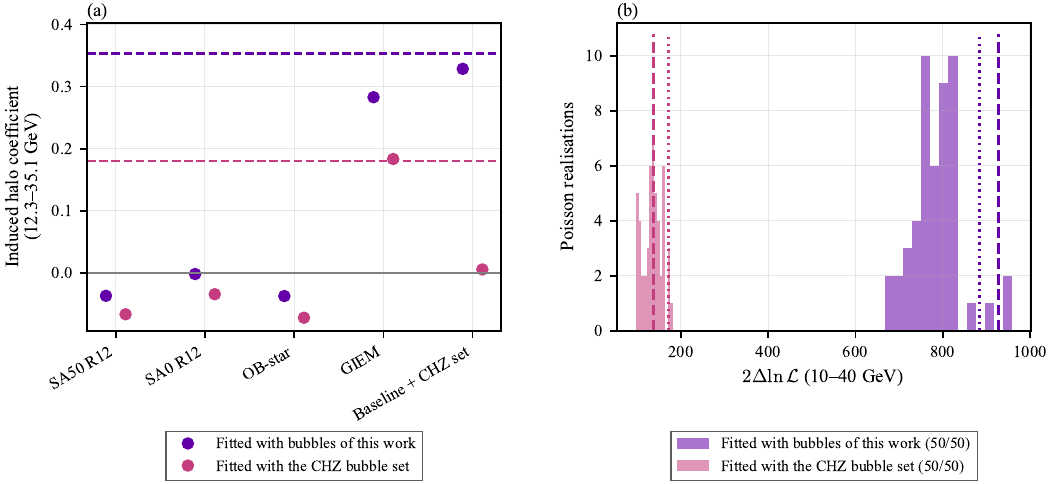}
\caption{\label{fig:nullmocks} Halo coefficient induced by foreground mismatch (Sec.~\ref{sec:nullres}). (a) Mean
halo coefficient over the 12.3--35.1~GeV bins (the 10--40~GeV range of the text) of noiseless mocks built without a
halo from each generating foreground (horizontal axis), fitted with our bubbles (purple) and with the CHZ set (pink).
Dashed lines: the value fitted to the data with each set. (b) $2\Delta\ln\mathcal{L}$ over 10--40~GeV from adding the
halo, for the Poisson realisations of the mock that induces the largest coefficient in each fitter, the baseline with
the CHZ set fitted with our bubbles (purple) and the GIEM fitted with the CHZ set (pink). Dashed lines: the data.
Dotted lines: the 95th percentile of the realisations. The legend gives the realisations completed out of the number
run. Statistical only, template-level.}
\end{figure*}

\section{Posterior correlations of the remaining fits}
\label{app:covrows}
Table~\ref{tab:covariance_rows} gives the correlations at 20.8~GeV for the halo fits not drawn in
Fig.~\ref{fig:covariance}(c).
\input{f_posterior_template_covariance_removed_rows}
\end{rtwoblock}
\clearpage

\bibliography{bib}
\end{document}

%% file: envelope_table.tex
\begin{table*}
\caption{\label{tab:envelope} The fitted $\rho^2$ halo coefficient (pole convention, posterior median; at 20.8~GeV with the 16--84\% interval) for \rtwo{every combination} of foreground and bubble template set, $2\Delta\ln\mathcal{L}$ over 10--40~GeV against the same combination without the halo (a diagnostic, not a significance) and the smallest $N/\tau$ of the halo fit. \rtwo{Every template construction is tested on noiseless mocks built from its own best fit (Sec.~\ref{sec:protocol}). The OB-star, interstellar-emission grid and two three-dimensional CHZ rows use a tested construction on another foreground and have no mock with the halo of their own (Sec.~\ref{sec:envres}).} Only the \rtwo{first two groups enter} the quoted range. \rtwo{The GIEM row is shown outside it} (Sec.~\ref{sec:families}). \rtwo{The three-dimensional models reproduce the morphology but not the amplitude of the published reference comparison (Sec.~\ref{sec:galprop3d}) and are an additional check.} \rtwo{$^\dag$CHZ geometry re-fitted by us.} With our implementation of their method we did not reproduce their best-fit edges (Sec.~\ref{sec:chzres}). \rtwo{The interstellar-emission grid rows are the $|b| \geq 10\degree$ fits of the disk-cut scan with our bubbles re-extracted for each model. They cover only the ten lowest energy bins, so their $N/\tau$ is the smallest over those bins, and they enter the quoted range with the other two-dimensional members.}}
\begin{ruledtabular}
\begin{tabular}{llccccc}
Foreground & Bubbles & 12.3~GeV & 20.8~GeV & 35.1~GeV & $2\Delta\ln\mathcal{L}$ & $N/\tau$ \\
\hline
\multicolumn{7}{l}{\textit{Two-dimensional (quoted range)}} \\
Baseline (Lorimer, $z_h = 6$~kpc) & This work & 0.356 & 0.397$_{-0.022}^{+0.021}$ & 0.307 & 927 & 52 \\
Baseline (Lorimer, $z_h = 6$~kpc) & CHZ, published & 0.172 & 0.236$_{-0.029}^{+0.028}$ & 0.133 & \rtwo{136.5} & 53 \\
OB-star, $z_h = 6$~kpc & This work & 0.392 & 0.402$_{-0.020}^{+0.019}$ & 0.309 & 1236 & 56 \\
OB-star, $z_h = 6$~kpc & CHZ, published & 0.298 & 0.317$_{-0.025}^{+0.026}$ & 0.206 & 419 & 51 \\
\multicolumn{7}{l}{\textit{\rtwo{Two-dimensional interstellar-emission grid (quoted range, ten lowest energy bins)}}} \\
Lorimer, $z_h = 4$~kpc & This work & 0.352 & 0.387$_{-0.021}^{+0.021}$ & 0.296 & 930 & 57 \\
Lorimer, $z_h = 6$~kpc \rtwo{(baseline set-up, grid build)} & This work & 0.357 & 0.398$_{-0.021}^{+0.021}$ & 0.306 & 935 & 54 \\
Lorimer, $z_h = 6$~kpc, $E(B-V)$ cut 5 & This work & 0.345 & 0.386$_{-0.021}^{+0.021}$ & 0.295 & 870 & 61 \\
Lorimer, $z_h = 6$~kpc, $R = 20$~kpc & This work & 0.357 & 0.397$_{-0.022}^{+0.021}$ & 0.306 & 933 & 61 \\
Lorimer, $z_h = 6$~kpc, $T_S = 10^5$~K & This work & 0.358 & 0.396$_{-0.021}^{+0.022}$ & 0.305 & 919 & 59 \\
Lorimer, $z_h = 8$~kpc & This work & 0.359 & 0.403$_{-0.022}^{+0.021}$ & 0.312 & 922 & 56 \\
Lorimer, $z_h = 10$~kpc & This work & 0.360 & 0.406$_{-0.022}^{+0.021}$ & 0.315 & 915 & 53 \\
SNR, $z_h = 4$~kpc & This work & 0.388 & 0.402$_{-0.020}^{+0.020}$ & 0.308 & 1203 & 50 \\
SNR, $z_h = 6$~kpc & This work & 0.388 & 0.406$_{-0.020}^{+0.020}$ & 0.311 & 1194 & 53 \\
SNR, $z_h = 10$~kpc & This work & 0.386 & 0.409$_{-0.020}^{+0.020}$ & 0.314 & 1162 & 60 \\
Yusifov, $z_h = 4$~kpc & This work & 0.366 & 0.394$_{-0.020}^{+0.020}$ & 0.301 & 1036 & 55 \\
Yusifov, $z_h = 6$~kpc & This work & 0.368 & 0.401$_{-0.021}^{+0.020}$ & 0.308 & 1027 & 62 \\
Yusifov, $z_h = 10$~kpc & This work & 0.369 & 0.408$_{-0.021}^{+0.021}$ & 0.314 & 1002 & 57 \\
\multicolumn{7}{l}{\textit{Two-dimensional, outside the range}} \\
GIEM (\texttt{gll\_iem\_v07}) & none & 0.128 & 0.206$_{-0.022}^{+0.022}$ & 0.172 & 191 & 131 \\
\multicolumn{7}{l}{\textit{Three-dimensional (additional check)}} \\
SA0 & This work & 0.373 & 0.386$_{-0.020}^{+0.021}$ & 0.291 & 1032 & 50 \\
SA0 F98 & This work & 0.366 & 0.381$_{-0.020}^{+0.021}$ & 0.287 & 976 & 54 \\
SA0 R12 & This work & 0.370 & 0.384$_{-0.020}^{+0.021}$ & 0.290 & 1006 & 51 \\
SA0 R12 & CHZ, published & 0.296 & 0.316$_{-0.026}^{+0.027}$ & 0.198 & 371 & 50 \\
SA0 R12 & CHZ, re-fitted$^\dag$ & 0.171 & 0.227$_{-0.025}^{+0.026}$ & 0.143 & 162 & 50 \\
SA100 & This work & 0.418 & 0.415$_{-0.019}^{+0.019}$ & 0.321 & 1415 & 51 \\
SA100 F98 & This work & 0.420 & 0.418$_{-0.019}^{+0.019}$ & 0.323 & 1428 & 51 \\
SA100 R12 & This work & 0.419 & 0.415$_{-0.020}^{+0.019}$ & 0.320 & 1415 & 52 \\
SA50 & This work & 0.401 & 0.403$_{-0.020}^{+0.020}$ & 0.309 & 1249 & 52 \\
SA50 F98 & This work & 0.398 & 0.404$_{-0.020}^{+0.020}$ & 0.308 & 1223 & 53 \\
SA50 F98 & CHZ, published & 0.304 & 0.319$_{-0.026}^{+0.027}$ & 0.204 & 405 & 50 \\
SA50 F98 & CHZ, re-fitted$^\dag$ & 0.166 & 0.218$_{-0.026}^{+0.026}$ & 0.137 & 149 & 56 \\
SA50 R12 & This work & 0.397 & 0.401$_{-0.020}^{+0.020}$ & 0.307 & 1221 & 53 \\
SA50 R12 & CHZ, published & 0.309 & 0.321$_{-0.026}^{+0.026}$ & 0.205 & 419 & 52 \\
SA50 R12 & CHZ, re-fitted$^\dag$ & 0.169 & 0.218$_{-0.025}^{+0.025}$ & 0.142 & 162 & 55 \\
SA50, 2D gas & This work & 0.404 & 0.411$_{-0.020}^{+0.020}$ & 0.316 & 1267 & 51 \\
\end{tabular}
\end{ruledtabular}
\end{table*}

%% file: dm_table.tex
\begin{table*}
\caption{\label{tab:dmfits} \rtwo{Dark matter fit to the pixelwise $\rho^2$ halo spectrum of each member, for prompt $s$-wave annihilation spectra (PPPC) to $b\bar b$, $W^+W^-$ and $\tau^+\tau^-$. The fits are statistical only and template-level. $\mathcal{R}$ is the best-fit cross section divided by the dwarf upper limit at the same mass (the convention of Sec.~\ref{sec:tension}), as a point estimate, with the band from the $J$-factor uncertainties given in Sec.~\ref{sec:tension}. $^\ddag$ The published baseline fit. Values of $\chi^2/\nu$ are in parentheses for members fitted over ten or fewer bins, whose $\chi^2/\nu$ is not comparable with the others. $^\dag$CHZ geometry re-fitted by us. The GIEM row has no separate bubble template. $\leq$ marks a best-fit mass at the lower edge of the scanned range.}}
\begin{ruledtabular}
\begin{tabular}{llccccccccc}
& & & \multicolumn{3}{c}{$b\bar b$} & \multicolumn{3}{c}{$W^+W^-$} & \multicolumn{2}{c}{$\tau^+\tau^-$} \\
\cline{4-6}\cline{7-9}\cline{10-11}
Foreground & Bubbles & Bins & $m$ (GeV) & $\mathcal{R}$ & $\chi^2/\nu$ & $m$ (GeV) & $\mathcal{R}$ & $\chi^2/\nu$ & $m$ (GeV) & $\chi^2/\nu$ \\
\hline
\multicolumn{11}{l}{\textit{Two-dimensional (quoted range)}} \\
Baseline (Lorimer, $z_h = 6$~kpc)$^\ddag$ & This work & 12 & 716 & 5.17 & 7.5 & 557 & 4.23 & 7.4 & 70 & 14.6 \\
Baseline (Lorimer, $z_h = 6$~kpc) & CHZ, published & 11 & 697 & 2.77 & 1.8 & 527 & 2.25 & 1.6 & 60 & 3.8 \\
OB-star, $z_h = 6$~kpc & This work & 13 & 388 & 6.96 & 7.8 & 299 & 5.57 & 5.2 & 40 & 40.5 \\
OB-star, $z_h = 6$~kpc & CHZ, published & 13 & 149 & 6.77 & 12.0 & 139 & 5.11 & 8.6 & $\leq 20$ & 40.0 \\
\multicolumn{11}{l}{\textit{Two-dimensional interstellar-emission grid (quoted range)}} \\
Lorimer, $z_h = 4$~kpc & This work & 9 & 657 & 4.93 & (7.7) & 498 & 3.90 & (8.1) & 70 & (19.0) \\
Lorimer, $z_h = 6$~kpc (grid build) & This work & 9 & 707 & 5.10 & (9.4) & 527 & 4.08 & (9.9) & 70 & (19.3) \\
Lorimer, $z_h = 6$~kpc, $E(B-V)$ cut 5 & This work & 9 & 726 & 5.01 & (10.3) & 557 & 4.04 & (10.8) & 70 & (17.9) \\
Lorimer, $z_h = 6$~kpc, $R = 20$~kpc & This work & 9 & 707 & 5.09 & (9.5) & 537 & 4.12 & (10.0) & 70 & (19.1) \\
Lorimer, $z_h = 6$~kpc, $T_S = 10^5$~K & This work & 9 & 677 & 4.91 & (8.5) & 508 & 4.00 & (8.9) & 70 & (18.8) \\
Lorimer, $z_h = 8$~kpc & This work & 9 & 736 & 5.30 & (11.3) & 557 & 4.20 & (11.8) & 80 & (19.8) \\
Lorimer, $z_h = 10$~kpc & This work & 9 & 746 & 5.35 & (12.2) & 577 & 4.29 & (12.7) & 80 & (19.3) \\
SNR, $z_h = 4$~kpc & This work & 10 & 448 & 6.76 & (9.7) & 348 & 5.19 & (7.7) & 50 & (42.8) \\
SNR, $z_h = 6$~kpc & This work & 10 & 488 & 6.92 & (10.5) & 368 & 5.07 & (9.1) & 60 & (36.3) \\
SNR, $z_h = 10$~kpc & This work & 10 & 557 & 6.38 & (13.0) & 428 & 4.85 & (12.3) & 60 & (29.7) \\
Yusifov, $z_h = 4$~kpc & This work & 9 & 577 & 5.95 & (5.4) & 428 & 4.59 & (5.1) & 60 & (23.0) \\
Yusifov, $z_h = 6$~kpc & This work & 9 & 617 & 5.60 & (6.2) & 458 & 4.39 & (6.1) & 60 & (22.0) \\
Yusifov, $z_h = 10$~kpc & This work & 9 & 667 & 5.15 & (8.1) & 508 & 4.17 & (8.3) & 70 & (21.0) \\
\multicolumn{11}{l}{\textit{Outside the quoted range}} \\
SA0 & This work & 13 & 438 & 6.40 & 7.6 & 348 & 4.99 & 5.9 & 50 & 31.0 \\
SA0 F98 & This work & 13 & 488 & 6.55 & 7.6 & 368 & 4.80 & 6.2 & 60 & 26.2 \\
SA0 R12 & This work & 13 & 438 & 6.38 & 7.4 & 338 & 4.89 & 5.7 & 50 & 30.2 \\
SA0 R12 & CHZ, published & 13 & 129 & 7.21 & 13.2 & 119 & 5.42 & 9.9 & $\leq 20$ & 43.4 \\
SA100 & This work & 13 & 269 & 7.80 & 19.4 & 219 & 6.22 & 12.8 & 30 & 81.6 \\
SA100 F98 & This work & 13 & 279 & 7.95 & 18.2 & 229 & 6.17 & 11.8 & 30 & 79.4 \\
SA100 R12 & This work & 13 & 269 & 7.82 & 19.6 & 219 & 6.24 & 12.9 & 30 & 83.3 \\
SA50 & This work & 13 & 328 & 7.59 & 11.7 & 259 & 5.69 & 7.4 & 40 & 55.0 \\
SA50 F98 & This work & 13 & 348 & 7.45 & 10.6 & 269 & 5.69 & 6.8 & 40 & 51.2 \\
SA50 F98 & CHZ, published & 13 & 119 & 7.77 & 15.6 & 110 & 5.48 & 11.9 & $\leq 20$ & 52.4 \\
SA50 R12 & This work & 13 & 318 & 7.57 & 11.9 & 259 & 5.67 & 7.6 & 40 & 55.5 \\
SA50 R12 & CHZ, published & 13 & 110 & 8.01 & 17.1 & 110 & 5.71 & 13.2 & $\leq 20$ & 58.9 \\
SA50, 2D gas & This work & 13 & 368 & 7.35 & 9.4 & 289 & 5.79 & 5.8 & 40 & 46.8 \\
SA0 R12 & CHZ, re-fitted$^\dag$ & 5 & 438 & 3.37 & (2.0) & 309 & 2.70 & (1.5) & 60 & (6.8) \\
SA50 F98 & CHZ, re-fitted$^\dag$ & 5 & 408 & 3.24 & (1.9) & 289 & 2.70 & (1.5) & 60 & (6.3) \\
SA50 R12 & CHZ, re-fitted$^\dag$ & 4 & 547 & 3.01 & (1.0) & 388 & 2.28 & (1.0) & 90 & (2.8) \\
GIEM (\texttt{gll\_iem\_v07}) & none & 13 & 169 & 3.10 & 17.3 & 169 & 2.36 & 14.6 & $\leq 20$ & 31.4 \\
\end{tabular}
\end{ruledtabular}
\end{table*}

%% file: f_posterior_template_covariance_removed_rows.tex
\begin{table*}
\caption{\label{tab:covariance_rows} \rtwo{Posterior correlation $r$ between the halo normalisation and each template at 20.8~GeV for the halo fits that Fig.~\ref{fig:covariance}(a) does not show: the three-dimensional GALPROP foreground models with our bubbles other than the three drawn there, and every three-dimensional fit with the CHZ structures (``CHZ'' at the published geometry, $\dagger$ re-fitted by us, a check only, Sec.~\ref{sec:families}). A dash marks a template that is not in that fit. Same convention and source as the figure.}}
\begin{ruledtabular}
\begin{tabular}{lccccccc}
Fit & ICS & Gas & Bubbles (+) & Bubbles ($-$) & Flat (CHZ) & Inner (CHZ) & Cocoons (CHZ) \\
\hline
SA0, ours & $-$0.40 & $-$0.16 & $-$0.43 & $-$0.08 & -- & -- & -- \\
SA0 R12, ours & $-$0.41 & $-$0.17 & $-$0.44 & $-$0.06 & -- & -- & -- \\
SA0 R12, CHZ & $-$0.46 & $-$0.05 & -- & -- & $-$0.65 & 0.49 & $-$0.20 \\
SA0 R12, CHZ$^\dagger$ & $-$0.49 & $-$0.05 & -- & -- & $-$0.54 & 0.25 & $-$0.40 \\
SA50, ours & $-$0.36 & $-$0.19 & $-$0.42 & $-$0.06 & -- & -- & -- \\
SA50 F98, ours & $-$0.38 & $-$0.17 & $-$0.41 & $-$0.06 & -- & -- & -- \\
SA50 F98, CHZ & $-$0.45 & $-$0.03 & -- & -- & $-$0.67 & 0.52 & $-$0.20 \\
SA50 F98, CHZ$^\dagger$ & $-$0.48 & $-$0.08 & -- & -- & $-$0.53 & 0.23 & $-$0.44 \\
SA50 R12, ours & $-$0.37 & $-$0.18 & $-$0.42 & $-$0.07 & -- & -- & -- \\
SA50 R12, CHZ & $-$0.45 & $-$0.04 & -- & -- & $-$0.67 & 0.51 & $-$0.19 \\
SA50 R12, CHZ$^\dagger$ & $-$0.48 & $-$0.07 & -- & -- & $-$0.50 & 0.17 & $-$0.42 \\
SA100, ours & $-$0.33 & $-$0.20 & $-$0.39 & $-$0.02 & -- & -- & -- \\
SA100 R12, ours & $-$0.34 & $-$0.20 & $-$0.40 & $-$0.05 & -- & -- & -- \\
\end{tabular}
\end{ruledtabular}
\end{table*}